\documentclass[twocolumn]{aastex631}
\usepackage[T1]{fontenc}
\usepackage{graphicx}
\usepackage[]{natbib}
\usepackage{amsmath, multirow}
\usepackage{hyperref}
\usepackage{longtable}
\usepackage[colorinlistoftodos]{todonotes}

\newcommand{\gtsimeq}{\raisebox{-0.6ex}{$\,\stackrel
        {\raisebox{-.2ex}{$\textstyle >$}}{\sim}\,$}}
\newcommand{\hst}{\textit{HST}}
\newcommand{\jwst}{\textit{JWST}}

\newcommand{\um}{$\mu$m}
\newcommand{\lya}{Lyman-$\alpha$}

\newcommand{\Oiii}{[O~{\sc iii}]}

\newcommand{\Ha}{H$\alpha$}
\newcommand{\Hb}{H$\beta$}

\begin{document}
\title{A Steep Decline in the Galaxy Space Density Beyond Redshift 9 in the CANUCS UV Luminosity Function}

\shorttitle{The CANUCS UV Luminosity Function}
\shortauthors{Willott et al.}

\author[0000-0002-4201-7367]{Chris J. Willott}
\affiliation{NRC Herzberg, 5071 West Saanich Rd, Victoria, BC V9E 2E7, Canada}
\email{chris.willott@nrc.ca}

\author[0000-0001-8325-1742]{Guillaume Desprez}
\affiliation{Department of Astronomy and Physics and Institute for Computational Astrophysics, Saint Mary's University, 923 Robie Street, Halifax, Nova Scotia B3H 3C3, Canada}

\author[0000-0003-3983-5438]{Yoshihisa Asada}
\affiliation{Department of Astronomy and Physics and Institute for Computational Astrophysics, Saint Mary's University, 923 Robie Street, Halifax, Nova Scotia B3H 3C3, Canada}
\affiliation{Department of Astronomy, Kyoto University, Sakyo-ku, Kyoto 606-8502, Japan}

\author[0000-0001-8830-2166]{Ghassan T. E. Sarrouh}
\affiliation{Department of Physics and Astronomy, York University, 4700 Keele St. Toronto, Ontario, M3J 1P3, Canada}

\author[0000-0002-4542-921X]{Roberto Abraham}
\affiliation{David A. Dunlap Department of Astronomy and Astrophysics, University of Toronto, 50 St. George Street, Toronto, Ontario, M5S 3H4, Canada}

\author[0000-0001-5984-0395]{Maru\v{s}a Brada{\v c}}
\affiliation{University of Ljubljana, Department of Mathematics and Physics, Jadranska ulica 19, SI-1000 Ljubljana, Slovenia}
\affiliation{Department of Physics and Astronomy, University of California Davis, 1 Shields Avenue, Davis, CA 95616, USA}

\author[0000-0003-2680-005X]{Gabe Brammer}
\affiliation{Cosmic Dawn Center (DAWN), Denmark}
\affiliation{Niels Bohr Institute, University of Copenhagen, Jagtvej 128, DK-2200 Copenhagen N, Denmark}

\author[0000-0001-8489-2349]{Vince Estrada-Carpenter}
\affiliation{Department of Astronomy and Physics and Institute for Computational Astrophysics, Saint Mary's University, 923 Robie Street, Halifax, Nova Scotia B3H 3C3, Canada}

\author[0000-0001-9298-3523]{Kartheik G. Iyer}
\affiliation{Columbia Astrophysics Laboratory, Columbia University, 550 West 120th Street, New York, NY 10027, USA}

\author[0000-0003-3243-9969]{Nicholas S. Martis}
\affiliation{University of Ljubljana, Department of Mathematics and Physics, Jadranska ulica 19, SI-1000 Ljubljana, Slovenia}
\affiliation{Department of Astronomy and Physics and Institute for Computational Astrophysics, Saint Mary's University, 923 Robie Street, Halifax, Nova Scotia B3H 3C3, Canada}
\affiliation{NRC Herzberg, 5071 West Saanich Rd, Victoria, BC V9E 2E7, Canada}

\author[0000-0002-7547-3385]{Jasleen Matharu}
\affiliation{Cosmic Dawn Center (DAWN), Denmark}
\affiliation{Niels Bohr Institute, University of Copenhagen, Jagtvej 128, DK-2200 Copenhagen N, Denmark}

\author[0000-0002-8530-9765]{Lamiya Mowla}
\affiliation{Whitin Observatory, Department of Physics and Astronomy, Wellesley College, 106 Central Street, Wellesley, MA 02481, USA}

\author[0000-0002-9330-9108]{Adam Muzzin}
\affiliation{Department of Physics and Astronomy, York University, 4700 Keele St. Toronto, Ontario, M3J 1P3, Canada}

\author{Gaël Noirot}
\affiliation{Department of Astronomy and Physics and Institute for Computational Astrophysics, Saint Mary's University, 923 Robie Street, Halifax, Nova Scotia B3H 3C3, Canada}

\author[0000-0002-7712-7857]{Marcin Sawicki}
\affiliation{Department of Astronomy and Physics and Institute for Computational Astrophysics, Saint Mary's University, 923 Robie Street, Halifax, Nova Scotia B3H 3C3, Canada}

\author[0000-0002-6338-7295]{Victoria Strait}
\affiliation{Cosmic Dawn Center (DAWN), Denmark}
\affiliation{Niels Bohr Institute, University of Copenhagen, Jagtvej 128, DK-2200 Copenhagen N, Denmark}

\author[0009-0009-4388-898X]{Gregor Rihtar\v{s}i\v{c}}
\affiliation{University of Ljubljana, Department of Mathematics and Physics, Jadranska ulica 19, SI-1000 Ljubljana, Slovenia}

\author[0009-0000-8716-7695]{Sunna Withers}
\affiliation{Department of Physics and Astronomy, York University, 4700 Keele St. Toronto, Ontario, M3J 1P3, Canada}


\begin{abstract}
We present a new sample of 158 galaxies at redshift $z>7.5$ selected from deep \jwst\ NIRCam imaging of five widely-separated sightlines in the CANUCS survey. Two-thirds of the pointings and 80\% of the galaxies are covered by 12 to 14 NIRCam filters, including seven to nine medium bands, providing accurate photometric redshifts and robustness against low redshift interlopers. A sample of 28 galaxies at $z>7.5$ with spectroscopic redshifts shows a low systematic offset and scatter in the difference between photometric and spectroscopic redshifts. We derive the galaxy UV luminosity function at redshifts 8 to 12, finding a slightly higher normalization than previously seen with \hst\ at redshifts 8 to 10. We observe a steeper decline in the galaxy space density from $z=8$ to $12$ than found by most \jwst\ Cycle 1 studies. In particular, we find only eight galaxies at $z>10$ and none at $z>12.5$, with no $z>10$ galaxies brighter than F277W AB=28 or $M_{\rm UV}=-20$ in our unmasked, delensed survey area of 53.4 square arcminutes. We attribute the lack of bright $z>10$ galaxies in CANUCS compared to GLASS and CEERS to intrinsic variance in the galaxy density along different sightlines. The evolution in the CANUCS luminosity function between $z=8$ and $12$ is comparable to that predicted by simulations that assume a standard star formation efficiency, without invoking any special adjustments.  
\end{abstract}


\section{Introduction}
\label{sec:intro}

Tracking the rise of the first galaxies to emerge from the earliest matter overdensities is a key goal of modern cosmology. The early evolution of dark matter halos massive enough to allow gas to cool and form stars is well established. Much more uncertain, however, are the details of the baryonic processes that enable star formation within these halos. With the launch of the {\it James Webb Space Telescope} (\jwst, \citealt{Gardner2023}) we are now in a position to make imaging and spectroscopic observations of galaxies out to unprecedented distances to investigate this topic.

There have been a flurry of new discoveries during the first year of \jwst\ observations. From multi-band NIRCam imaging, galaxies with photometric redshift estimates as high as 18 have been found \citep{Atek2023, Castellano2023, Donnan2023, Yan2023, Harikane2023a, Finkelstein2023a, Bouwens2023a, Hainline2023arXiv}. The highest redshift spectroscopically-confirmed \jwst-selected galaxy is at a somewhat lower $z=13.2$ \citep{Robertson2023, Curtis-Lake2023}. 

Pure photometric selection by the Lyman break color and/or photometric redshift techniques must be treated with caution. This was highlighted by the widely different samples determined by different teams from the first public \jwst\ datasets (see e.g. \citealt{Bouwens2023a} and \citealt{Adams2023arXiv}). The poster child for the dangers of such selection is the galaxy CEERS-93316 that was determined by several independent teams to be a plausible bright galaxy at a redshift between 15 and 17. An alternative low-redshift solution for this galaxy had been suggested based on a tentative sub-millimetre detection and a $z\approx 5$ overdensity in the field \citep{Zavala2023, Naidu2022arXiv}. Subsequent NIRSpec spectroscopy showed it to be a $z=4.9$ galaxy with a red continuum and strong emission lines \citep{ArrabalHaro2023a}. At this particular redshift, all four NIRCam Long-Wave filters (F277W, F356W, F410M and F444W) contained a strong emission line that mimicked a $z>15$ blue ultraviolet (UV) continuum. Another potential issue with some of the early \jwst\ surveys is they did not include key filters (e.g. F115W and F410M in SMACS J0723.3-7327, F090W in CEERS, F090W, F200W, F356W and F410M in COSMOS-Web) whose absence may cause increased uncertainty in photometric redshifts and provide weaker discrimination from low redshift interlopers.

Despite the uncertainties associated with photometric high-redshift galaxy selection, these samples have been used to derive the galaxy UV luminosity function \citep{Naidu2022, Castellano2023, Donnan2023, Harikane2023a, Morishita2023a, Finkelstein2023a, Bouwens2023a, Adams2023arXiv, Perez-Gonzalez2023, McLeod2024, Leung2023, Franco2023arXiv, Casey2023arXiv, Finkelstein2023arXiv}. It is generally agreed that there are more galaxies at $z>10$, particularly UV-luminous ones, than had been expected from studies with the {\it Hubble Space Telescope} (\hst) and from simulations that assume a constant star-formation efficiency from gas accretion onto dark matter halos \citep{Bouwens2021}.

Various explanations for this excess have been proposed. Two of the earliest fields to be observed, GLASS and CEERS, may be over-dense in bright $z \gtsimeq 10$ galaxies compared to other fields. This is certainly true for GLASS \citep{Castellano2023}, but has not yet been reported for CEERS. These two fields feature prominently in a large fraction of the luminosity function studies mentioned in the previous paragraph. As shown by \cite{Steinhardt2021}, cosmic variance can be very significant at these redshifts and many independent sightlines are required to get a true picture of the Universe. Another issue pointed out by \cite{Bouwens2023a} is  that because photometric redshifts are often overestimated in noisy data \citep{Fujimoto2023, ArrabalHaro2023b, Serjeant2023}, offsets in both luminosity and redshift can have a significant impact on the derived luminosity function. There is also the issue of whether the UV luminosity from these bright galaxies is powered by star formation or may include a significant contribution from an Active Galactic Nucleus (AGN), as is hypothesized for two of the most luminous known $z>10$ galaxies \citep{Maiolino2023arXiv, Bogdan2024}. However, the majority of $z>10$ galaxy candidates in CEERS are spatially-resolved, suggesting minimal AGN contribution to their UV luminosities \citep{Finkelstein2023arXiv}. 

If there is a real increase in the space density of $z>10$ UV-luminous star-forming galaxies, several explanations have been proposed. There may be increased scatter in the halo mass -- star-formation rate relation due to bursty star formation episodes \citep{Endsley2023arXiv, Sun2023}, or a bias towards observing rapidly-formed halos \citep{Mason2023}. The Initial Mass Function (IMF) of stars may shift to be more top-heavy at these redshifts due to extremely low metallicities \citep{Inayoshi2022}. The lack of dust at $z>10$ may also be a factor in enhancing the UV luminosity without increasing the star-formation rate \citep{Mason2023, Ferrara2023, Ferrara2023arXiv} with observations showing extremely blue UV slopes at $z>10$ \citep{Topping2023arXiv}.

In this paper we present a new derivation of the UV luminosity function from $z=8$ to $12$. We use data from the CAnadian NIRISS Unbiased Cluster Survey (CANUCS, \citealt{Willott2022}), that has not been used in any of the previous luminosity function studies\footnote{The sample of $z>7.5$ galaxies used by \cite{Adams2023arXiv} includes MACS0416 NIRCam data from the PEARLS survey that has a small overlap ($\approx 3$ square arcmin) with the CANUCS area used in this paper.}. Observing ten fields along five independent sightlines, the CANUCS data set is far less prone to effects of  clustering variance, referred to hereafter as cosmic variance, than previous studies whose samples were dominated by one or two fields. The total NIRCam imaging dual filter integration time in CANUCS was 123 hours, making it one of the most comprehensive extragalactic NIRCam imaging projects executed in \jwst\ Cycle 1.

The unique aspect of CANUCS amongst Cycle 1 programs is the use in some observations of 14 NIRCam filters including nine medium bands. These medium bands sample the spectral energy distribution (SED) more finely than the Wide filters, producing higher quality photometric redshifts and masses \citep{Sarrouh2024arXiv}. As shown by \cite{Withers2023}, the medium bands are excellent at detecting strong emission lines from extreme emission line galaxies (EELGs), one of the potential low-redshift contaminants of photometric high-redshift galaxy selection.

In Section 2 we present the CANUCS data used in this work. Section 3 describes the selection of high-redshift galaxies including potential contaminants, spectroscopic verification and completeness. In Section 4 we present our results. Section 5 discusses these in the context of previous work and theoretical simulations. In Section 6 we give our conclusions. We assume a flat $\Lambda$CMD cosmology, with $\Omega_\Lambda=0.7$, $\Omega_{\rm m}=0.3$, and $H_0=70~{\rm km\,s^{-1}\,Mpc^{-1}}$. Magnitudes are given in the AB system \citep{Oke1983}.  

\section{Data}
\label{sec:data}

\subsection{Imaging}
\label{sec:imaging}

We briefly describe the CANUCS imaging data here, noting that the data used are exactly the same as described by \cite{Desprez2023arXiv}, where the reader may find further information. The CANUCS survey targets five strong gravitational-lensing clusters: Abell 370, MACS0416, MACS0417, MACS1149 and MACS1423. For each cluster there are three pointings: the central cluster field (CLU), a NIRCam Flanking Field (NCF) and a NIRISS Flanking Field (NSF). The CLU fields were observed with both NIRISS \citep{Doyon2023} and NIRCam \citep{Rieke2023}, whereas the NCF and NSF fields were observed with only one of the two instruments. In this work we search for high-redshift galaxies in the CLU and NCF fields of all five clusters. Each CLU and NCF field contains both NIRCam modules A and B, so in total CANUCS observed 20 NIRCam modules, the same as the CEERS survey, but CANUCS uses longer exposures per filter and more filters per field.

The NCF fields used 14 NIRCam filters: F090W, F115W, F140M, F150W, F162M, F182M, F210M, F250M, F277W, F300M, F335M, F360M, F410M and F444W. The exposure times are mostly longer than in the CLU fields with 10 filters having $10.3\,{\rm ks}$ and 4 filters (F140M, F162M, F250M, F300M) having $5.7\,{\rm ks}$. 
For MACS1149 a program definition error caused the F162M and F250M filters to not be observed; however this is compensated for by deeper observations in F150W and F277W at similar wavelengths. The NIRCam 3$\sigma$ point source limiting magnitudes in the NCF fields are between 29.4 and 30.0 for most filters. The NCF NIRCam imaging used the NIRISS WFSS Large dither pattern, which has dithers of about 1 arcsec. This provides the most uniform NIRCam depth, but does not dither over the SW detector gaps. Consequently, each NCF field has an area covered in the SW and LW filters of 8.8 square arcmin. Table \ref{tab:datancf} lists the limiting flux-densities for all the filters in the NCF fields.

The CLU fields were observed with NIRCam F090W, F115W, F150W, F200W, F277W, F356W, F410M, and F444W filters with exposure times of $6.4\,{\rm ks}$ each. The CLU NIRCam module B fields (centred on the core of the lensing clusters) were also covered by NIRISS with the F115W, F150W and F200W filters with direct imaging ($2.3\,{\rm ks}$ per filter) and the GR150R and GR150C grisms ($19.2\,{\rm ks}$ per filter). The NIRCam 3$\sigma$ point source limiting magnitudes in the CLU fields are between 29.1 to 29.8. The CLU NIRCam imaging used the INTRAMODULEX-6 dither pattern to dither over the Short-Wave (SW) detector gaps and image each field with area 11.1 square arcmin in the SW and Long-Wave (LW) filters. Table \ref{tab:dataclu} lists the limiting flux-densities for all the filters in the CLU fields.

\begin{deluxetable*}{c c c c c c}
\hspace{-0.5cm}
\tablewidth{220pt}
\tablecaption{\label{tab:datancf} Summary of \jwst\ and \hst\ data in CANUCS NCF Fields}
\tablehead{Filter & Abell 370 & MACS0416 & MACS0417 & MACS1149 & MACS1423}
\startdata
\hline
F090W &	3.9 & 4.2 & 4.8   & 3.8 & 4.4   \\
F115W &	3.9 & 4.2 & 4.6   & 3.9 & 4.3 	\\
F140M &	7.7 & 7.6 & 8.8   & 7.8 & 7.5 	\\
F150W &	3.4 & 3.5 & 3.8   & 2.8 & 3.5 	\\
F162M &	7.1 & 7.4 & 8.2   & --  & 7.2 	\\
F182M &	4.0 & 4.4 & 4.5   & 4.2 & 4.3 	\\
F210M &	4.9 & 4.9 & 5.6   & 4.8 & 4.8 	\\
F250M &	7.3 & 7.8 & 8.8   & --  & 7.0 	\\
F277W &	2.1 & 2.0 & 2.5   & 1.7 & 1.9 	\\
F300M &	5.0 & 4.9 & 6.2   & 5.1 & 4.7 	\\
F335M &	3.9 & 3.9 & 4.1   & 3.4 & 3.5 	\\
F360M &	3.9 & 3.9 & 4.1   & 3.8 & 3.5 	\\
F410M &	4.1 & 4.0 & 4.4   & 4.0 & 3.9 	\\
F444W &	3.3 & 3.1 & 3.8   & 3.2 & 3.1 	\\
\hline
F435W &	3.4 & 3.1 & 13.3* & 3.2 & 14.3*	\\
F606W &	3.0 & 4.0 & 6.9*  & 2.9 & 7.8* 	\\
F814W &	3.0 & 2.6 & --    & 2.8 & --	\\
\hline
Area (sq. arcmin)\dag & 7.11 & 6.72 & 7.01 & 8.00 & 7.31\\
\% HST\ddag           & 60   &  50  &  74   & 61   &  66 \\
\enddata
\tablenotetext{}{For each filter, limiting flux-densities are quoted in nJy. These are 3$\sigma$ in a 0.3 arcsec diameter aperture, determined by measuring the distribution of fluxes in apertures in empty sky regions on the F444W-PSF-convolved images. All \jwst\ data are from NIRCam. All \hst\ filters are ACS except those marked with an asterisk are WFC3/UV. This instrument does not have a F435W filter, so the F438W filter was used instead.\\ 
\dag The unmasked areas of each field are quoted. These are not corrected for gravitational lensing.\\
\ddag The percentage of the NIRCam area covered by deep HST optical (ACS or WFC3/UV) imaging. This does not include area covered by the BUFFALO survey that is too shallow to be relevant for this work.}
\end{deluxetable*}

\begin{deluxetable*}{c c c c c c}
\hspace{-0.5cm}
\tablewidth{220pt}
\tablecaption{\label{tab:dataclu} Summary of \jwst\ and \hst\ data in CANUCS CLU Module A Fields}
\tablehead{Filter & Abell 370 & MACS0416 & MACS0417 & MACS1149 & MACS1423}
\startdata
\hline
F090W &	5.7 & 6.7  &  8.4  & 6.2 & 7.2 \\
F115W &	6.0 & 7.0  &  6.1  & 6.1 & 6.6 \\
F150W &	5.4 & 5.5  &  6.4  & 5.3 & 5.2 \\
F200W &	4.9 & 4.6  &  5.8  & 4.6 & 4.4 \\
F277W &	3.0 & 3.1  &  4.2  & 3.0 & 3.0 \\
F356W &	3.1 & 2.9  &  3.7  & 3.0 & 2.9 \\
F410M &	5.9 & 5.7  &  7.8  & 5.7 & 5.5 \\
F444W &	4.8 & 4.7  &  5.3  & 4.6 & 4.3 \\
\hline 	      	      
F435W &	3.7 & 3.0  & 7.0  & -- & 6.1 \\
F606W &	3.5 & 2.6  & 5.0  & -- & 4.3 \\
F814W &	2.8 & 2.4  & 19.0 & -- & 8.9 \\
\hline
Area (sq. arcmin)\dag & 4.77 & 4.95 & 4.23 & 5.30 & 4.79 \\
\% HST\ddag           &  1  &  8  &  24    &  0  & 32 \\
\enddata
\tablenotetext{}{For each filter, limiting flux-densities are quoted in nJy. These are 3$\sigma$ in a 0.3 arcsec diameter aperture, determined by measuring the distribution of fluxes in apertures in empty sky regions on the F444W-PSF-convolved images. All \jwst\ data are from NIRCam. All \hst\ filters are ACS.\\
\dag The unmasked areas of each field are quoted. These are not corrected for gravitational lensing.\\
\ddag The percentage of the NIRCam area covered by deep HST ACS optical imaging. This does not include area covered by the BUFFALO survey that is too shallow to be relevant for this work.} \end{deluxetable*}

We include the available \hst\ optical observations in ACS F435W, F606W, F814W or WFC3/UV F438W, F606W. However, we note that the \hst\ coverage is quite inhomogeneous in fields outside the central cluster field. All the NCF fields have at least partial overlap from either the Hubble Frontier Fields Parallels \citep{Lotz2017} or program HST-GO-16667 PI Brada\v{c}. Whilst CANUCS does include the available WFC3/IR data, these data are not used in this work since the \jwst\ data is superior. The typical depth and fractional coverage of the \hst\ optical observations are given in Tables \ref{tab:datancf} and \ref{tab:dataclu}. For the CLU fields, only module A is considered in Table \ref{tab:dataclu}, since only that module is used for the luminosity function as explained below.

Full details of the CANUCS image data processing and photometry will be presented in a future work. The imaging reductions are similar to as described in \citet{Noirot2023} and photometry is as described in \cite{Asada2024}. The images were processed with a combination of the official STScI \jwst\ pipeline and {\tt grizli}. All images were aligned to Gaia DR3 astrometry and drizzled onto the same pixel scale of 40 milliarcsec/pixel. An empirical point spread function for each filter was generated from isolated stars in the field. The images were convolved with kernels to match the PSFs in all images to that of the lowest resolution image, F444W.

Source detection was performed on $\chi_{\rm mean}$-detection images \citep{Drlica-Wagner2018} made by combining all available images without any PSF convolution. Sources were detected using a two-mode {\it hot+cold} detection strategy. Photometry was corrected for Galactic extinction using the \citet{Fitzpatrick1999} dust-attenuation law with $R_V=3.1$. For galaxy colors we used 0.3 arcsec diameter apertures measured from the PSF-matched images. The total flux is determined from a scaled Kron-like elliptical aperture \citep{Kron1980} in the F277W band. The aperture corrections using the Kron-like aperture were deemed unreliable for a small number of galaxies. Flux corrections greater than a factor of 3.3 were set to 3.3. Flux corrections lower than 1.4 were set to 1.4, since this is the aperture correction for a point source in F444W. Low S/N detections ($<15$ in F277W 0.3 arcsec aperture) that showed flux corrections in the range 2.5 to 3.3 were set to 2.0, a typical value for faint galaxies. Flux uncertainties were determined by the distribution of values in 0.3 arcsec apertures placed in empty regions in the images.

We search for $z>7.5$ galaxies in the full CANUCS NIRCam area. However, in the present luminosity function analysis we do not include the regions of highest gravitational lensing magnification near the cores of the lensing clusters (CLU NIRCam module B). There are three reasons for this. Firstly, the gravitational lensing models are still being improved based on the CANUCS data. Secondly, to fully interpret the luminosity function from strongly-lensed regions requires simulations in the source plane with an accurate model for the intrinsic galaxy size distribution (e.g. \citealt{Atek2018}). The intrinsic size distribution of low-luminosity galaxies in \jwst\ lensed fields is a work in progress, so in this study we only carry out completeness simulations in the image plane. Thirdly, due to the decrease in cosmological volume in high-magnification regions, the total volume in CLU module B fields is significantly less than in the other CANUCS fields. 

The main source of incompleteness for bright sources in the CANUCS NIRCam imaging is a loss of area due to masking by detector artifacts, wisps, other galaxies, stars and diffraction spikes. We determined the effective unmasked area by inserting and recovering artificial sources as will be described in Section \ref{sec:completeness}. The unmasked area in this work is 36.15 square arcmin in the NCF fields (Table \ref{tab:datancf}) and 24.04 square arcmin in the CLU module A fields (Table \ref{tab:dataclu}). 

Despite not using CLU module B centered on the clusters, there is weak gravitational lensing from the clusters in our fields that decreases the effective survey area. For the NCF fields the magnification is typically $\mu < 1.1$. For the CLU module A fields there is a significant area with  $\mu \approx 2$ for Abell 370, but mostly $\mu < 1.5$ for the other four clusters.
We use strong lensing models built with \texttt{Lenstool} \citep{Kneib1993,Jullo2007} constrained by previous results and our new CANUCS \jwst\ data. The models use cluster-size and galaxy-size halos, described as double Pseudo-Isothermal Elliptical (dPIE) profiles \citep{Eliasdottir2007arXiv}, optimized by $\chi^2$ minimisation of the distance between observed and predicted multiple images. The lensing models are presented in forthcoming papers: Abell 370 in \cite{Gledhill2024arXiv}, MACS0416 and MACS1149 in Rihtar\v{s}i\v{c} et al. (in prep.), MACS0417 and MACS1423 in Desprez et al. (in prep.). 
We used these lensing models to determine the lensing magnification at each point in our mosaics and reduce the sky area by the inverse of the magnification. The unmasked, delensed area is 33.81 square arcmin in the NCF fields and 19.55 square arcmin in the CLU module A fields. 

To summarize, with the exception of Section \ref{sec:specver}, we do not include galaxies from the CLU module B fields in this work and instead only use CLU module A and both modules of the NCF fields with a total unmasked, delensed sky area of 53.4 square arcmin. For comparison, the unmasked sky area of the CEERS survey (including the area reduction due to other galaxies and stars) is $\approx 80$ square arcmin \citep{Finkelstein2023arXiv}.
 
\subsection{Spectroscopy}

The CANUCS program includes NIRSpec low-resolution prism multi-object spectroscopic follow-up using the the Micro-Shutter Assembly (MSA, \citealt{Ferruit2022}). Targets for spectroscopy were drawn from a wide range of science cases using the Cycle 1 NIRCam and NIRISS data in conjunction with \hst\ data. Galaxies with high photometric redshifts were placed into one of the highest priority categories when designing the MSA configurations. At the time of writing, four of the five CANUCS clusters have been observed with NIRSpec (MACS1149 is scheduled for late 2023). Each observation was split into three distinct MSA configurations to enable the large gaps between the four MSA quadrants to be dithered across. The total exposure time per configuration is 2900s and high-priority targets were sometimes observed in two or three configurations. Details of the NIRSpec processing are given in \cite{Desprez2023arXiv}.

\section{Selection}
\label{sec:selection}
 
\subsection{Photometric Redshifts}
\label{sec:photoz}

The primary method to identify high-$z$ galaxies in this work uses photometric redshift codes that compare the observed photometry with a suite of templates of galaxies and stars/brown dwarfs. We use the EAZY-py implementation of the EAZY code \citep{Brammer2008}. A set of 12 galaxy templates from FSPS \citep{Conroy2010} are supplemented by three templates from \cite{Larson2023}. The latter are young (1 to 10 million year old), blue galaxies with strong rest-frame optical emission lines that provide a good match to the properties of high-redshift galaxies. The template set includes dust reddening with a \cite{Calzetti2000} attenuation law with maximum E(B-V)=1. The templates are redshifted on a grid from $z=0$ to $z=20$ and subject to IGM attenuation of Lyman continuum and \lya\ photons \citep{Inoue2014}. 

A standard luminosity function based magnitude-redshift prior (EAZY-py prior\_F160W\_TAO.dat) is used. To check whether our use of this prior biases against high-redshift solutions for relatively bright sources, we repeat the selection without the prior. No extra, bright, high-redshift sources are selected in this test. A small number of additional possible high-$z$ galaxies, close to the F277W S/N limit, with a redshift distribution similar to our main sample, are selected. Since this prior is used in the selection of sources from the completeness simulations (Section \ref{sec:completeness}), the luminosity function derivation is not biased by the use of this prior.

The use of a large number of NIRCam filters (14 in NCF and 8 in CLU fields) provides good constraints on photometric redshifts when S/N is high and galaxies have spectra that resemble the templates. The strong optical emission lines observed in \jwst\ spectra of galaxies in the redshift range $5<z<9$ can dominate medium-band \citep{Williams2023, Withers2023} photometry, providing tight constraints on photometric redshifts, particularly at certain redshifts. At higher redshifts, the strong optical lines move out of the NIRCam filters and photometric redshifts are purely based on defining the position of the spectral break at \lya. 

The high-$z$ galaxy sample presented here is defined to lie at redshift $z>7.5$. This limit was chosen because all such galaxies should be undetected in the F090W filter that has a red edge transmission cutoff at 1.005\,\um, corresponding to the wavelength of \lya\ at $z=7.27$. The \hst\ optical coverage of the CANUCS survey area is quite uneven. Therefore, we rely on non-detections in F090W as the primary method to distinguish high-$z$ galaxies from low redshift interlopers. \hst\ optical data, when available, was included in the photometric redshift fitting and visually inspected to ensure it is consistent (i.e. non-detections) for galaxies classified as high redshift. We do not implement a S/N cut on the \hst\ optical data due to non-Gaussian noise and artifacts in some of the outer regions of the  \hst\ mosaics that are constructed from only a few dithers.

To minimize the contamination of the high-$z$ galaxy sample by lower redshift galaxies we adopt the following selection criteria:
\begin{equation}
\label{eqn:selection}
    \begin{split}
        z_{\rm phot\_ML} > 7.5, \\
        z_{\rm phot\_16} > 6, \\
        S/N_{\rm F090W} < 2,\\
        S/N_{\rm F277W} > 8.\\
    \end{split}
\end{equation}
where $z_{\rm phot\_ML}$ is the maximum likelihood best-fit redshift and $z_{\rm phot\_16}$ is the redshift below which 16\% of the redshift probability distribution function (PDF) lies. A limit of $z_{\rm phot\_16}>6.0$ is applied since the primary low-$z$ interlopers are at redshifts substantially below this. The S/N constraint on the F277W filter ensures that only well-detected objects are included. This is necessary to avoid including a large number of galaxies with large flux uncertainties that have poorly constrained photometric redshifts.

There is no criterion for a low $\chi^2$ in our selection. This is to guard against removing good high-redshift galaxies with one or two bad photometry points. Only two galaxies in the sample have reduced $\chi^2>3$. In both cases, visual inspection shows the candidates to be good with one bad data point; CANUCS-4214058 has a F162M flux contaminated by persistence and CANUCS-3211653 has an artifact in the ACS F814W image.

In addition to EAZY-py, we also use  the Phosphoros code (\citealt{Desprez2020}, Paltani et al., in prep.). The configuration is similar to the one using optical and near-infrared photometry prescribed by \citet{Desprez2023}, with the addition of the \citet{Larson2023} templates and the extension of the explored redshift range to $z=0$--$15$ to better account for high-redshift solutions. A top-hat prior on the rest-frame $r$-band absolute magnitude between $-24$ and $-7$ is adopted.  The selection of high-$z$ candidates is done using three point estimates: the mode and the median of the redshift PDF, and the best-fit redshift. Sources for which all three estimates are at $z>7.5$ are selected as candidates and visually inspected for confirmation. The samples selected with Phosphoros and EAZY-py overlap in a vast majority of cases, thus providing reassurance of the high-$z$ nature of the galaxies present in the two selections. The main class of disagreements are a small fraction of faint, Phosphoros-selected galaxies that fail to pass the EAZY-py $z_{\rm phot\_16} > 6$ criterion because EAZY-py identifies a substantial $1.5 <z< 3$ Balmer break solution. This is likely a consequence of the different magnitude-redshift priors used for the two codes, with the EAZY-py prior applying more weight at lower redshift. In our experiment with turning off the EAZY-py prior, some of these faint Phosphoros-selected galaxies were recovered. For the sake of simplicity in the sample completeness analysis, only the more conservative EAZY-py with prior sample is considered in the rest of this work for the derivation of the luminosity function.

\begin{figure}
    \centering
    \includegraphics[width=\linewidth]{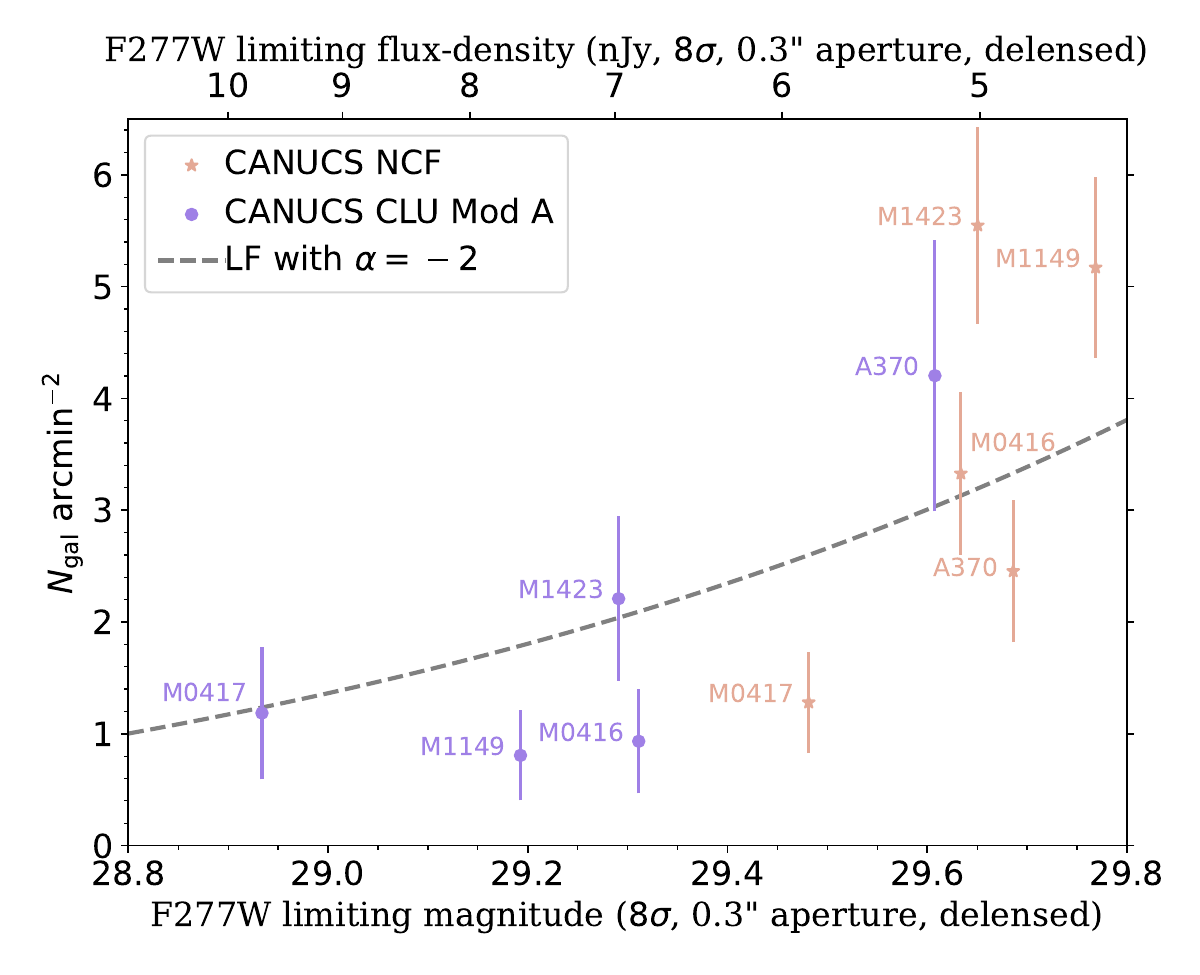}
    \caption{Sky density of galaxies at $z>7.5$ in the ten CANUCS fields used in this paper. The vertical axis shows the number of detected galaxies divided by the unmasked, delensed sky area of the field. Error bars show Poisson uncertainties only. The horizontal axes indicate the limiting F277W magnitude (or flux-density) limits, where the limits have been modified to account for the median gravitational lensing in the field. This lensing correction is mostly small with magnification $\mu \leq 1.2$ for all fields except A370 CLU module A that has $\mu = 1.5$. There is a trend for the fields with the deepest data to have the highest galaxy density, as expected. To model this effect we show the expected variation with limiting magnitude for a luminosity function with faint-end slope $\alpha=-2$. Even after accounting for the different depths of the fields, the field-to-field scatter is considerably larger than the Poisson uncertainties, with half the fields lying $\approx 2\sigma$ from the curve.}
    \label{fig:perfield}
\end{figure}

The final galaxy sample for the luminosity function contains 158 galaxies satisfying the selection criteria in Equation \ref{eqn:selection}. We note that if we replace $z_{\rm phot\_ML} > 7.5$ with the median of the PDF, $z_{\rm phot\_50}>7.5$, only four of these 158 galaxies would be eliminated. The NCF fields cover two-thirds of the pointings in this study and contain 125 (80\%) of the galaxies. The increased galaxy density in NCF is largely a consequence of the deeper imaging. The maximum likelihood photometric redshifts range from 7.50 to 12.38 with a median of 8.31. It is interesting to note that despite searching up to $z=20$, our data that includes medium bands such as F140M, F162M, F182M, and F210M in the NCF fields does not find any convincing candidates at $z>13$. Only eight of our galaxies are found at $z_{\rm phot\_ML}>10$, a much lower fraction than some other studies. Example SEDs are shown in Appendix \ref{sec:seds} and the full table of galaxy properties is given in Appendix \ref{sec:galaxytables} Table \ref{tab:sample}. 

We find significant variation in the number of $z>7.5$ galaxies in each field. For the NCF fields there are 15, 21, 8, 41 and 40 galaxies in the fields Abell 370, MACS0416, MACS0417, MACS1149, MACS1423, respectively. For the CLU module A fields the numbers are 12, 4, 4, 4 and 9. There are small differences in the depths of the fields, their effective sky areas and their median gravitational lensing magnifications (which both boosts fluxes and decreases sky area). In order to account for these differences, we calculate the sky density of sample galaxies per field, using the unmasked, delensed area. The sky densities range from 0.8 to 5.5 galaxies per square arcminute. In Figure \ref{fig:perfield} we compare these sky densities with the magnitude limits of each field. We use the F277W flux limits from Tables \ref{tab:datancf} and \ref{tab:dataclu}, adjusting to a $8\sigma$ limit as used in the sample selection (Equation \ref{eqn:selection}). These magnitude limits are further adjusted for gravitational lensing magnification. This lensing adjustment has the most impact for Abell 370 CLU with median $\mu=1.5$.  

Figure \ref{fig:perfield} shows the highest galaxy densities are seen in the deepest fields, which is not that surprising. To estimate the expected effect of variable depth, we calculate the increase in counts as a function of magnitude limit for a typical high-$z$ luminosity function. This simple model fits the observed trend quite well. However, there is still an excess of field-to-field scatter compared to this model. Five of the ten fields deviate from this curve at the $\approx 2\sigma$ level when considering only Poisson variance. MACS0416 and MACS1149 CLU and MACS0417 NCF all lie significantly below the curve, whereas MACS1149 and MACS1423 NCF lie significantly above it.

These statistics highlight the importance of surveying several widely-separated fields to obtain a true picture of the distant universe (e.g. \citealt{Steinhardt2021, Desprez2023arXiv}). The expected cosmic variance over $7.5<z<12.5$ in each NCF field is 20\%, similar to the expected Poisson variance \citep{Trenti2008}. For a mean of 25 galaxies, $\sigma = 7$ from Poisson and cosmic variance combined. Three of the five NCF fields show number counts more than $2\sigma$ from the mean, which could suggest the cosmic variance is even higher than expected. However, the galaxies are not evenly spread throughout the volume between $z=7.5$ and $z=12.5$, with more than half in the range $7.5<z<8.5$. Restricting the cosmic variance calculation to $7.5<z<8.5$, we find that none of the fields are $>2\sigma$ outliers, but four of the five are $>1\sigma$ outliers, providing a more modest disagreement between the observed and expected cosmic variance.

The F277W S/N $>8$ selection criterion was chosen because a limit at F200W or shorter wavelengths would bias against $z>13.5$ galaxies when \lya\ enters this filter. However, at the lower redshift range of the sample, F277W probes rest-frame wavelengths significantly longer than 1500\AA, the wavelength typically adopted for UV luminosity functions. To test whether the F277W S/N limit biases against very blue UV slopes at $z<10$, we repeated the selection using a limit of F150W S/N$>8$. With this criterion only four additional galaxies with photometric redshifts between 7.8 and 8.9 are selected. Notably, all four have F277W S/N in the range between 7 and 8 showing they narrowly miss the original F277W criterion. We conclude we are not missing a large population of galaxies with very blue UV slopes.

\subsection{Potential Contaminants}

There are two main types of possible contaminants: galaxies at intermediate redshifts that have a strong Balmer break and/or strong emission lines, possibly combined with a dusty continuum and brown dwarfs in our galaxy. For the galaxies, we rely on the EAZY-py selection constraint of $z_{\rm phot\_16}>6$. This eliminates possible galaxies at $1 < z < 5$ that have any of the above features, since these features are well captured by the range of galaxy templates and variable dust attenuation. The suite of 12 to 14 NIRCam filters in the NCF fields that comprise two-thirds of our pointings are especially good at isolating strong emission lines in lower redshift galaxies that may mimic a high-$z$ blue UV continuum and sharp \lya\ break \citep{Eisenstein2023arXiv}. For example, the $z=4.9$ dusty, strong emission line galaxy CEERS-93316 was  detected with four emission-line-contaminated filters above the potential $z\sim 16$ solution Lyman break \citep{ArrabalHaro2023a}. Such a galaxy would have been detected in eight filters above 2\,\um\ in CANUCS NCF, with four of those filters not including any of the \Hb, \Oiii\ or \Ha\ emission lines. Indeed, a similar galaxy, CANUCS-1205006, was found in MACS0417 NCF photometry and identified as a strong line emitter with dusty red continuum at $z_{\rm phot\_ML}=5.73$, a hypothesis subsequently confirmed with NIRSpec spectroscopy giving $z_{\rm spec}=5.64$ \citep{Withers2023}. However, as seen in Figure \ref{fig:perfield}, we do not observe an excess of galaxies in the CLU fields compared to NCF that would be expected if the data without the large suite of medium filters suffered from extensive contamination by low-redshift, strong-emission-line galaxies.

Brown dwarfs have long been known to be potential contaminants of high-redshift galaxy samples and the exceptional 1 to 5 \um\ sensitivity of \jwst\ makes it very effective for finding brown dwarfs in the disk or halo of our galaxy \citep{Wilkins2014}. All potential high-redshift galaxies are checked to ensure they are not brown dwarfs in three ways. Firstly, for the small subset of objects detected in previous WFC3/IR imaging, typically observed 5 to 10 years ago, positional offsets would indicate a proper motion. Secondly, most of the high-redshift galaxies are spatially-resolved, but a small fraction is consistent with being point-like. Finally, all high-redshift galaxies are fit with the Sonora brown dwarf models that cover a wide range of effective temperature and gravity \citep{Marley2021}. None of the galaxies have photometry that is better fit by a brown dwarf model than by the best fit $z>7.5$ galaxy template. Several very cool ($T_{\rm eff}<1000$\,K) brown dwarfs have been identified in the CANUCS data (e.g. \citealt{Desprez2023arXiv}), but none of these pass the $z>7.5$ galaxy selection criteria adopted in this work.

As a further check on the reliability of our galaxy sample we stack the galaxy images to see how well their overall properties match those expected for high-redshift galaxies. In particular, due to complete absorption of flux shortward of Lyman-$\alpha$, we expect to see zero flux in the F090W filter for all the galaxies and zero flux in the F115W filter for the sub-sample at $z>9.6$. We stack galaxies in 3 groups: A $7.5<z<9.6$ NCF, B $7.5<z<9.6$ CLU, C $z>9.6$ CLU+NCF. We separate CLU and NCF for the lower redshift sample because the filter set differs, with CLU having fewer NIRCam filters (8 vs 14). In the highest redshift bin there are not enough galaxies to make two separate stacks. Appendix \ref{sec:stacks} provides more details of the stacking method and shows the stacked images in each filter. There is no significant flux in the F090W filter for either median or mean stacked images for all three groups and also no significant flux in the median or mean F115W images for group C. This supports little to no low-redshift contamination of the CANUCS high-redshift sample. 

\begin{figure}
    \centering
    \includegraphics[width=0.99\linewidth]{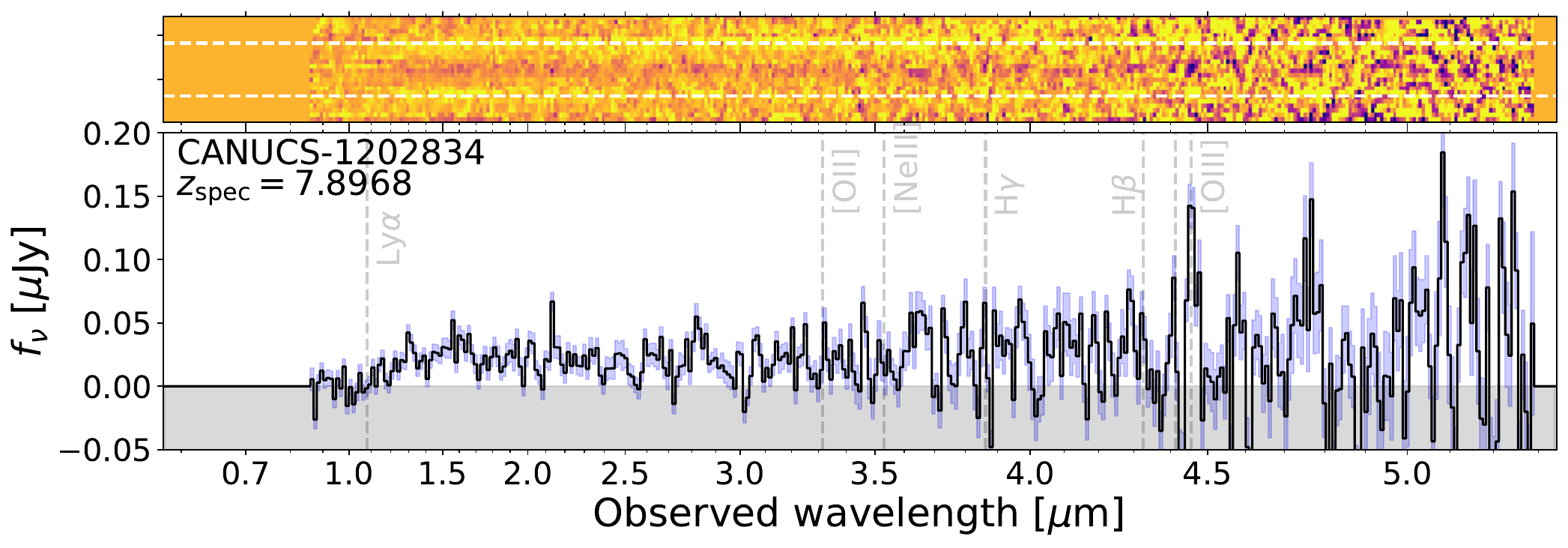}\\
    \includegraphics[width=0.99\linewidth]{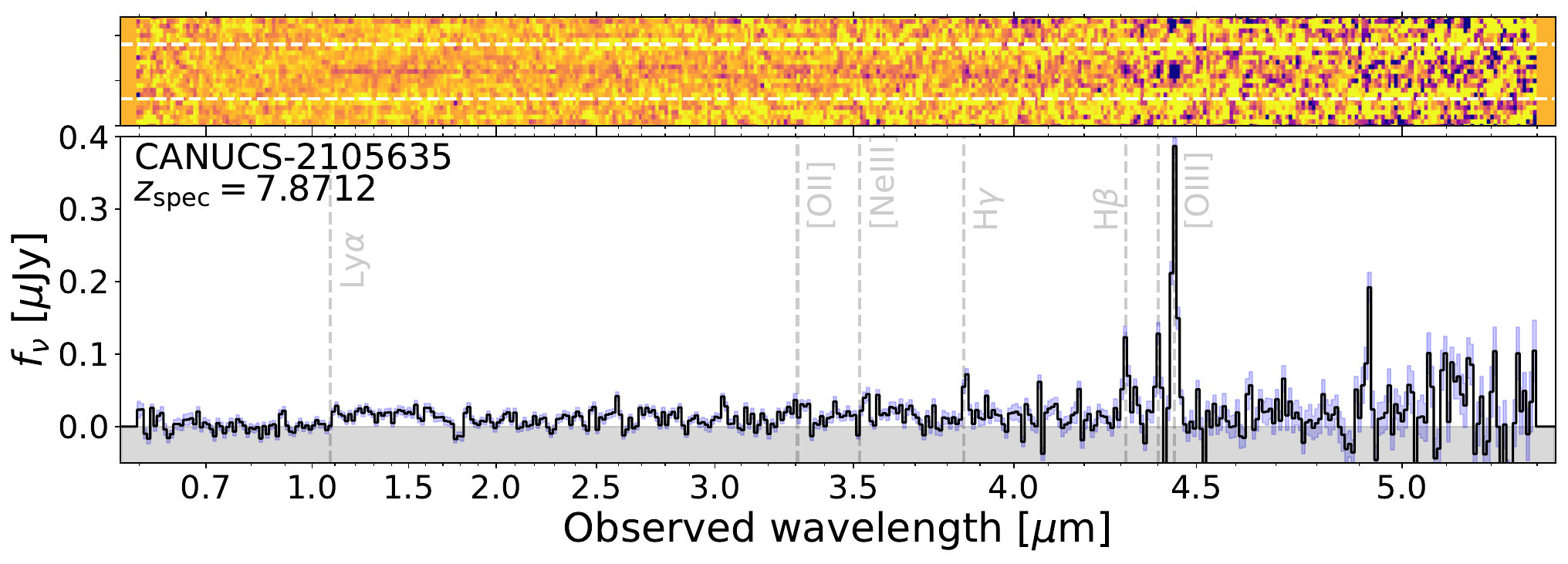}\\
    \includegraphics[width=0.99\linewidth]{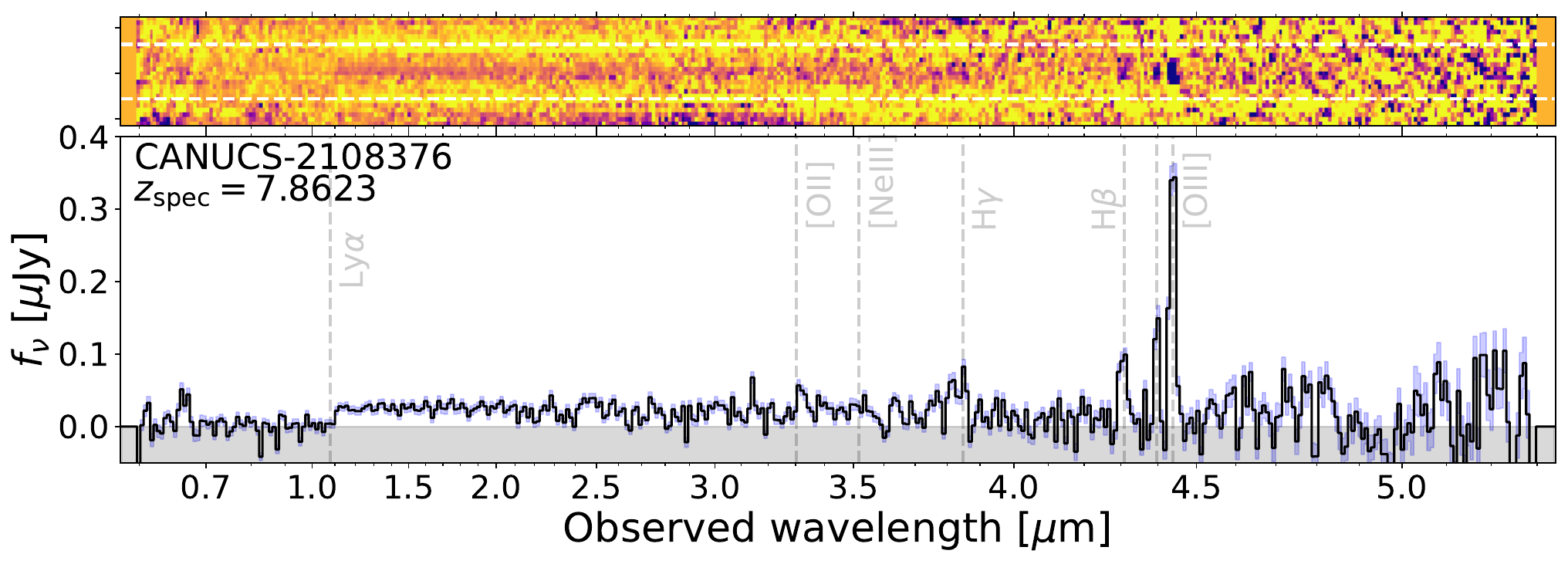}\\
    \includegraphics[width=0.99\linewidth]{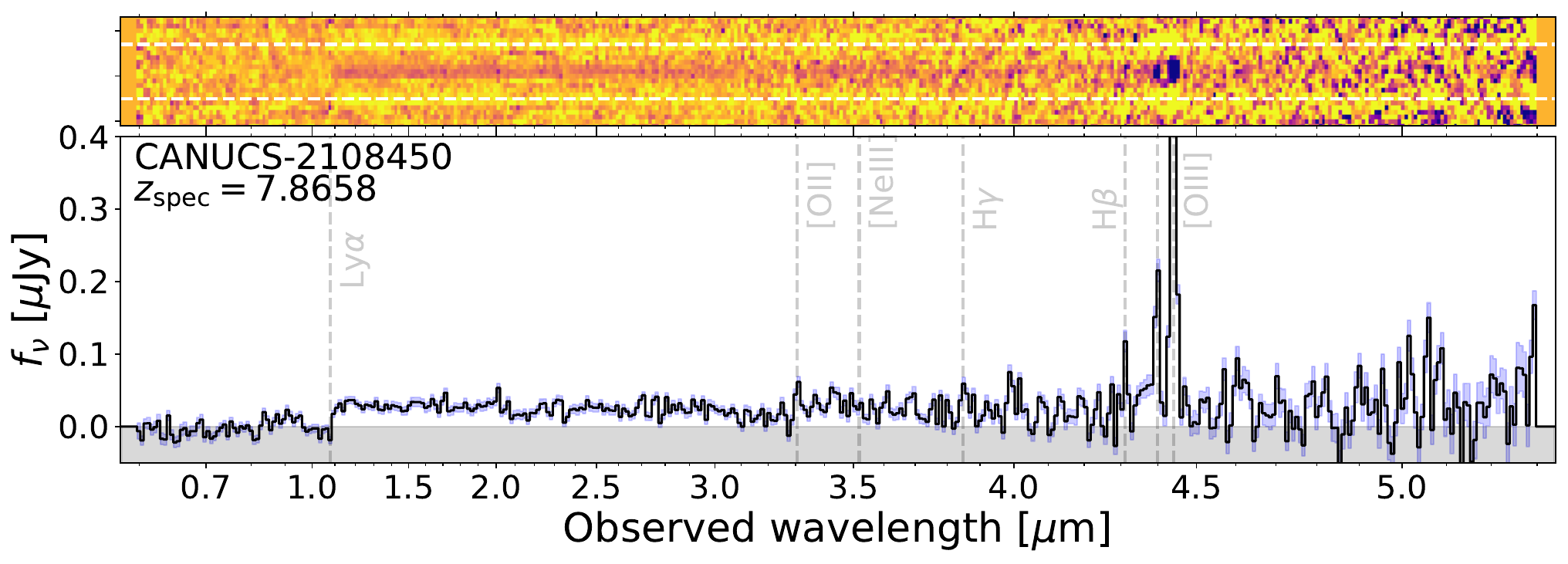}\\
    \includegraphics[width=0.99\linewidth]{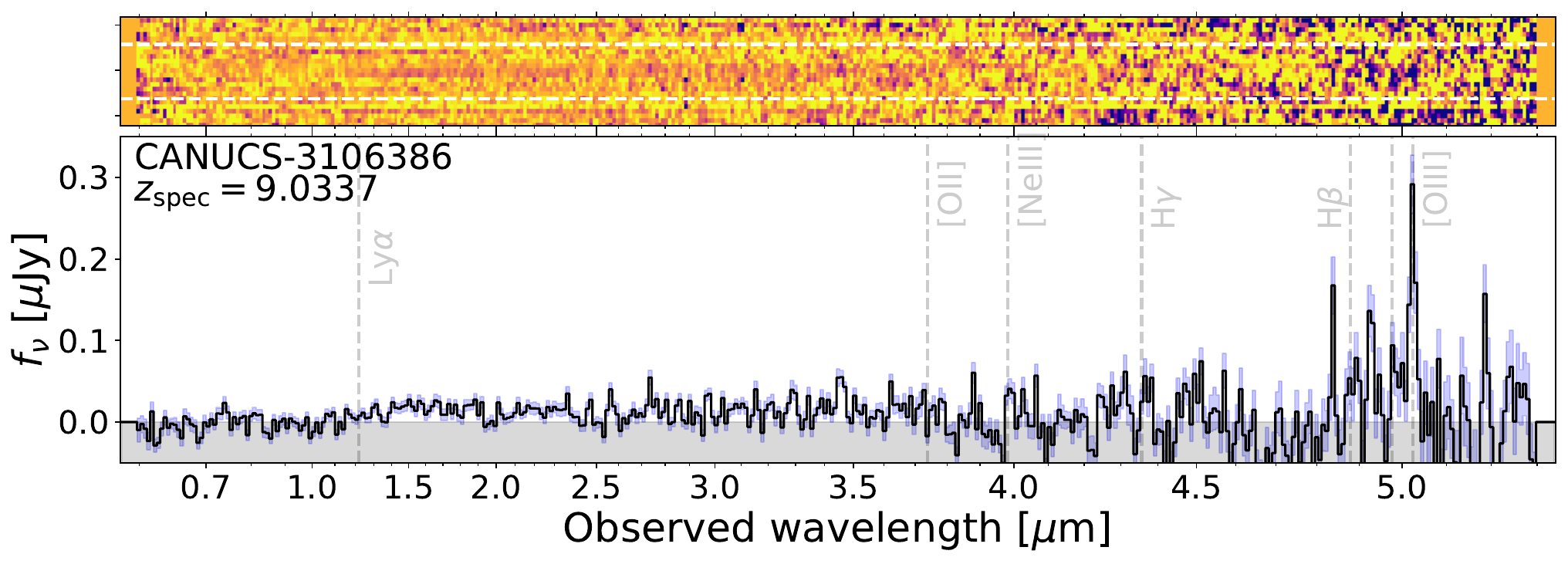}\\
    \includegraphics[width=0.99\linewidth]{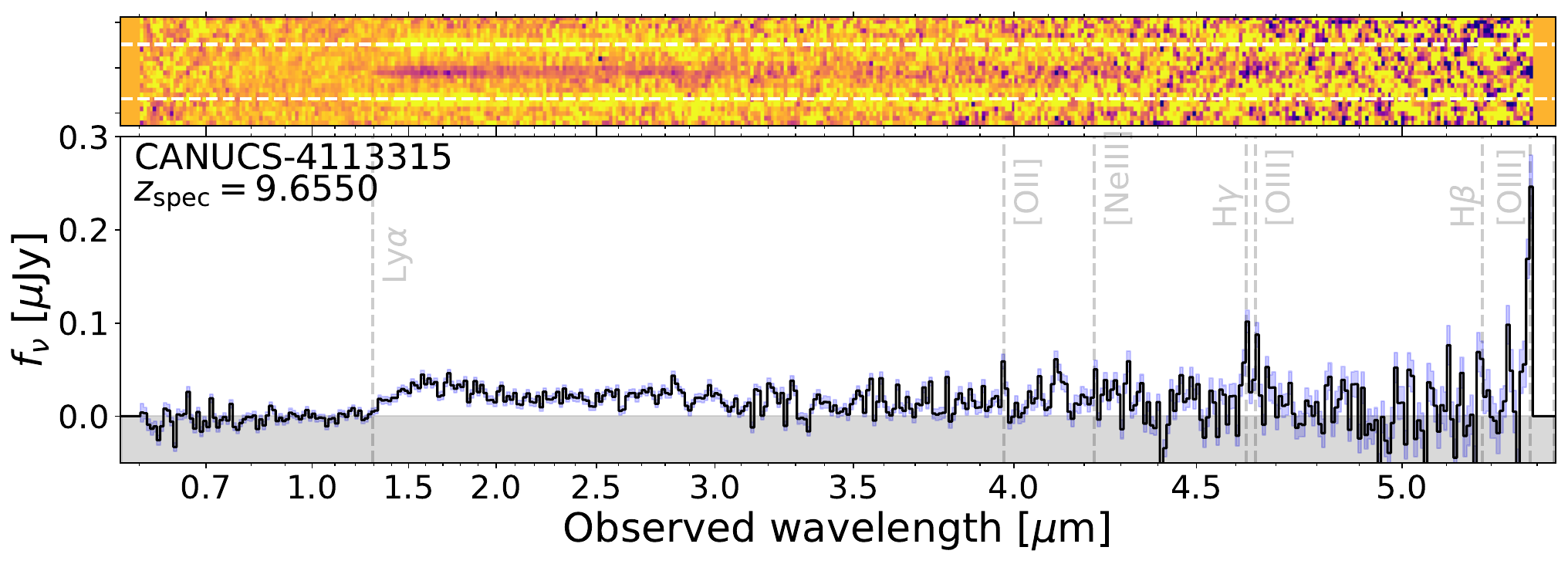}
    \caption{NIRSpec prism spectra for the six CANUCS galaxies in the sample used for the luminosity function in this paper. Each upper panel shows the nodded, background-subtracted two-dimensional spectrum on an inverse scale, so positive flux appears dark in the central region and light in the two offset positions marked with white dashed lines. The lower panels show the background-subtracted one-dimensional spectra with $\pm 1 \sigma$ uncertainty shaded in blue. The expected locations of strong emission lines and the \lya\ transition based on the emission line redshifts are marked.}
    \label{fig:spectra}
\end{figure}

\subsection{Spectroscopic Verification}
\label{sec:specver}

In this section we use spectroscopic redshifts to check the reliability of our sample selection and photometric redshift estimation.
The focus of the CANUCS NIRSpec followup observations is on the lensing clusters and strongly-lensed galaxies. Therefore, the vast majority of NIRSpec targets lie in the CLU field NIRCam module B that is not used in the luminosity function analysis in this paper. However, these spectra still provide a useful check on our photometric redshift reliability. A small fraction of the NIRSpec targets lie in the CLU field NIRCam module A, and for the MACS0417 cluster only, one of the three MSA configurations targeted the NCF field. A total of 27 galaxies matching the selection criteria of Equation \ref{eqn:selection} have been observed in all CANUCS NIRSpec observations and resulted in spectra with confirmed redshifts. All of these redshifts are found to be above 7.5. There are no low-$z$ interlopers or brown dwarfs found. Six of these galaxies are in the sample in this paper (CLU module A or NCF) and 21 were selected in an identical manner, but are not included in this paper because they lie in CLU module B, eight of which are presented in \cite{Desprez2023arXiv} and \cite{Mowla2024arXiv} and the others will be the subject of future papers. Additionally, there is one galaxy in CLU module B with $z_{\rm spec}=7.57$ and $z_{\rm phot\_ML}=7.49$ that narrowly fails the $z_{\rm phot\_ML}>7.5$ criterion. A further five faint targets meeting our selection criteria were observed with NIRSpec, but the resulting spectra failed to provide a definite redshift. There is no evidence from these spectra that they are low-redshift interlopers. Indeed, if their photometry was dominated by emission lines at lower redshift such as CEERS-93316, then such lines would have easily been identified. The spectra of the six confirmed $z>7.5$ galaxies in CLU module A or NCF are shown in Figure \ref{fig:spectra}. All six galaxies show strong emission lines in their spectra leading to secure redshifts. There is also zero flux below the wavelength of \lya.

\begin{figure}
    \centering
    \includegraphics[width=\linewidth]{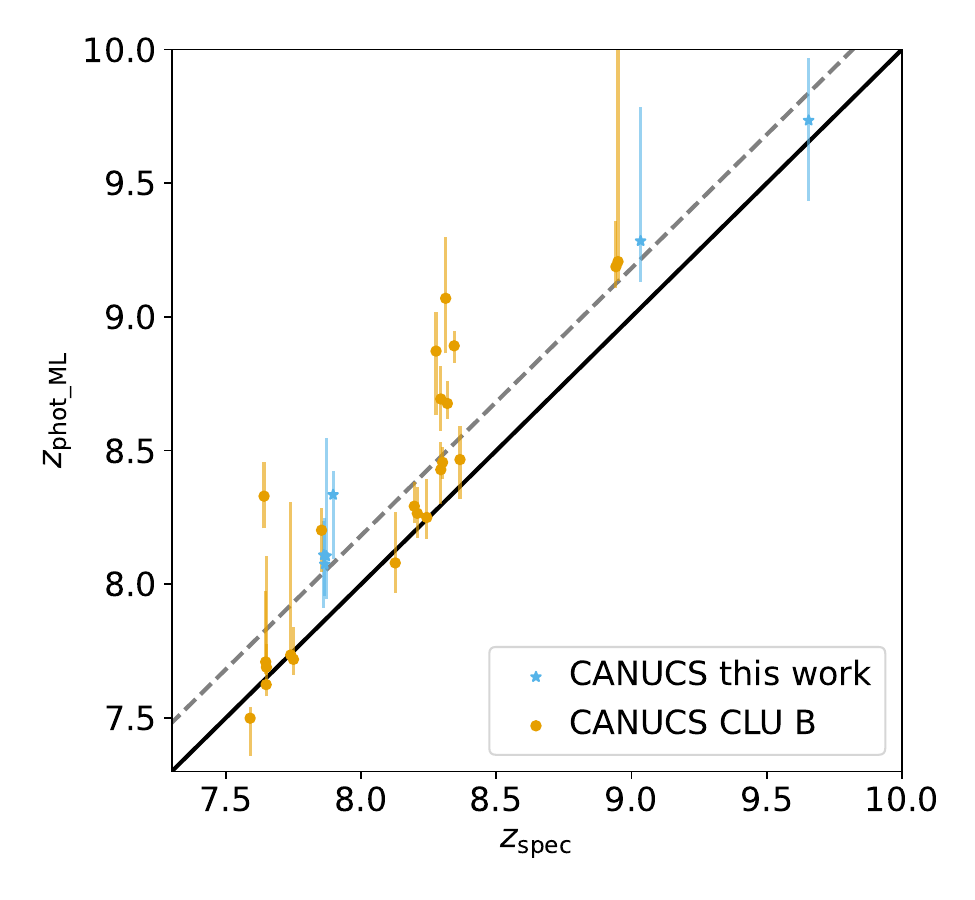}
    \caption{NIRSpec spectroscopic redshifts for 28 CANUCS galaxies with expected photometric redshifts $z_{\rm phot\_ML} \gtsimeq 7.5$. Only 6 of the 28 galaxies (blue symbols) are from the NCF or CLU NIRCam module A fields and are included in this paper. The remaining 22 galaxies are from CLU NIRCam module B covering the cores of the lensing clusters (orange symbols); this region being more heavily targeted by CANUCS spectroscopy. There is good agreement between the photometric and spectroscopic redshifts. The median offset from the one-to-one line (solid) is $\delta z = 0.21$ (dashed gray line), with the photometric redshifts slightly overestimating the true redshifts, as has been reported in other studies.}
    \label{fig:specphotz}
\end{figure}

In Figure \ref{fig:specphotz} we show the photometric versus spectroscopic redshifts for the full CANUCS sample of 28 $z_{\rm phot\_ML} \gtsimeq 7.5$ galaxies. The photometric redshifts are derived as described in Section \ref{sec:photoz} using templates that included the very blue star-forming galaxies in \cite{Larson2023} with Lyman-$\alpha$ line emission removed, since only a small fraction of $z>7.5$ galaxies show Lyman-$\alpha$ line emission. The confirmed spectroscopic redshifts match very closely the photometric estimates used to select the candidates. There are no `catastrophic' outliers (adopting the typical criterion of $|\delta z|/(1+z)>0.15$, where $\delta_z = z_{\rm phot\_ML}- z_{\rm spec}$). The median offset is $\delta z = 0.21$ (dashed gray line) with the photometric redshifts slightly overestimating the true redshifts. Several works have noticed similar, and in several cases much larger, offsets in the photometric redshifts of galaxies in the reionization era observed with \jwst\ \citep{Fujimoto2023, ArrabalHaro2023b, Hainline2023arXiv, Finkelstein2023arXiv}. This offset is likely due to the galaxy templates being too blue in the spectral region just redward of Lyman-$\alpha$. Spectra of many of the highest redshift galaxies show a departure from the blue UV slope in this region \citep{Curtis-Lake2023, Umeda2023} due to the Lyman-$\alpha$ damping wing from the IGM \citep{Miraldaescude1998}, from dense gas close to the galaxy \citep{Heintz2023arXiv} and/or two-photon nebular continuum \citep{Cameron2023arXiv,Mowla2024arXiv}. As seen in Figure \ref{fig:spectra}, a similar flux deficit just redward of Lyman-$\alpha$ is apparent in five of the six CANUCS galaxies (the exception is CANUCS-2105635), consistent with their slightly over-estimated photometric redshifts in Figure \ref{fig:specphotz}.

The normalized median absolute deviation around the median offset of $\delta z = 0.21$ is extremely small at only 0.024, similar to the value found by \cite{Hainline2023arXiv} for JADES when using similar templates with no Lyman-$\alpha$ emission. We note that our NIRCam filter sets are similar to JADES (8 filters for CANUCS CLU, versus 9 for JADES), so it is encouraging that we find very similar median offset and dispersion. This shows that our inhomogeneous \hst\ coverage is not an issue when dealing with $z>7.5$ galaxies as there is no extra information in the \hst\ bands compared to F090W. One might expect the dispersion in NCF to be even lower, due to the 14 NIRCam filter coverage in those fields, however we do not have sufficient NCF galaxy spectra to quantify this. We do not apply a statistical correction for this small redshift offset since it likely varies with redshift in an undetermined way. 

\begin{figure*}
    \centering
    \includegraphics[width=\linewidth]{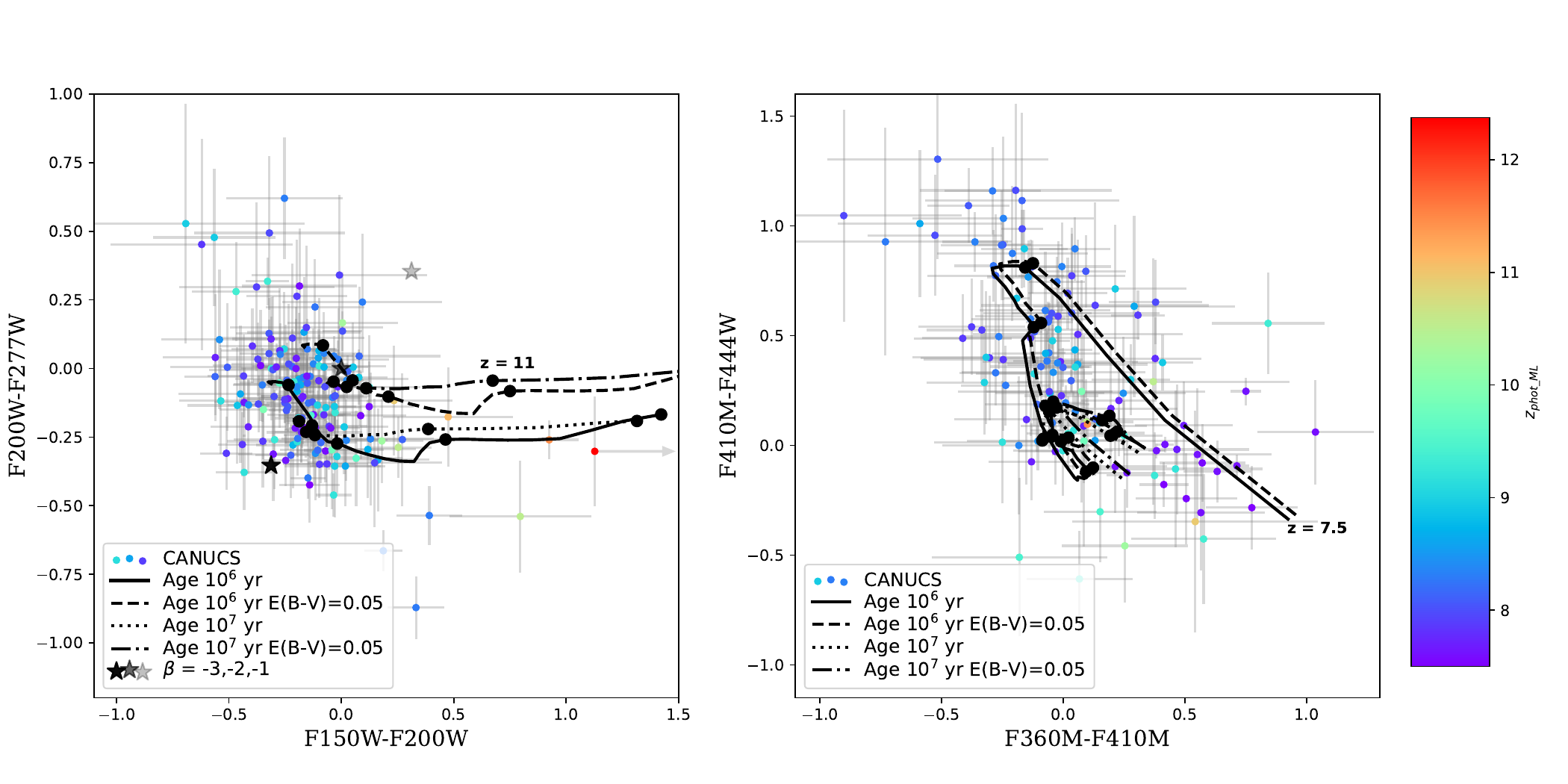}
    \caption{Color-color diagrams showing the CANUCS $z_{\rm phot\_ML}>7.5$ galaxy sample and redshift tracks for the four template galaxy SEDs used in the completeness simulations. Black filled circles are shown on the template tracks at $z=8,9,10,11,12$. The left panel shows F200W$-$F277W against F150W$-$F200W. These colors both trace the UV continuum slope $\beta$ at these redshifts, although F150W-F200W starts to redden due to absorption shortward of \lya\ at $z>9.7$. Values of the UV slope $\beta=-3,-2,-1$ are shown with star symbols. Most of the CANUCS galaxies lie in the range $-3<\beta<-2$. The right panel shows F360M$-$F410M against F410M$-$F444W for galaxies from the NCF fields (F360M is not used in the CLU fields). These colors are sensitive to strong \Oiii+\Hb\ emission lines from $z=7.5$ to $z=9.2$. The spread in these colors of the CANUCS galaxies are explained well by strong line emission in a large fraction of the $z<9.2$ galaxies.}
    \label{fig:colcol}
\end{figure*}

In the rest of this paper we adopt the spectroscopic redshift if measured, otherwise we use $z_{\rm phot\_ML}$. The absolute magnitude of each galaxy, $M_{\rm UV}$, is calculated by integrating the best fit template spectrum between rest-frame 1450\,\AA\ and 1550\,\AA, correcting from aperture to total magnitude (since the SED fitting was performed using the 0.3" aperture fluxes) and applying a lensing correction.

\subsection{Completeness}
\label{sec:completeness}

In order to derive the luminosity function we need to know how successfully we are able to identify high-$z$ galaxies with appropriate photometric redshifts as a function of their intrinsic luminosity. The completeness of a photometric-redshift selected sample can be complex due to the way in which features both within the target redshift range and at contaminating lower redshifts move through the filter set. Since we are using data from 15 separate NIRCam modules with variations in depth and filter set between the NCF and CLU fields and, to a lesser extent, a range of zodiacal and scattered light background in the direction of the five CANUCS clusters, the only way to determine our completeness is a full simulation of inserting artificial sources into all of the data. 

Images of artificial galaxies are simulated using Sersic profiles. A Gaussian distribution of ellipticities (defined as $1-b/a$) centered at 0.3 with $\sigma=0.2$ is adopted, where values below 0 and above 0.7 are not allowed. The Sersic index is a Gaussian distribution centered at 1.5 with $\sigma=0.3$, with negative values not allowed. The assumed galaxy size distribution of the simulations is very important as it can have a large effect on the resulting faint end of the luminosity function  \citep{Oesch2015, Bouwens2017b, Atek2018}. We use recent size-luminosity-redshift relations derived from a compilation of \jwst\ Cycle 1 fields by \cite{Morishita2023barXiv}. For each galaxy with simulated absolute magnitude and redshift we use the \cite{Morishita2023barXiv} Appendix A relation with the parameter values in their Table 4 to assign a size based on the best-fit relation and its dispersion. For the 5\% of cases with effective radii $> 1$\,kpc, we draw again from the distribution, because there are no galaxies this large in our observational sample. 

\begin{figure*}
    \centering
    \includegraphics[width=\linewidth]{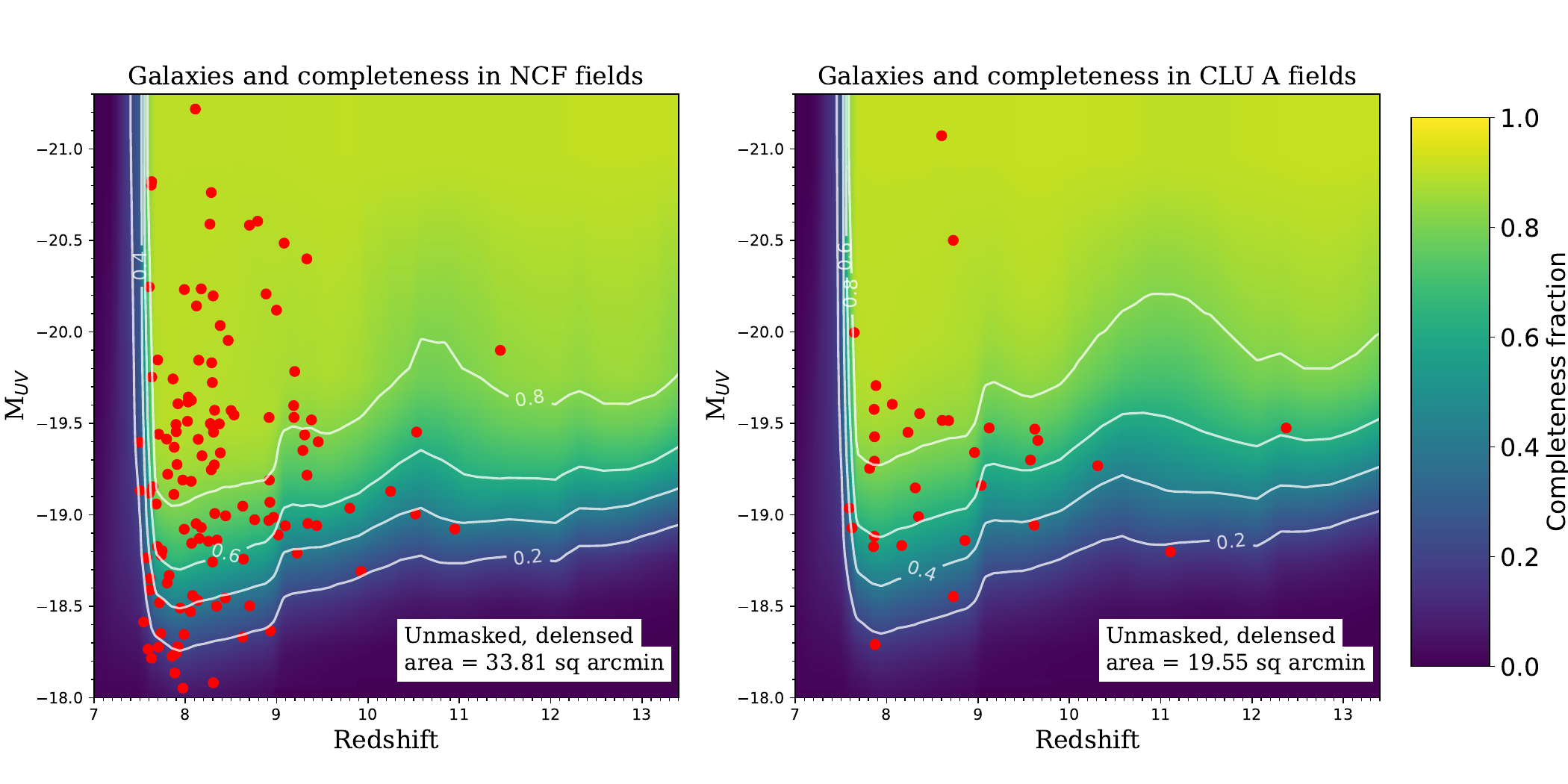}
    \caption{The absolute magnitudes and photometric redshifts for the galaxies in our $z>7.5$ sample are shown with red circles. The left panel shows galaxies in the NCF fields and the right panel galaxies in the CLU module A fields. The color shading indicates the selection completeness fraction as a function of these two parameters. Contours are shown in white at fractions 0.2, 0.4, 0.6 and 0.8. Note that there is a spread in completeness from cluster-to-cluster due to variable imaging depth and (especially in CLU) lensing magnification, so it is not surprising to find some galaxies below the ensemble 0.2 contour.}
    \label{fig:completeness}
\end{figure*}

The spectral energy distributions for the artificial galaxies are based on the \cite{Larson2023} galaxy templates with no \lya\ emission, specifically two ages of $10^6$ and $10^7$ years, subjected to dust attenuation of E(B-V)=0 and 0.05 each. These four models bracket the observed properties of our sample galaxies. Figure \ref{fig:colcol} shows two color-color diagrams for our sample galaxies overlaid with redshift tracks for the four galaxy models. The F150W-F200W-F277W color-color diagram shows most CANUCS $z>7.5$ galaxies have UV slopes in the range $-3<\beta<-2$. The F360M-F410M-F444W diagram shows a large fraction of CANUCS $7.5<z<9$ galaxies have very strong \Oiii+\Hb\ emission lines.  In general, the CANUCS galaxies are well-described by these four models, despite the presence of a few outliers. For the purpose of the completeness simulations, the match is adequate. 

A grid of absolute UV magnitude ($-21.5<M_{\rm UV}<-18$) versus redshift ($7<z<13.5$) is defined and at each point of this grid a set of randomly simulated galaxies is inserted into each of our images. A total of 2.5 million simulated galaxies are inserted across all runs, with $\sim 500$ galaxies per NIRCam module at a time. The images with the added simulated galaxies are then processed using the exact same detection and photometry pipeline described in Section\,\ref{sec:data} to determine the recovery fraction. It is found that 10 to $20\%$ of bright simulated galaxies are not recovered by the detection process, due to coinciding or blending with a brighter object in the field. We consider this as masked area and quote the remaining unmasked area in Section \ref{sec:imaging}. The detection fraction remains quite high even for faint high-redshift galaxies (e.g. $\approx 50$\% at $M_{\rm UV}=-18.25, z=11$ in NCF fields), because the detection is performed on the deep $\chi_{\rm mean}$ image constructed from many bands. 

For the luminosity function calculations, the important quantity is not only whether the object would be detected and entered into our source catalog, but whether it would also be selected by the photometric redshift routine based on the measured photometry. Faint sources with high photometric uncertainties will not be selected because their photometric redshifts are often ambiguous with multi-peaked PDFs and may not satisfy the condition that less than 16\% of the PDF is at $z<6$. Therefore, for all artificial galaxies that are recovered from the detection image we run their measured photometry through EAZY-py with exactly the same parameters as those in Section \ref{sec:photoz} and apply the selection criteria of Equation \ref{eqn:selection}. This results in substantially lower completeness for faint objects compared to that of the detection image, as expected, but guarantees a high purity sample containing few low-redshift interlopers.

Figure \ref{fig:completeness} shows the location of our sample in the absolute magnitude versus redshift plane. Separate panels are shown for the NCF and CLU subsamples because of their different filter sets, depths and typical lensing magnification. Also shown on the figure are the results of the completeness simulations, including the recovery of sources into our sample with photometric redshift fitting. The sharp vertical contours at $z\approx 7.5$ are due to the $z_{\rm phot\_ML}>7.5$ requirement. There is a spread in completeness from cluster-to-cluster due to variable imaging depth across the field and, to a small extent, lensing magnification, as we are not using highly-magnified regions. The distribution of observed galaxies matches well the expectations based on the completeness simulations, e.g. a few galaxies selected close to the 0.2 contours. We inspect similar completeness plots for each cluster individually and find that, as expected, the lower luminosity sources tend to come from the clusters with higher completeness at low luminosity and only three galaxies are below the 0.1 completeness contour within their field. Therefore we are confident the completeness simulations reflect well the chance of a galaxy appearing in our sample. The dip in completeness at $z\approx 11$ can be understood as where the \lya\ break is in the middle of the F150W filter so galaxies are more easily confused with lower redshift Balmer break galaxies. A similar dip occurs at $z\approx 9$  where the \lya\ break is in the middle of F115W, but the effect is minimized  at $z<9$ by \Oiii\ emission in F444W distinguishing these galaxies from lower redshift Balmer break galaxies.

\section{Results}
\label{sec:results}

\begin{deluxetable}{c c c c c}
\hspace{-1.0cm}
\tablewidth{240pt}
\tablecaption{\label{tab:binnedlf} Binned Galaxy Luminosity Function}
\tablehead{$M_{\rm UV}$ & $M_{\rm UV}$ & $M_{\rm UV}$ & $z_{\rm phot\_ML}$ & $\Phi$  \\
low & high & median & median & (Mpc$^{-3}$\,mag$^{-1}$)}
\startdata
\multicolumn{5}{c}{Redshift bin: $7.5<z<8.5$}\\
\hline
$-21.0$ & $-20.5$ & $-20.78$ & $7.95$ & $8.8^{+7.1}_{-4.4}\times 10^{-5}$  \\
$-20.5$ & $-20.0$ & $-20.21$ & $8.15$ & $13\pm 6\times 10^{-5}$ \\
$-20.0$ & $-19.5$ & $-19.68$ & $8.03$ & $42\pm 11\times 10^{-5}$  \\
$-19.5$ & $-19.0$ & $-19.29$ & $7.90$ & $80\pm 18\times 10^{-5}$  \\
$-19.0$ & $-18.5$ & $-18.83$ & $8.07$ & $116\pm 28\times 10^{-5}$  \\
$-18.5$ & $-18.0$ & $-18.27$ & $7.89$ & $253\pm 79 \times 10^{-5}$ \\
\hline
\multicolumn{5}{c}{Redshift bin: $8.5<z<9.5$}\\
\hline
$-21.0$ & $-20.5$ & $-20.58$ & $8.73$ & $7.1^{+7.0}_{-4.0}\times 10^{-5}$  \\
$-20.5$ & $-20.0$ & $-20.30$ & $9.04$ & $9.5^{+7.7}_{-4.8}\times 10^{-5}$ \\
$-20.0$ & $-19.5$ & $-19.53$ & $8.92$ & $23\pm9\times 10^{-5}$  \\
$-19.5$ & $-19.0$ & $-19.34$ & $9.12$ & $31\pm11\times 10^{-5}$  \\
$-19.0$ & $-18.5$ & $-18.89$ & $8.86$ & $67\pm 24\times 10^{-5}$  \\
$-18.5$ & $-18.0$ & $-18.35$ & $8.78$ & $39^{+52}_{-26} \times 10^{-5}$ \\
\hline
\multicolumn{5}{c}{Redshift bin: $9.5<z<11$}\\
\hline
$-21.0$ & $-20.0$ & $-$      & $-$     & $<1.7\times 10^{-5}$  \\
$-20.0$ & $-19.0$ & $-19.28$ & $10.02$  & $12.4\pm4.9\times 10^{-5}$  \\
$-19.0$ & $-18.0$ & $-18.92$ & $9.92$ & $10.0^{+9.9}_{-5.7}\times 10^{-5}$  \\
\hline
\multicolumn{5}{c}{Redshift bin: $11<z<12.5$}\\
\hline
$-21.0$ & $-20.0$ & $-$      & $-$     & $<2.0\times 10^{-5}$  \\
$-20.0$ & $-19.0$ & $-19.69$ & $11.91$ & $2.6^{+3.4}_{-1.7}\times 10^{-5}$  \\
$-19.0$ & $-18.0$ & $-18.80$ & $11.11$ & $4.8^{+11.1}_{-4.1}\times 10^{-5}$  
\enddata
\tablenotetext{}{Upper limits are 1$\sigma$. All small number ($N<5$) uncertainties use the Poisson results of \cite{Gehrels1986}. Both Poisson and cosmic variance are included in the uncertainties.}
\end{deluxetable}

\subsection{Binned luminosity function}
\label{sec:binned}
For a first estimate of the evolving luminosity function, the galaxy sample is binned in both redshift and absolute magnitude. The bins of both axes are of variable width with smaller bins at $z<9.5$ than at higher redshifts, due to the paucity of $z>9.5$ galaxies in the sample. We use the binned $1/V_{a}$ method of \cite{Avni1980} where
$V_{a}^{i}$ is the co-moving volume available for a source $i$ in a bin with sizes $\Delta M_{\rm UV}$ and $\Delta z$. The available volume takes into
account the completeness, $p(M_{\rm UV},z)$, of Section\,\ref{sec:completeness} and is
\begin{displaymath}
V_{a}^{i} ~= \int \hspace{-0.15cm} \int p(M_{\rm UV},z)~ \frac{dV}{dz} ~dz ~dM_{\rm UV}~{\rm Mpc}^{3}.
\end{displaymath}
The combination of $p(M_{\rm UV},z)$ and volume element $dV/dz$ accounts for the effective sky area of all fields at this location in $M_{\rm UV},z$ space. The luminosity function in the bin is then
\begin{displaymath}
\Phi = \sum^{N}_{i=1} ~\frac{1}{V_{a}^{i}}~ (\Delta M_{\rm UV})^{-1} ~{\rm Mpc}^{-3}\,{\rm mag}^{-1}.
\end{displaymath}
Uncertainties on the space densities use Poisson statistics in the volume available integral with small number ($N<5$) corrections using \cite{Gehrels1986}. We add the uncertainty due to cosmic variance in quadrature using the cosmic variance calculator of \cite{Trenti2008}. The effect of cosmic variance is minor in most bins because CANUCS covers five widely separated fields on the sky. For the two bins that do not contain any galaxies we quote and plot $1\sigma$ upper limits.

\begin{figure*}
    \centering
    \includegraphics[width=\linewidth]{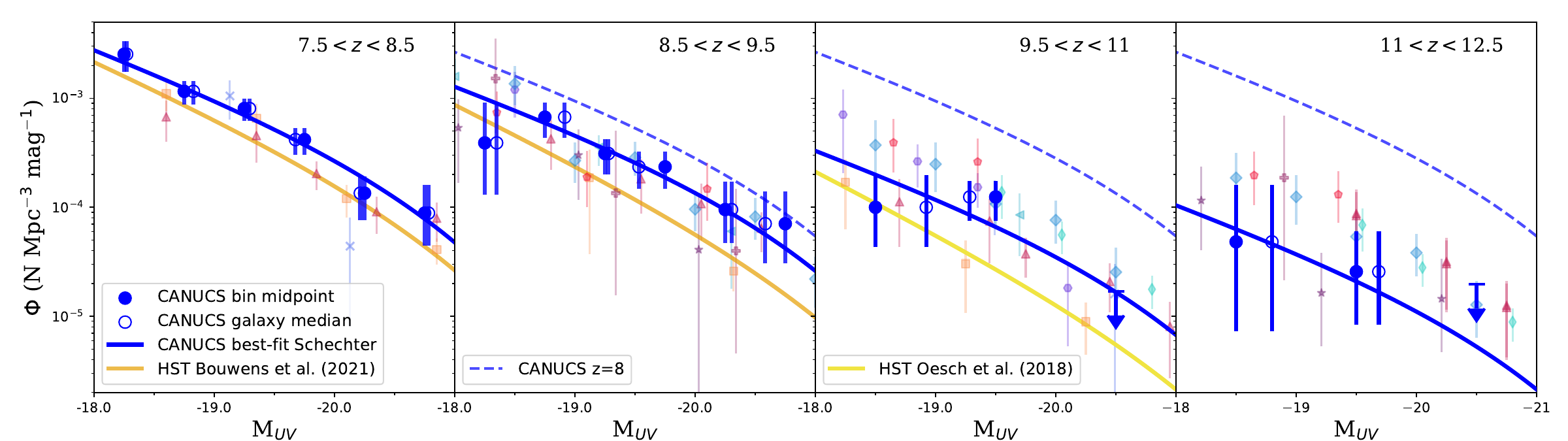}
    \caption{The galaxy luminosity function in four redshift bins. The CANUCS binned data from Section \ref{sec:binned} are plotted twice: filled blue circles are shown at the midpoint $M_{\rm UV}$ of the bin, whereas open blue circles are at the median $M_{\rm UV}$ of the galaxies in the bin. The difference between these can be substantial when there are few objects in the bin. For bins without any galaxies, $1\sigma$ upper limits are shown with blue arrows. The lowest luminosity bins at $z>8.5$ contain fairly low numbers of galaxies with a low overall completeness (Figure \ref{fig:completeness}), so those space densities should be taken with caution. The blue curves show the best-fit Schechter functions from Section \ref{sec:parametric}. The blue dashed curve in the three higher redshift bins shows the best-fit CANUCS $z=8$ Schechter function for reference. The orange and yellow curves shows the best-fit Schechter functions based on \hst\ data from \cite{Bouwens2021} and  \cite{Oesch2018}. Binned data from other studies are also plotted. For cases where these data use different redshift bin centers than our study, we make an evolutionary adjustment based on the typical evolution seen in previous \jwst\ studies (Equation \ref{eq:kevol} with $k=-0.2$). From left to right the literature data points are orange square \citep{Bouwens2021}, blue cross \citep{Bouwens2023a}, red triangle \citep{Adams2023arXiv}, blue triangle \citep{McLeod2016}, purple star \citep{Harikane2023a}, burgundy plus \citep{Morishita2023a}, red pentagon \citep{Leung2023}, blue diamond \citep{Finkelstein2023arXiv}, purple hexagon \citep{Donnan2023}, turquoise diamond \citep{McLeod2024}.}
    \label{fig:binned}
\end{figure*}

The binned results are presented in Table \ref{tab:binnedlf} and plotted in Figure \ref{fig:binned}. We observe consistent decreases in the galaxy space density going to higher redshifts. We find a somewhat higher density at $8<z<10$ than was found with \hst\ by \cite{Bouwens2021} and \cite{Oesch2018}, however we are consistent with the \hst\ results of \cite{McLeod2016}. In the bins at $z=8$ and $z=9$ we find good agreement with previous \jwst\ studies. In the two higher redshift bins, our space density values are comparable with the lower envelope of results from other studies with \jwst. We do not see a large population of $z>10$ galaxies.

At the two highest redshifts where there are few galaxies detected, there are sometimes substantial differences between the bin centers and median values of both $z$ and $M_{\rm UV}$. The median redshift in the bin is most often lower than the bin center due to the sharp decline in space density with redshift and the lower completeness at higher redshift for a fixed $M_{\rm UV}$. We note that this effect, combined with the fact that photometric redshift estimates in some studies are significantly biased towards higher redshifts (Section \ref{sec:specver}), means that in some studies the space density quoted for a redshift bin may actually correspond to a redshift substantially lower than the nominal bin center. 

\subsection{Parametric luminosity function}
\label{sec:parametric}

We use a Monte-Carlo Markov Chain (MCMC) method to fit a parametric luminosity function separately to the data in each of the four previously defined redshift bins. It is common to use either the \cite{Schechter1976} function or a double power-law. Whilst there is some evidence that double power-laws provide a better fit to galaxy luminosity functions at $z>7$ \citep{Bowler2020, Harikane2023a}, we adopt the Schechter function in this study since there is little difference between the two forms for the faint galaxies we find in CANUCS and the Schechter function uses one fewer parameter. 

The best-fit parameters and their uncertainties are derived using MCMC to minimize the function $S=-2 \ln \mathcal{L}$, where $\mathcal{L}$ is the likelihood;
\begin{eqnarray}
\label{eq:slike}
\lefteqn{\nonumber S ~=~ -2 \sum^{N}_{i=1} ~\ln~ [\Phi(M_{\rm UV\,i},z_{i})~ p(M_{\rm UV\,i},z_{i})]}\\
\nonumber & & +2 \int \hspace{-0.15cm} \int \Phi(M_{\rm UV},z)~ p(M_{\rm UV},z)~ \frac{dV}{dz}~ dz~dM_{\rm UV}
\end{eqnarray}
where the first term is a sum over each galaxy and the second is integrated over the full possible range of redshift and absolute magnitude \citep{Marshall1983}. 
Initial starting values of the three Schechter function parameters for the $7.5<z<8.5$ redshift bin are estimated from the binned luminosity function. Table \ref{tab:fitlf} and Figure \ref{fig:binned} show the results of the fitting. The best-fit function is very similar to that determined with \hst\ \citep{Bouwens2021}, with a marginally higher normalization.

For the three bins at higher redshifts, we do not have sufficient statistics and dynamic range in luminosity to fit all three parameters simultaneously. Based on the similar shapes of the binned luminosity functions at all redshifts in Figure \ref{fig:binned} and the results of \cite{Bouwens2021} showing that at $z>4$ the characteristic density, $\Phi^{*}$, varies much more than the characteristic absolute magnitude, $M^{*}_{\rm UV}$, we fix $M^{*}_{\rm UV}$ to $-20.8$, and the faint end slope, $\alpha$, to $-2.0$, consistent with our best-fit $z=8$ values and leave only $\Phi^{*}$ as a free parameter. The $z=8$ fit is also repeated with only $\Phi^{*}$ as a free parameter to obtain a meaningful uncertainty on $\Phi^{*}$, since it has high covariance with the other two parameters in the three parameter fit.

\begin{deluxetable}{c c c c}
\hspace{-0.5cm}
\tablewidth{220pt}
\tablecaption{\label{tab:fitlf} Best-fit Schechter Function Parameters}
\tablehead{Redshift & ${\rm Log}_{10}\,\Phi^{*}$ & $M^{*}_{\rm UV}$ & $\alpha$ \\
                    & (Mpc$^{-3}$\,mag$^{-1}$)   &                  & }
\startdata
\hline
$7.5<z<8.5$ & $-3.64^{+0.48}_{-0.59}$ & $-20.75^{+0.64}_{-0.81}$ & $-2.04^{+0.30}_{-0.24}$   \\
\hline
$7.5<z<8.5$ & $-3.63^{+0.04}_{-0.04}$ & $-20.8$ (fixed) & $-2.0$ (fixed)   \\
$8.5<z<9.5$ & $-3.95^{+0.07}_{-0.07}$ & $-20.8$ (fixed) & $-2.0$ (fixed) \\
$9.5<z<11$  & $-4.53^{+0.13}_{-0.14}$ & $-20.8$ (fixed) & $-2.0$ (fixed) \\
$11<z<12.5$ & $-5.03^{+0.24}_{-0.29}$ & $-20.8$ (fixed) & $-2.0$ (fixed)
\enddata
\tablenotetext{}{The first row gives the best-fit $z=8$ values when all three parameters are fit simultaneously. The next four rows are when $M^{*}_{\rm UV}$ and $\alpha$ are fixed to the best-fit $z=8$ values.}
\end{deluxetable}

The best-fit Schechter functions are plotted in Figure \ref{fig:binned}. They track the CANUCS binned data fairly well, showing that this parametric form is a good fit to the data. As is seen for the binned data, the space density decreases sharply beyond $z=9.5$, although at a slightly less rapid pace than previously seen with \hst.

\subsection{Luminosity density evolution}
\label{sec:evolution}
Integrating the luminosity function gives the UV radiation luminosity density. In the absence of dust, and neglecting variations due to stellar ages, this is proportional to the star formation rate density (e.g. \citealt{Madau2014}). As such, it is a quantity that can be closely compared with the results of simulations. 

\begin{figure*}
    \centering
    \includegraphics[width=\linewidth]{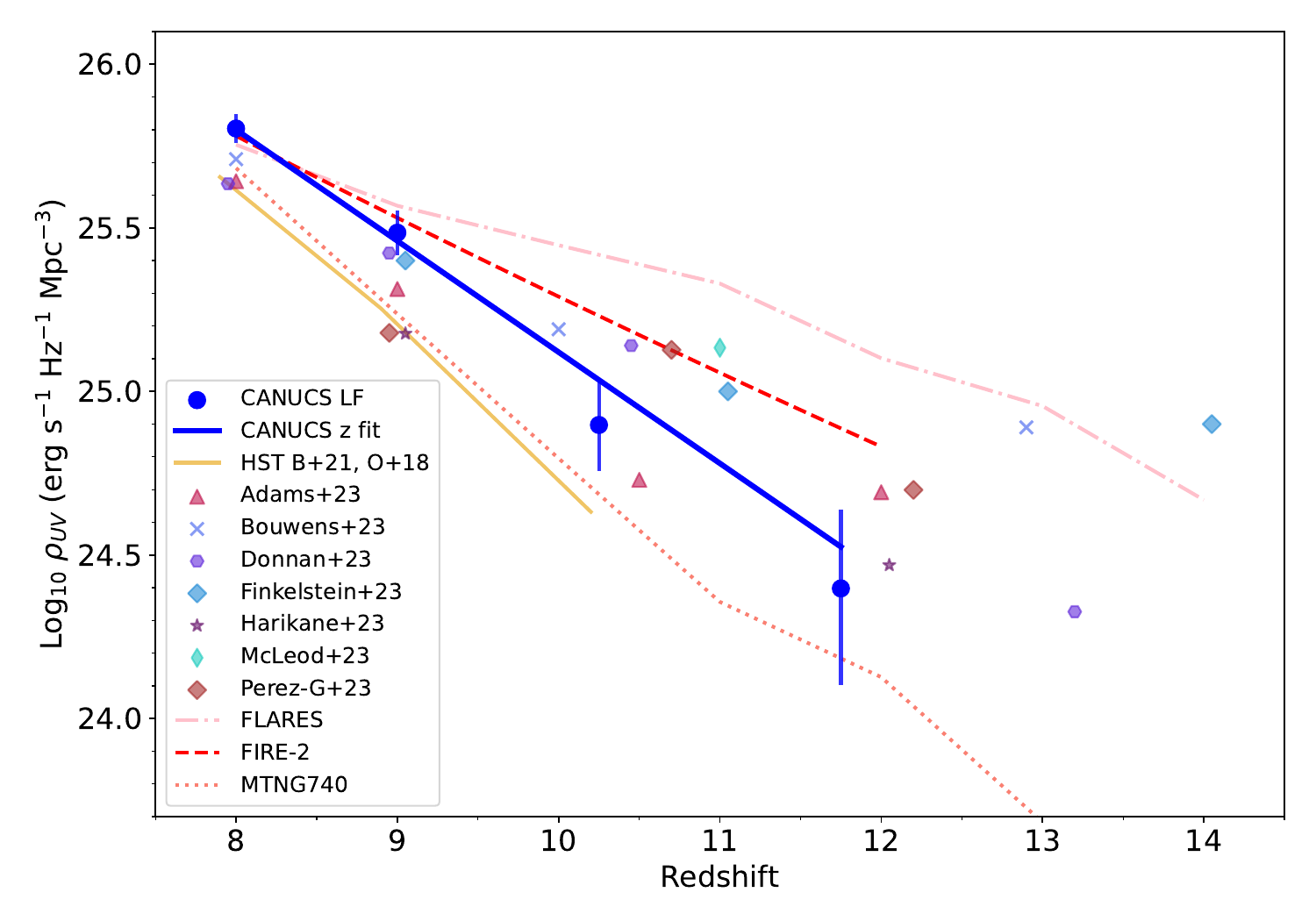}
    \caption{The evolution of the integrated UV luminosity density from the CANUCS luminosity functions of Table \ref{tab:fitlf} are shown as blue circles. The log-linear fit to these points ($k=-0.35$) is shown with the solid blue line. Results from other \jwst\ analyses are shown with the colored symbols labeled in the legend (error bars omitted for clarity). Results from previous works with \hst\ are shown with a solid orange line. Results from three simulations are shown with red/pink dashed and dotted lines (FLARES; \citealt{Vijayan2021,Wilkins2023a}, FIRE-2; \citealt{Sun2023},  MTNG740; \citealt{Kannan2023}). The evolution observed in CANUCS is only marginally shallower than was seen with \hst\ up to $z\approx 10$. The evolution is comparable with that predicted by the MTNG740 simulation and steeper than predicted by FLARES and FIRE-2. For all data shown here the luminosity function is integrated over the range $-17>M_{\rm UV}>-23$.}
    \label{fig:integrated}
\end{figure*}

We integrate the UV galaxy luminosity functions in Table \ref{tab:fitlf} over the range $-17>M_{\rm UV}>-23$ to measure the luminosity density in each of our redshift bins. The results are shown in Table \ref{tab:integrated} and Figure \ref{fig:integrated}. We fit the evolution in the CANUCS values with a simple log-linear evolutionary model in redshift $z$:
\begin{equation}
\label{eq:kevol}
{\rm log}_{10}\,\rho_{UV}\, ( {\rm erg \, s^{-1}\, Hz^{-1}\, Mpc}^{-3}) = A + k(z-8)
\end{equation}
where $A$ is the value at $z=8$ and $k$ is the evolutionary parameter. We find best-fit parameters of $A=25.80$ and $k=-0.35 \pm 0.11$. This log-linear fit is a reasonable approximation within the uncertainties, however the individual points show something of a steeping beyond $z=9$, reminiscent of the quadratic constant star formation efficiency curve that fits the \hst\ data up to $z=10$ \cite{Bouwens2021}. The CANUCS data are $\approx 0.2$ dex above the \hst\ curve at all redshifts, showing a similar evolution. The CANUCS data show a more rapid decline than most of the previous \jwst\ studies. The main exception is the EPOCHS study of \cite{Adams2023arXiv} which shows a comparable decline to CANUCS.

Figure \ref{fig:integrated} also compares the observed evolution with the predictions of three galaxy evolution simulations: FLARES \citep{Vijayan2021, Wilkins2023a}, FIRE-2 \citep{Sun2023} and Millennium-TNG (MTNG740, \citealt{Kannan2023}). The CANUCS results most closely match the evolution from MTNG740. This will be discussed further in the following section. 

\begin{deluxetable}{c c}
\hspace{-0.5cm}
\tablewidth{220pt}
\tablecaption{\label{tab:integrated} Evolution of UV Luminosity Density}
\tablehead{Redshift & ${\rm Log}_{10}\,\rho_{UV}$ (erg s$^{-1}$ Hz$^{-1}$ Mpc$^{-3}$)}
\startdata
\hline
$7.5<z<8.5$ & $25.80^{+0.04}_{-0.04}$   \\
$8.5<z<9.5$ & $25.49^{+0.07}_{-0.07}$\\
$9.5<z<11$  & $24.90^{+0.12}_{-0.14}$ \\
$11<z<12.5$ & $24.40^{+0.24}_{-0.29}$ 
\enddata
\tablenotetext{}{The luminosity function is integrated over the range $-17>M_{\rm UV}>-23$.}
\end{deluxetable}

\section{Discussion} \label{sec:discussion}

The CANUCS luminosity function shows a decline in the characteristic density, $\Phi^{*}$, and consequently the integrated UV luminosity density, at higher redshifts. Between $z=8$ and $z=11.75$ we observe a decrease of a factor of 25. We see marginal evidence for a steepening of this evolution with redshifts above $z=9$, as was also tentatively concluded based on \hst\ observations \citep{Oesch2014, Oesch2018, Bouwens2021}. In contrast, several studies with \jwst\ find a much shallower evolution between $z=8$ and $z=12$ with decreases in the range of 5 to 10 \citep{Donnan2023, Finkelstein2023a, Finkelstein2023arXiv, Bouwens2023a, McLeod2024, Perez-Gonzalez2023}. Only one of the previous \jwst\ studies \citep{Adams2023arXiv} finds as strong a decline in UV luminosity density between $z=8$ and $z=11$ as in CANUCS. \cite{Adams2023arXiv} do find a higher density at $z\approx 12$ than in CANUCS, although uncertainties are large in all studies at this redshift.

It is useful to consider which data were used by these various studies. \cite{Perez-Gonzalez2023} use only a single deep NIRCam pointing in GOODS-South, so the cosmic variance uncertainty is very high \citep{Steinhardt2021}. \cite{Finkelstein2023a} use the first four CEERS pointings and \cite{Finkelstein2023arXiv} use the full set of ten CEERS pointings. \cite{Donnan2023} and \cite{Bouwens2023a} use  the first four CEERS pointings, SMACS0723 and GLASS. Most of the area in these two studies is provided by CEERS and GLASS and this is indeed where most of the $z>10$ objects lie. \cite{Adams2023arXiv}  use these same fields, but also considerable additional area from the PEARLS and JADES surveys. \cite{McLeod2024} found a larger galaxy sample from many fields including the full 10 CEERS pointings, GLASS and UNCOVER (both in the Abell 2744 cluster) and several other lensing cluster or nearby galaxy fields. However, their sample of $z>10.5$ galaxies is still dominated (81\%) by galaxies from the directions of the CEERS and Abell 2744 fields. It is possible that both these fields are over-dense at certain redshifts. The Abell 2744 field, and in particular the GLASS field, is known to be over-dense in $z\approx 10$ galaxies \citep{Castellano2023}. It has not yet been reported that the CEERS field is also over-dense in $z>10$ galaxies. However we do find that CANUCS has considerably fewer bright galaxies at $z>10$ than are seen in most of these other studies.  For example, \cite{Finkelstein2023arXiv} show that the full CEERS survey contains eight galaxy candidates at $z>10$ with F277W magnitude $< 28$. This includes three galaxies spectroscopically confirmed at $z=11.416$, 11.043 and 10.01. By comparison, in CANUCS we find zero galaxies with $z>10$ and F277W magnitude $< 28$ in an effective sky area two-thirds as large as CEERS, where our simulations show high completeness of 70 to 90\%. We do find eight fainter $z>10$ galaxies in support of our high completeness at magnitude $< 28$. If the CEERS density of these bright galaxies was typical, we would have expected to find five in CANUCS, yet we find zero. Therefore it is quite possible that the CANUCS fields are under-dense compared to these other fields at $z>10$, but we await results from other large programs such as JADES and PRIMER to test this.

We now turn to the implications for theoretical simulations of galaxy formation. Based on the results of CANUCS and those of \cite{Adams2023arXiv}, only minor modifications to the standard assumptions of constant star formation efficiency and dust evolution are necessary. In Figure \ref{fig:integrated} we showed three simulation results for the integrated UV luminosity density evolution. The model predicting the shallowest evolution is FLARES. This is due to the very steep faint end slope ($\alpha=-3.1,-3.5$ at $z=10,12$, respectively, \citealt{Wilkins2023a}). Such a steep faint end slope has not been seen in any of the \jwst\ studies so far and if true would certainly have been noticed in some of the deeper fields \citep{Perez-Gonzalez2023, Leung2023}. \cite{Sun2023} showed that by considering increased scatter in the star-formation rate -- halo mass relation due to bursty, episodic star formation, the UV luminosity density at $z>10$ could be considerably enhanced. This would fit the evolution observed by some of the previous \jwst\ studies, but this model now significantly under-predicts the amount of evolution observed in CANUCS.
The model that is most consistent with the slope of the CANUCS evolution is the Millenium-TNG simulations of \cite{Kannan2023}. This model assumes a standard star-formation efficiency and yet provides a reasonably good fit to our data. We do observe a somewhat higher normalization than this model across all redshifts. 

\section{Conclusions} 
\label{sec:conclusions}

We use NIRCam imaging from five independent sightlines of the CANUCS survey to identify a new sample of $z>7.5$ galaxies and study the high-redshift evolution of the UV luminosity function from $z=8$ to $12$. Covering an effective unmasked, delensed area of 53.4 square arcminutes, we identify 158 galaxies with robust photometric redshifts of $z>7.5$. Two-thirds of our pointings and 80\% of the galaxies are covered by between 12 and 14 NIRCam filters, including many medium bands. These images provide accurate photometric redshifts and good discrimination against low-redshift interlopers. For a sub-sample of 28 CANUCS galaxies with NIRSpec spectroscopic redshifts, we find a very low scatter (normalized median absolute deviation $=0.024$) and low systematic offset of $\delta z = +0.21$. Most of the spectroscopic redshift galaxies are covered by only eight NIRCam filters in CLU fields, so we expect the performance for the galaxies with 12 or 14 filters in NCF fields to be even better. 

Our main conclusions are:
\begin{itemize}

\item{Most of galaxies have blue UV continuum slopes in the range $-3<\beta<-2$. The medium band filters F360M and F410M, in combination with F444W, show high equivalent width \Oiii+\Hb\ emission lines are very common out to at least $z=9$.}

\item{The $z=8$ to $12$ luminosity function is well fit by a Schechter function that has constant characteristic absolute magnitude and faint end slope $\alpha$ and only evolves in the characteristic space density, $\Phi$. We do not have enough bright galaxies to differentiate between a Schechter function and double power-law form.}

\item{The CANUCS luminosity functions derived at redshifts 8, 9 and 10 all have slightly higher normalization than those from some previous works with \hst\ \citep{Oesch2018, Bouwens2021}, although consistent with other \hst\ results \citep{McLeod2016}. We observe a steeper evolution from $z=8$ to $12$ than most recent studies with \jwst, with the exception of the work of \cite{Adams2023arXiv}. We find a clear deficit in UV-bright  $z>10$ galaxies compared to other studies, with no $z>10$ galaxies with F277W magnitude brighter than 28 in the CANUCS fields studied in this paper. There are various possible explanations for these differences including the number of NIRCam filters available in various fields, the robustness of photometric redshifts and intrinsic variance in the galaxy density along different sightlines.}

\item{The evolution in the UV luminosity density measured with CANUCS follows a log-linear relationship with redshift with coefficient $k=-0.35$. This rate of change is similar to that expected from modern hydrodynamical simulation codes that use standard assumptions about the efficiency of star formation when baryons accrete onto dark matter halos.} 
    
\end{itemize}

In future work we will extend our analysis to the low-luminosity population with $M_{\rm UV} \sim -15$ in the high-magnification regions of the CANUCS gravitational lensing clusters.

\section*{Acknowledgements}

We would like to thank the anonymous reviewer for their interesting and constructive suggestions.
This research was enabled by grant 18JWST-GTO1 from the Canadian Space Agency, and funding from the Natural Sciences and Engineering Research Council of Canada.
YA is supported by a Research Fellowship for Young Scientists from the Japan Society of the Promotion of Science (JSPS).
MB and GR acknowledge support from the ERC Grant FIRSTLIGHT and from the Slovenian national research agency ARRS through grants N1-0238, P1-0188 and the program HST-GO-16667, provided through a grant from the STScI under NASA contract NAS5-26555.
This research used the Canadian Advanced Network For Astronomy Research (CANFAR) operated in partnership by the Canadian Astronomy Data Centre and The Digital Research Alliance of Canada with support from the National Research Council of Canada the Canadian Space Agency, CANARIE and the Canadian Foundation for Innovation.

\section*{Data Availability}

Data presented in this paper will be made available upon request. The CANUCS DOI is \url{10.17909/ph4n-6n76}.

\facilities{\hst\ (ACS,WFC3), \jwst\ (NIRCam, NIRSpec)}

\bibliographystyle{aasjournal}
\bibliography{references} 

\appendix

\section{SEDs of CANUCS $z>7.5$ galaxy LF sample}
\label{sec:seds}

\begin{figure*}
    \centering
    \includegraphics[width=0.48\linewidth]{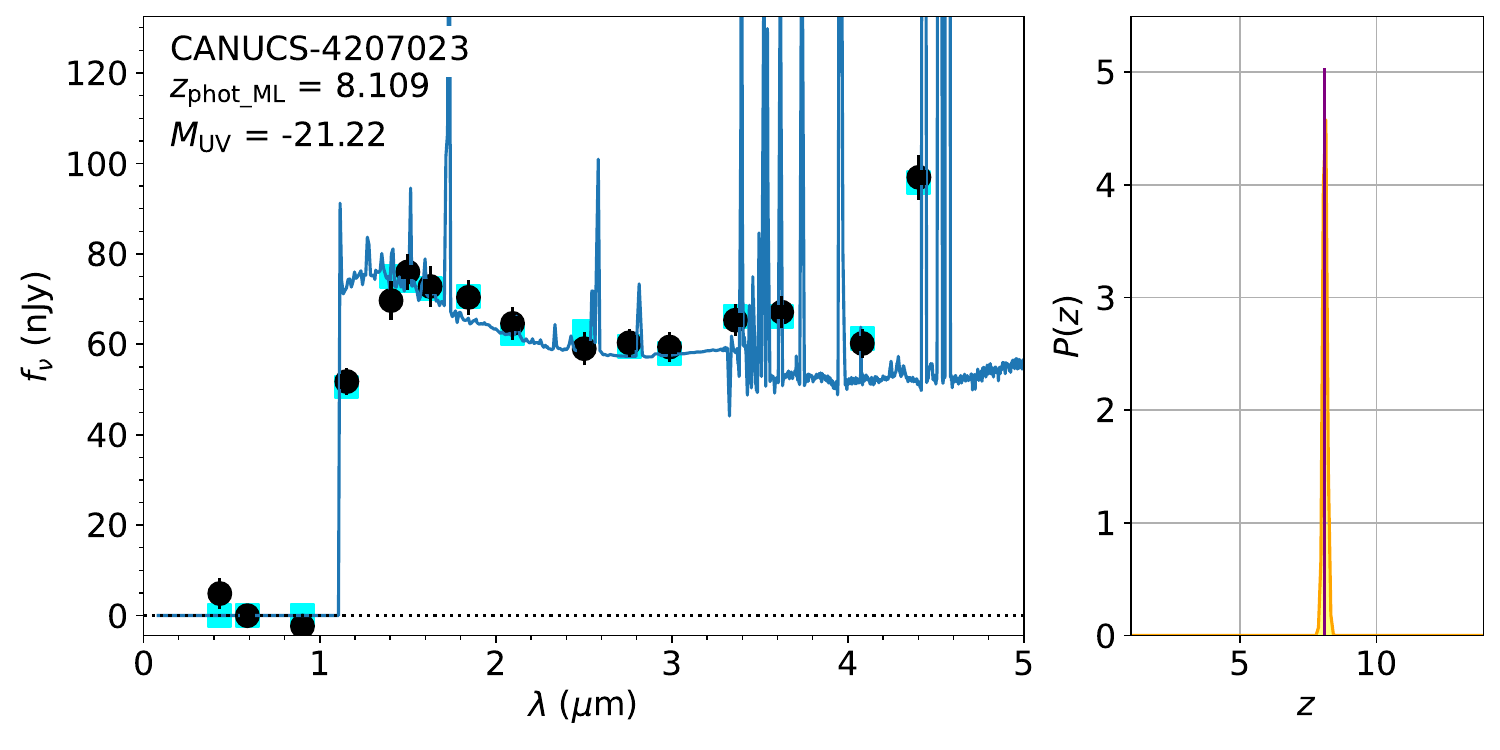}
    \includegraphics[width=0.48\linewidth]{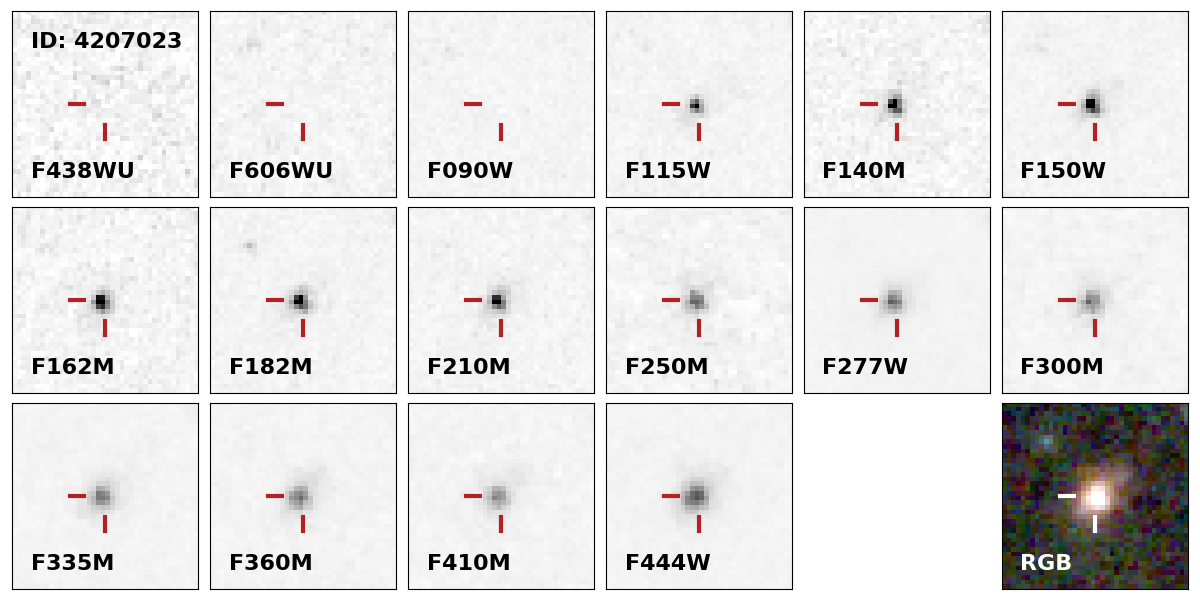}\\
    \includegraphics[width=0.48\linewidth]{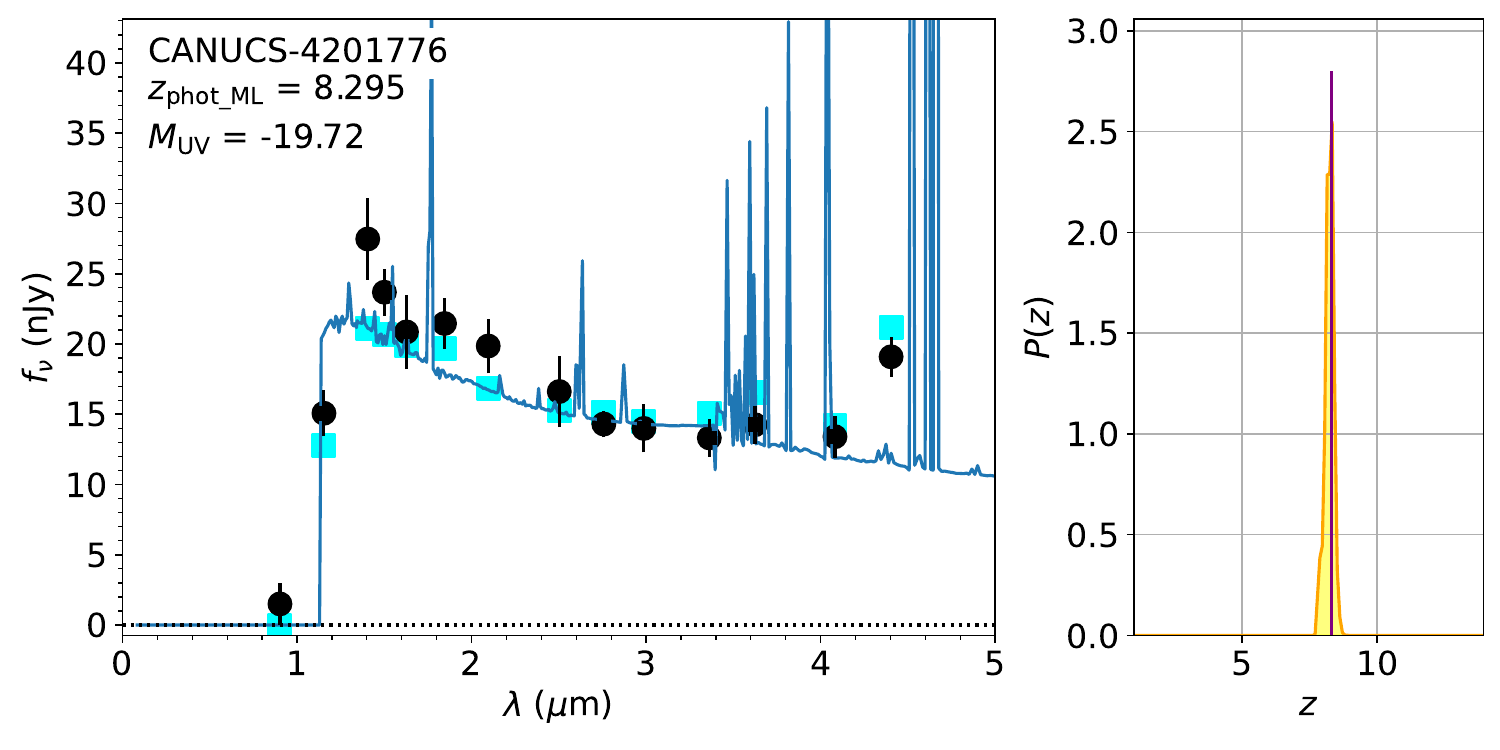}
    \includegraphics[width=0.48\linewidth]{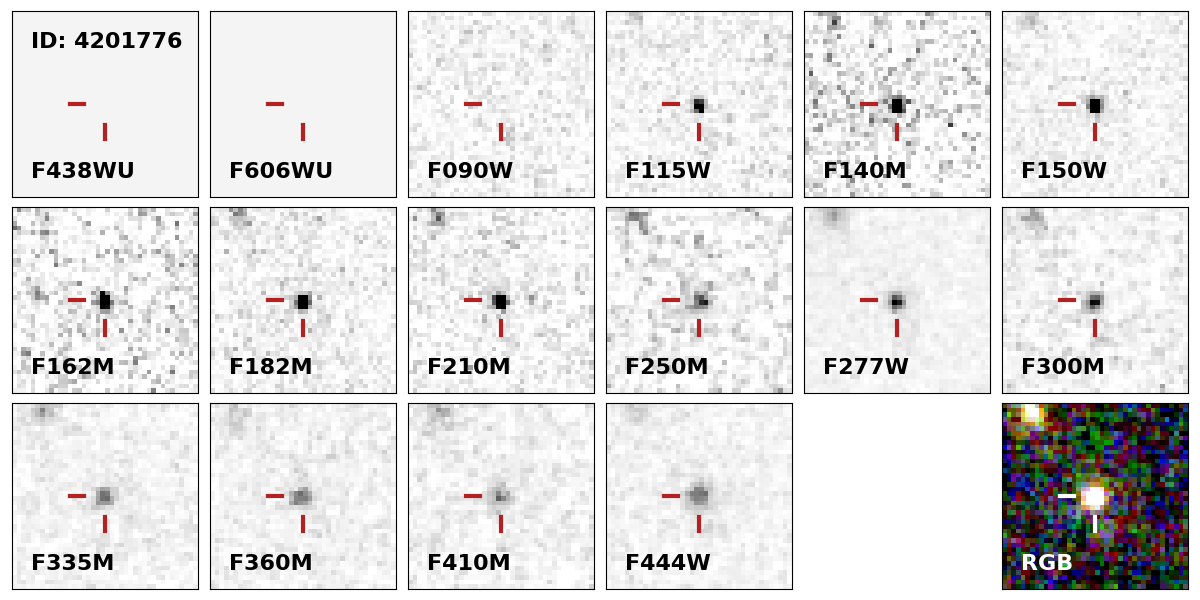}\\
    \includegraphics[width=0.48\linewidth]{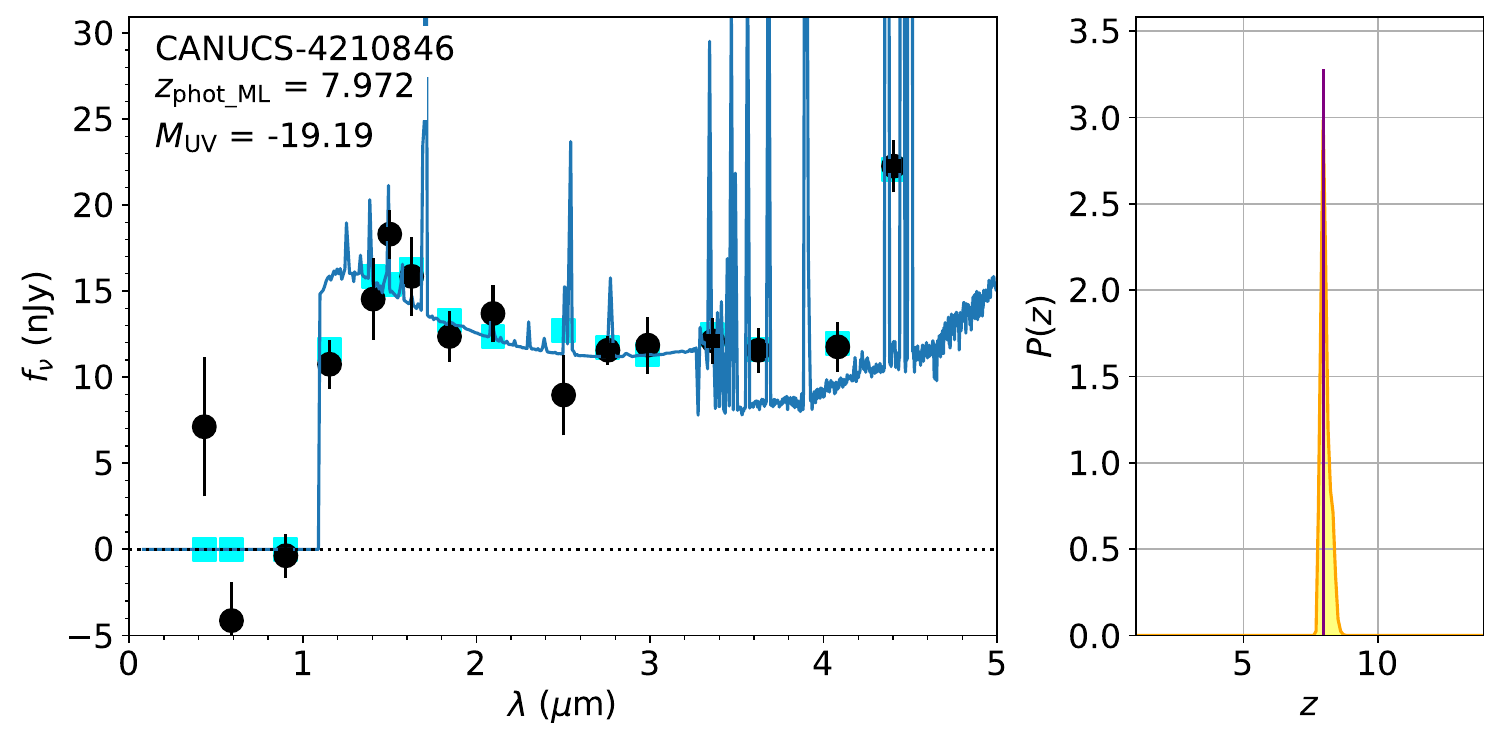}
    \includegraphics[width=0.48\linewidth]{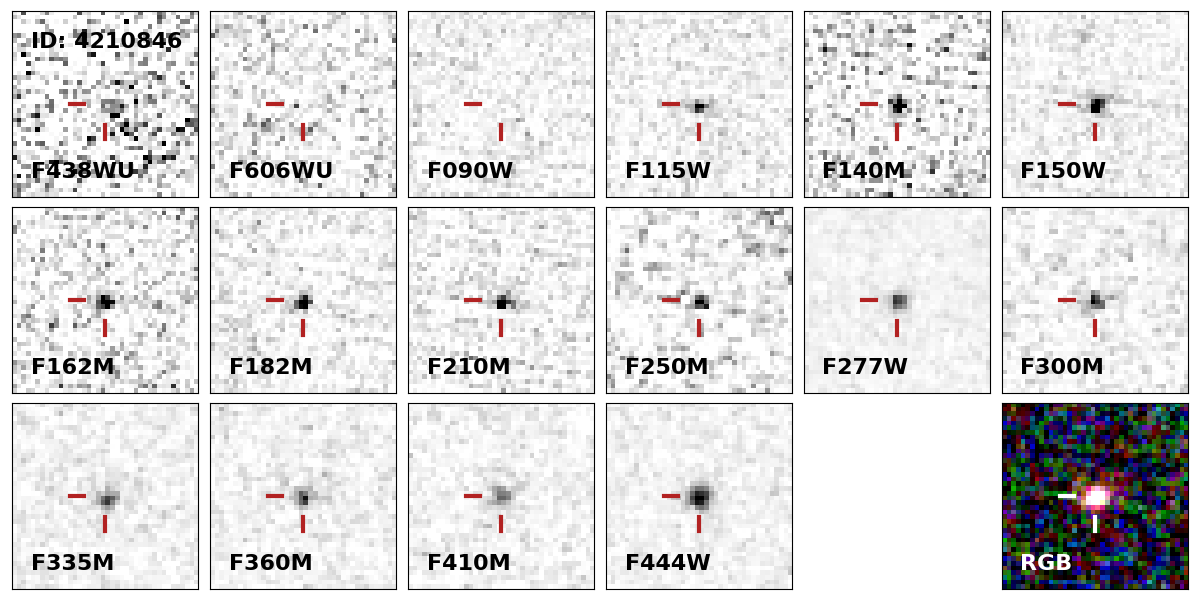}\\
    \includegraphics[width=0.48\linewidth]{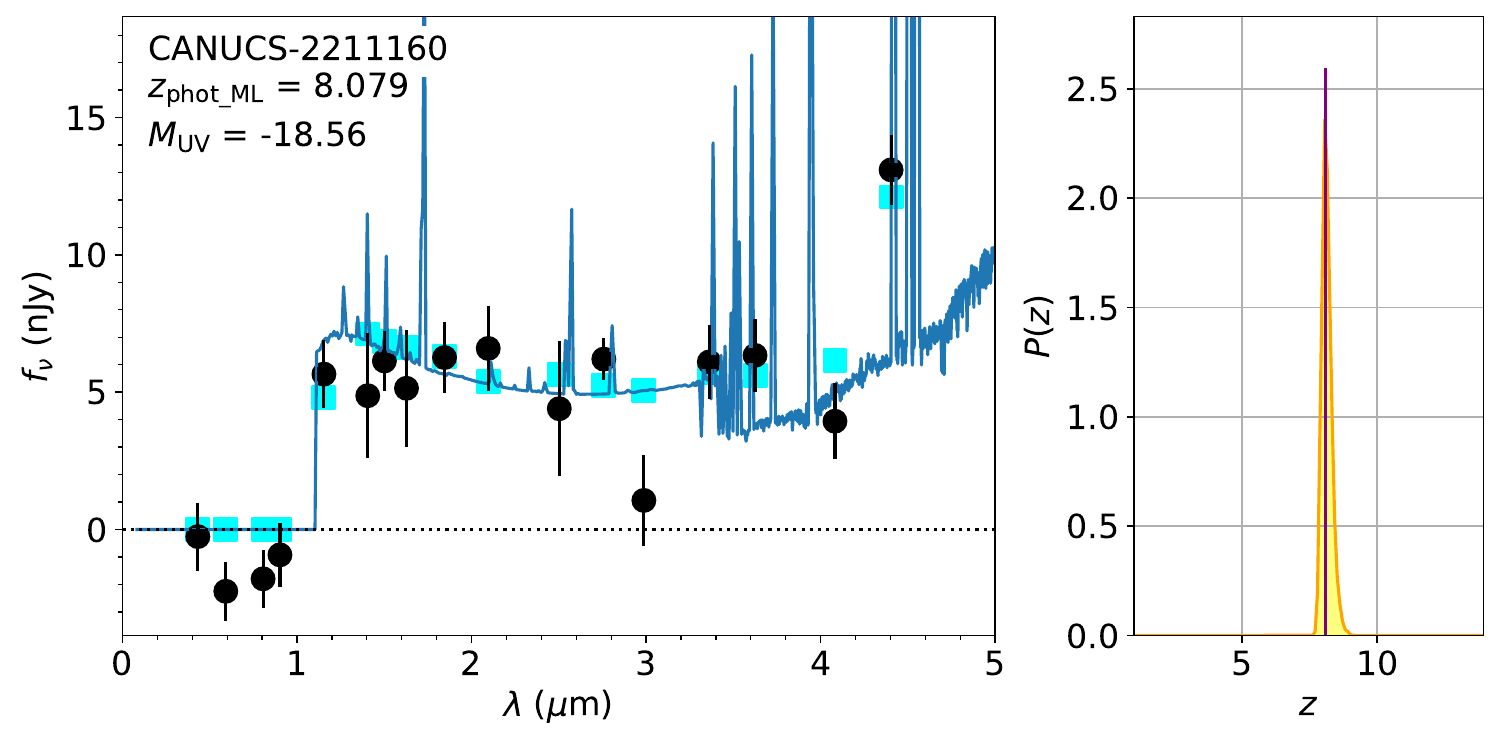}
    \includegraphics[width=0.48\linewidth]{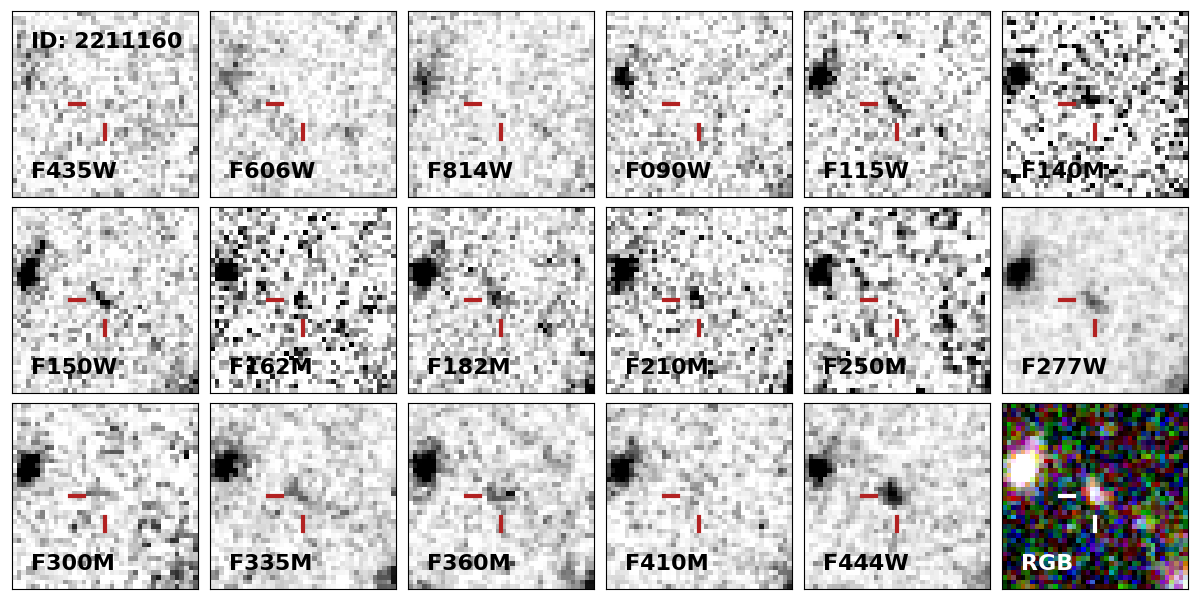}\\
    \includegraphics[width=0.48\linewidth]{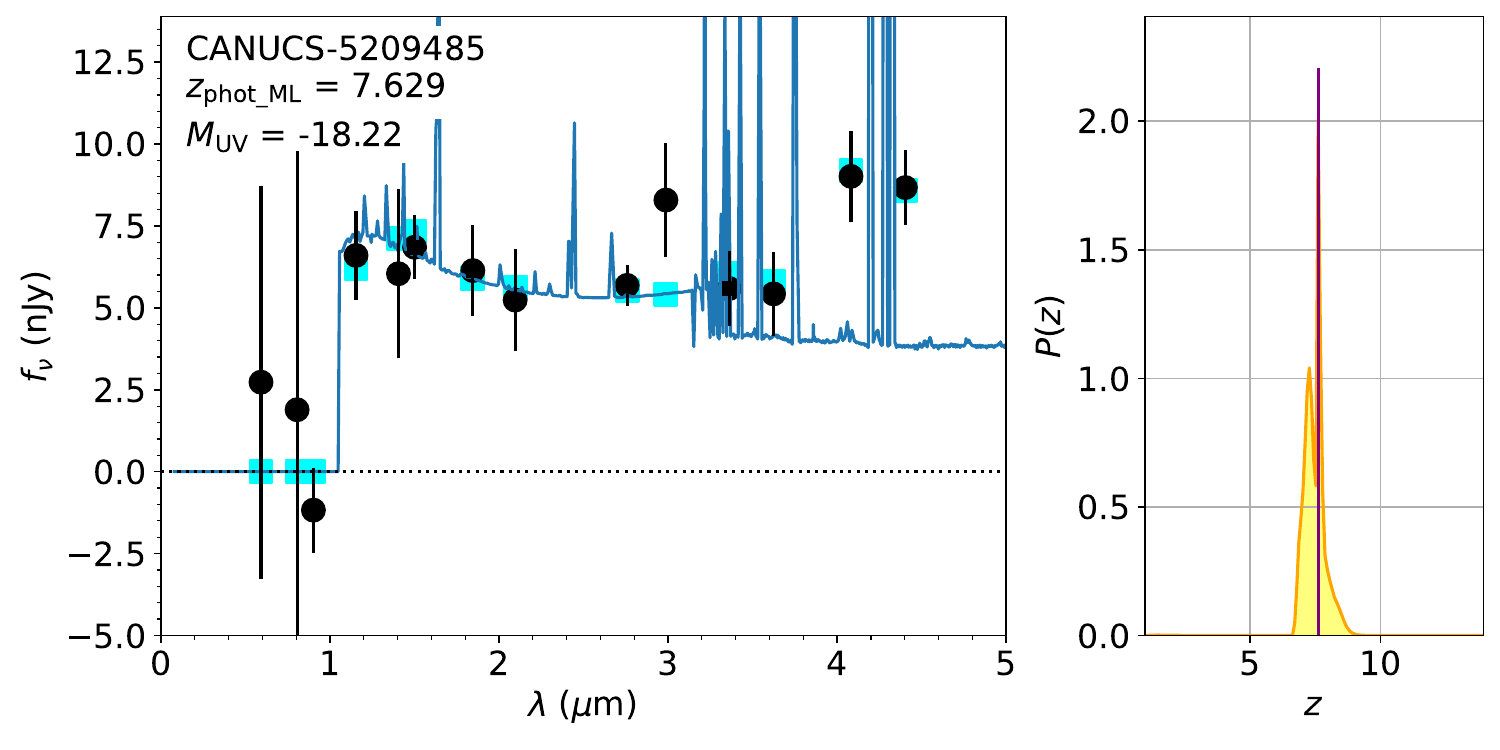}
    \includegraphics[width=0.48\linewidth]{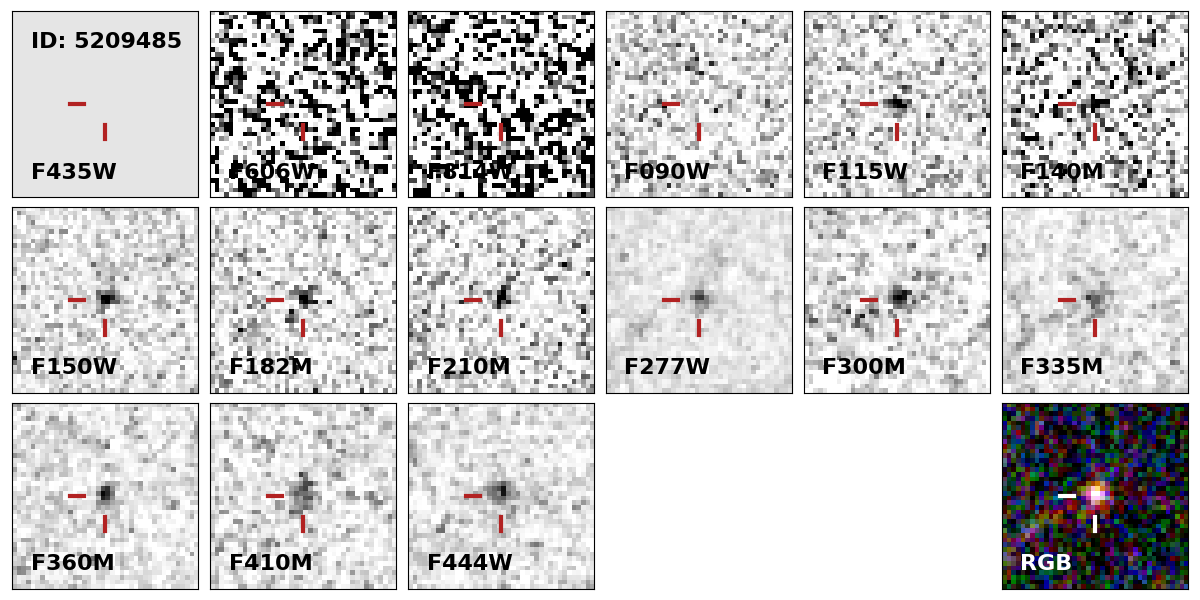}\\
    \caption{Example SEDs for CANUCS galaxies in the range $7.5<z_{\rm phot\_ML}<8.5$. The galaxies are ordered by increasing absolute magnitude. Left panels show the observed photometry (0.3 arcsec apertures on images convolved to the F444W PSF; black circles with errorbars), best-fit EAZY-py template (blue line) and template photometry integrated over filter bandpasses (cyan squares). Middle panels show the photometric redshift probability density with a purple line at $z_{\rm phot\_ML}$. Right panels show 1.6 arcsec cutouts in all the \hst\ optical (where available) and NIRCam filters. The RGB cutouts use colors R: all filters above $2.9\mu$m, G: all filters at $1.7 ~{\rm to}~ 2.9\mu$m, B: all NIRCam filters below $1.7\mu$m.}
    \label{fig:seds7p5}
\end{figure*}

\begin{figure*}
    \centering
    \includegraphics[width=0.48\linewidth]{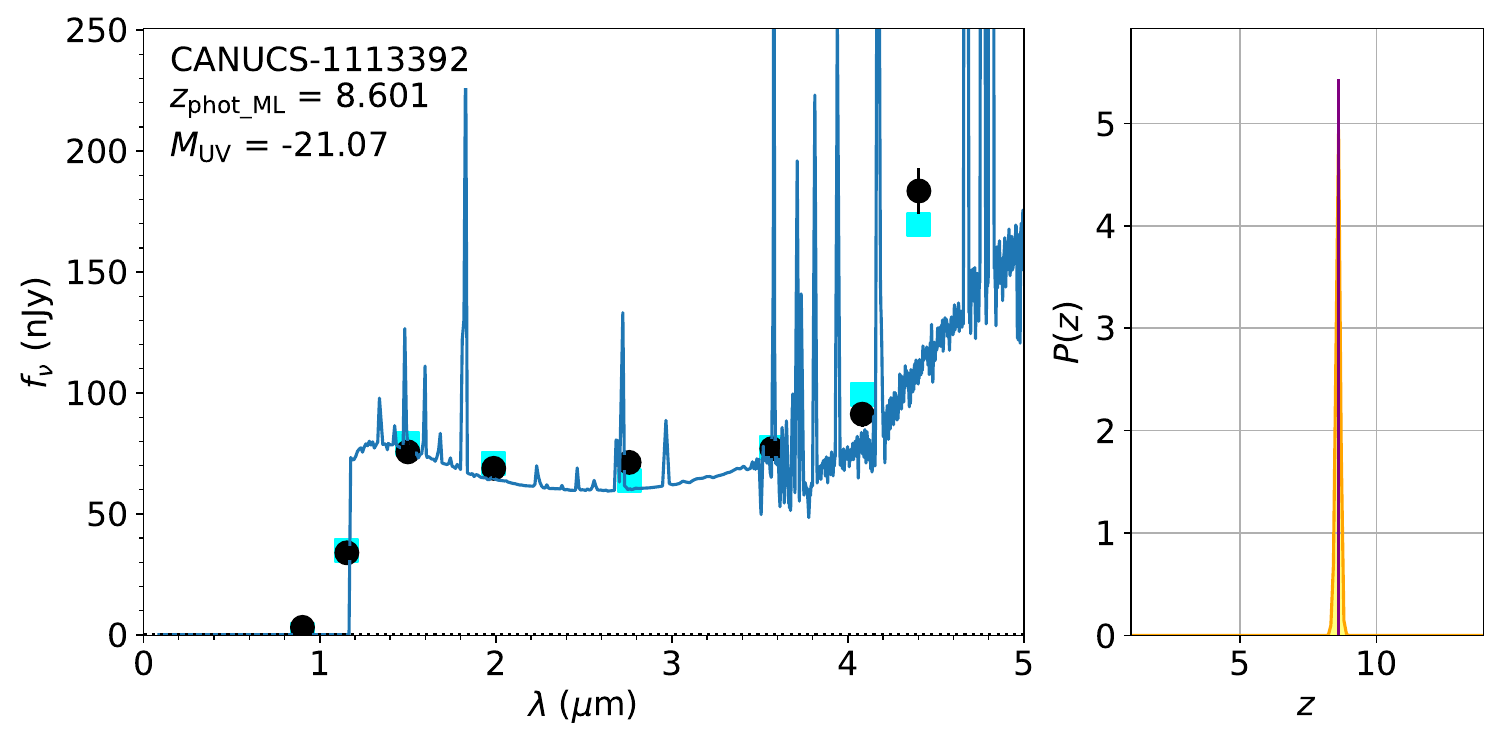}
    \raisebox{1.0cm}{\includegraphics[width=0.48\linewidth]{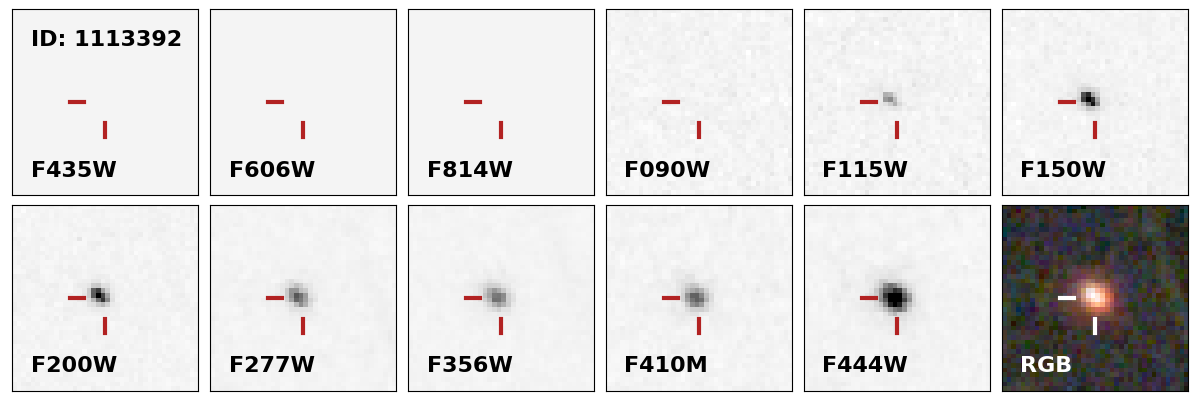}}\\
    \includegraphics[width=0.48\linewidth]{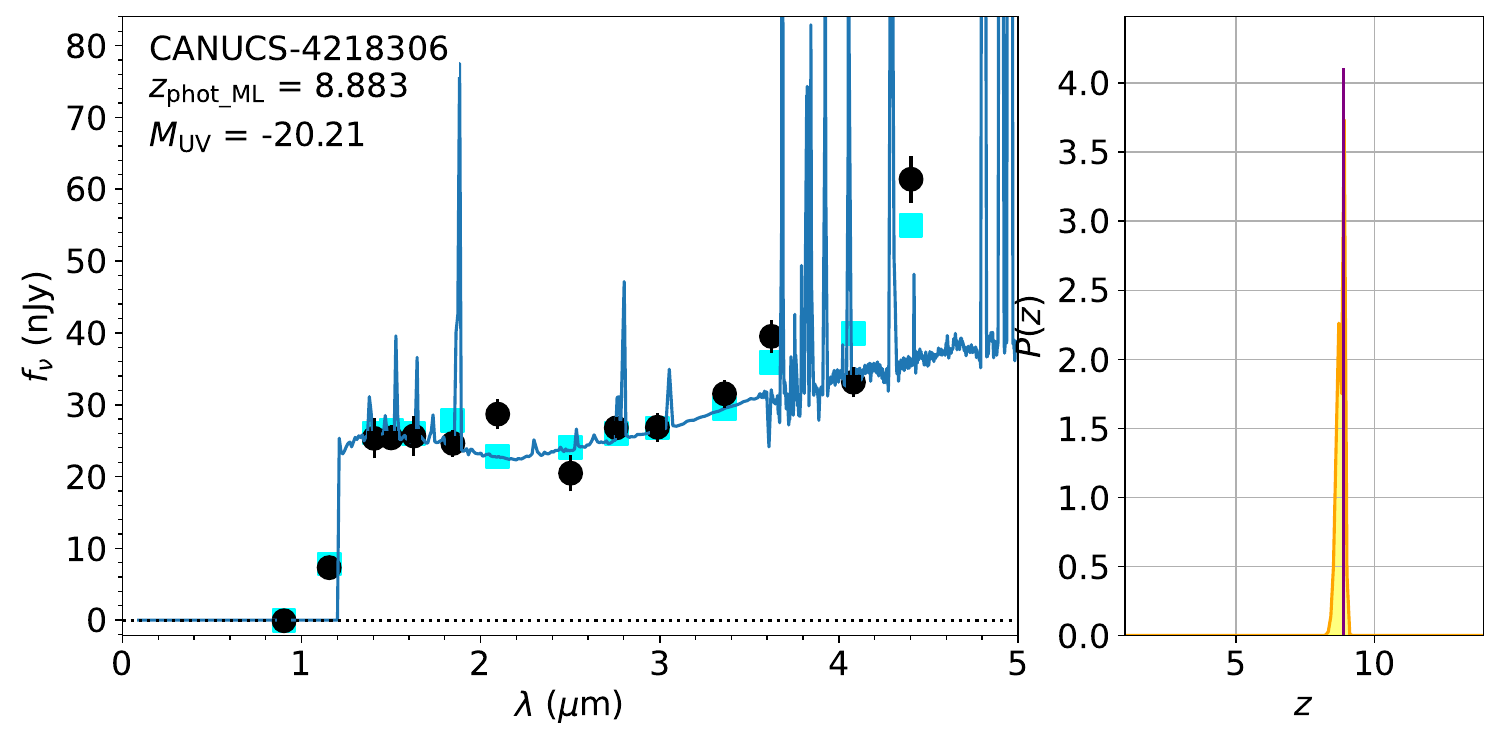}
    \includegraphics[width=0.48\linewidth]{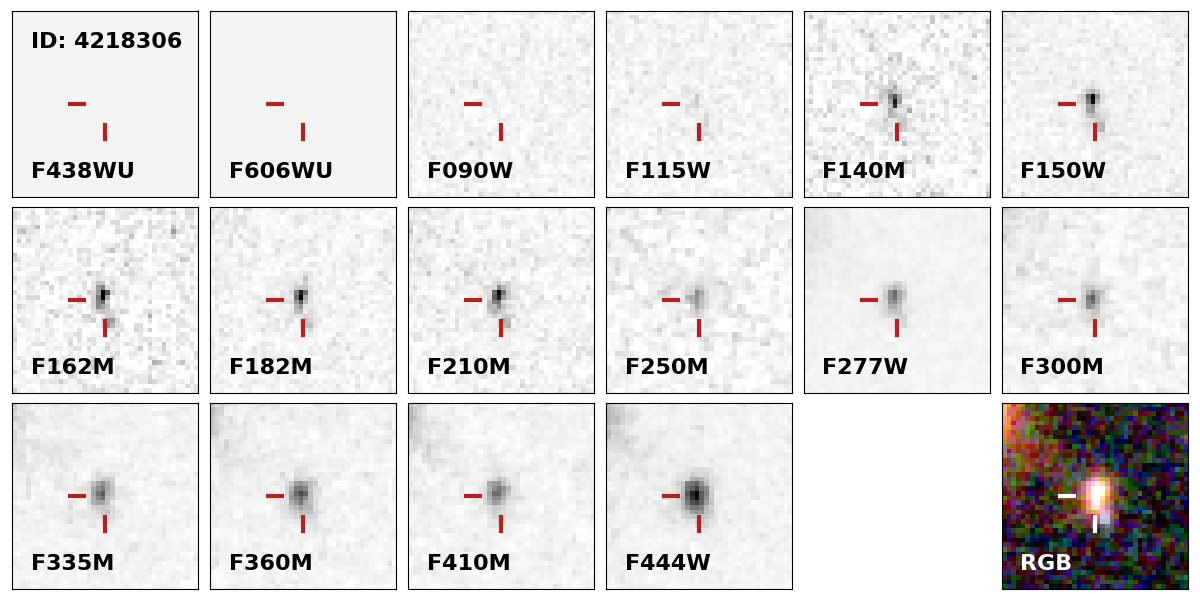}\\
    \includegraphics[width=0.48\linewidth]{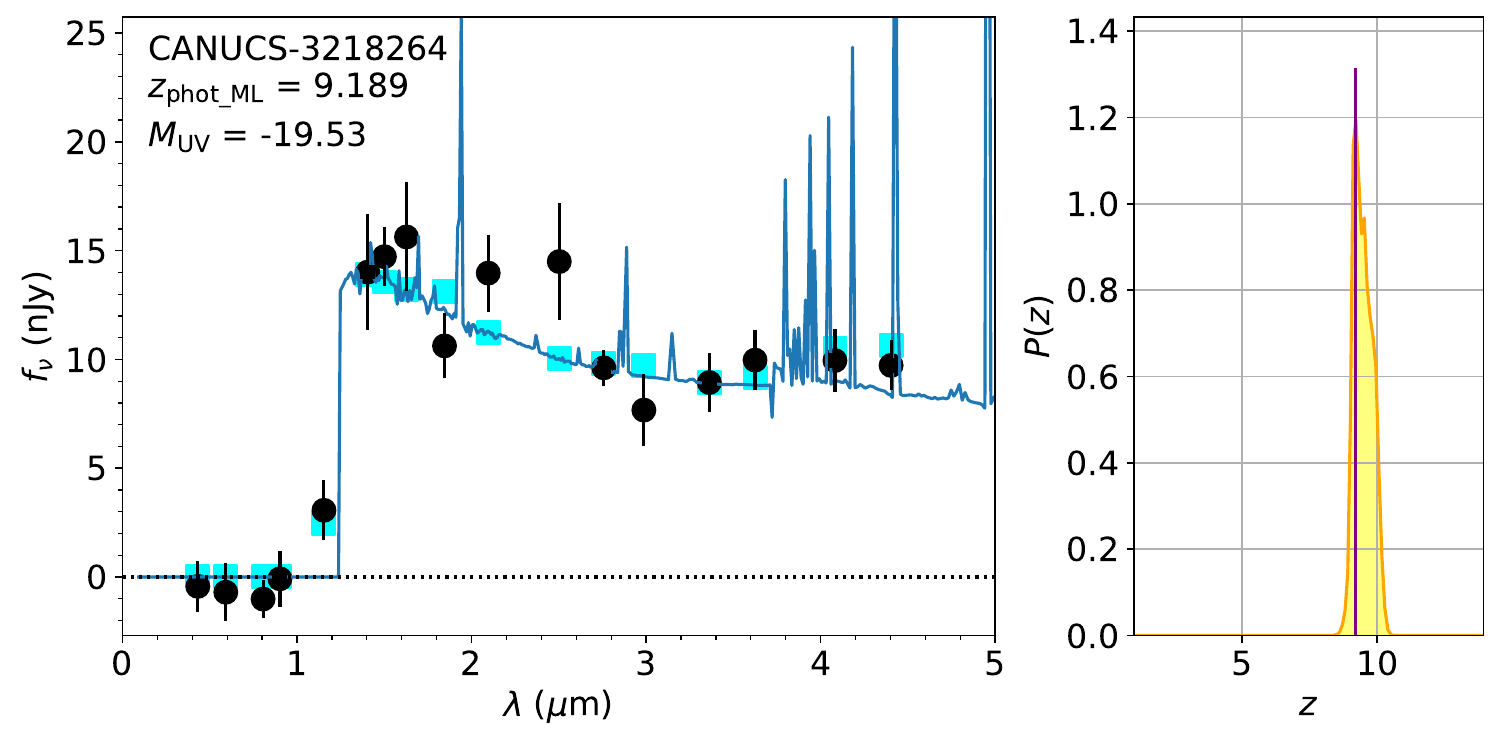}
    \includegraphics[width=0.48\linewidth]{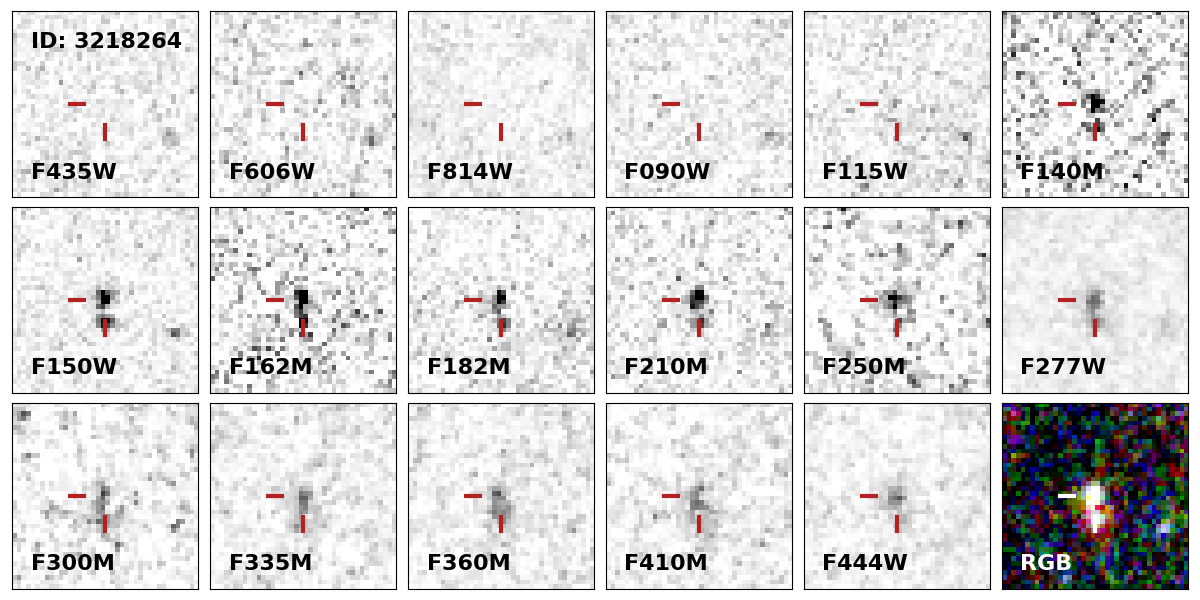}\\
    \includegraphics[width=0.48\linewidth]{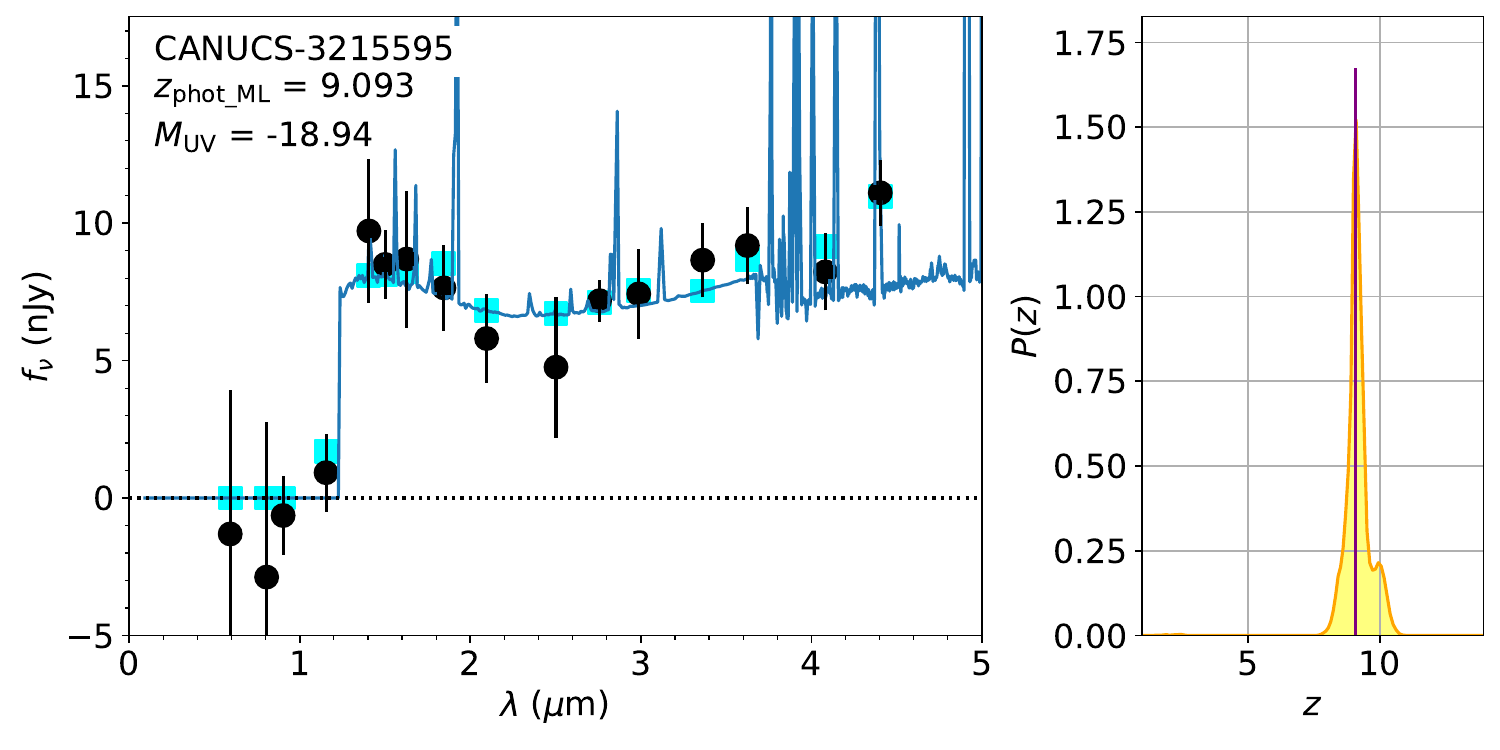}
    \includegraphics[width=0.48\linewidth]{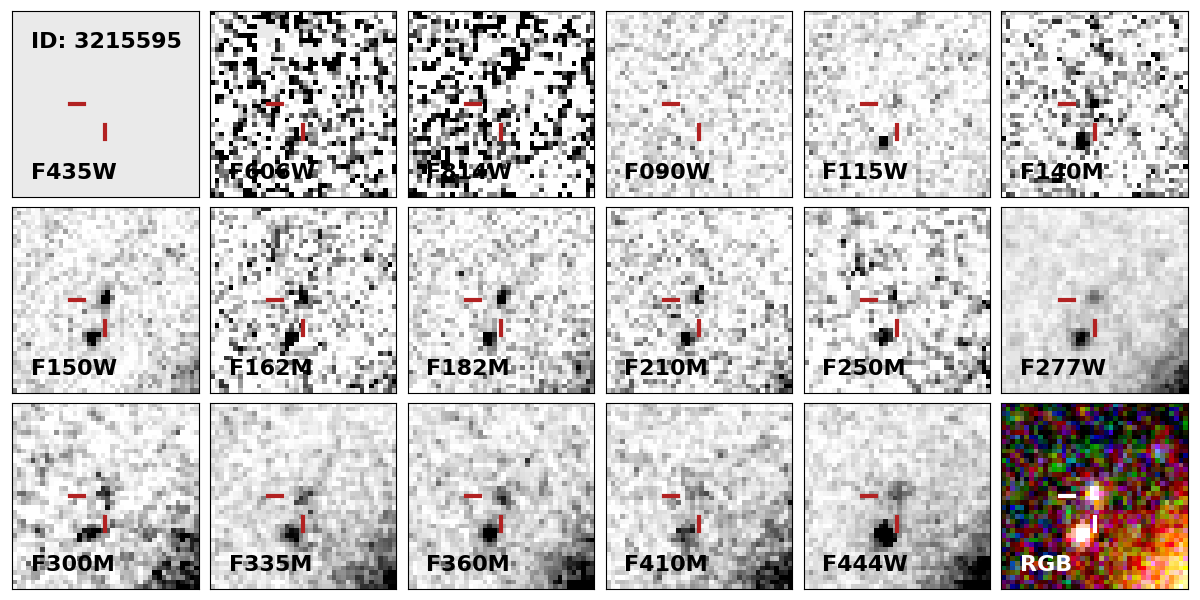}\\
    \includegraphics[width=0.48\linewidth]{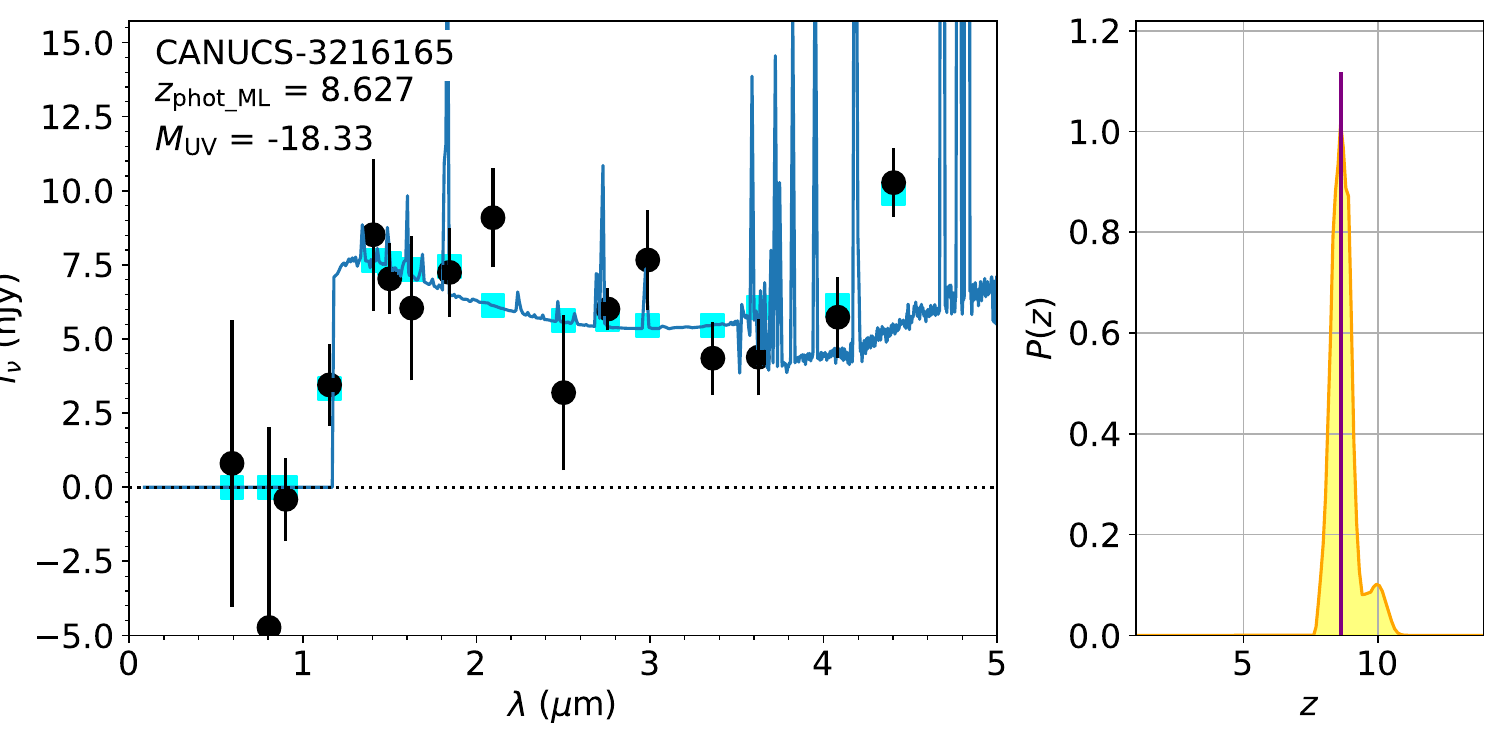}
    \includegraphics[width=0.48\linewidth]{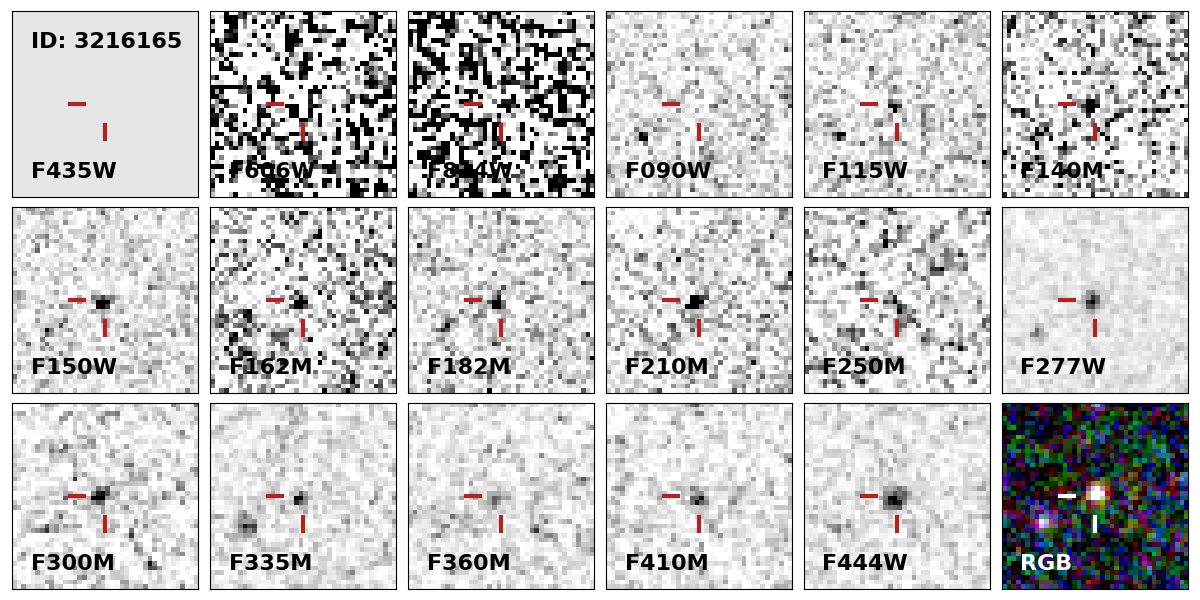}\\
    \caption{Example SEDs for CANUCS galaxies in the range $8.5<z_{\rm phot\_ML}<9.5$. Details as in Figure \ref{fig:seds7p5}.}
    \label{fig:seds8p5}
\end{figure*}

\begin{figure*}
    \centering
    \includegraphics[width=0.48\linewidth]{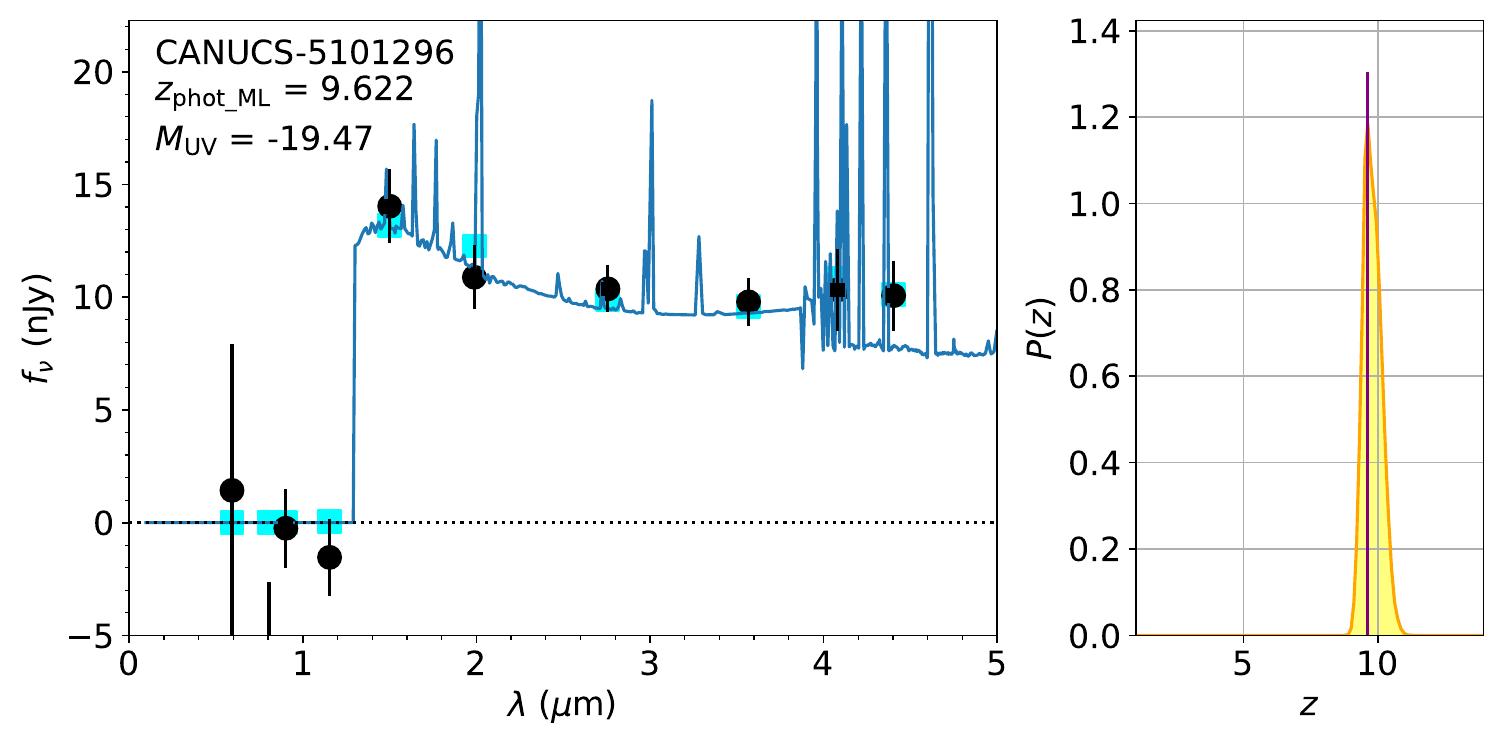}
    \raisebox{1.0cm}{\includegraphics[width=0.48\linewidth]{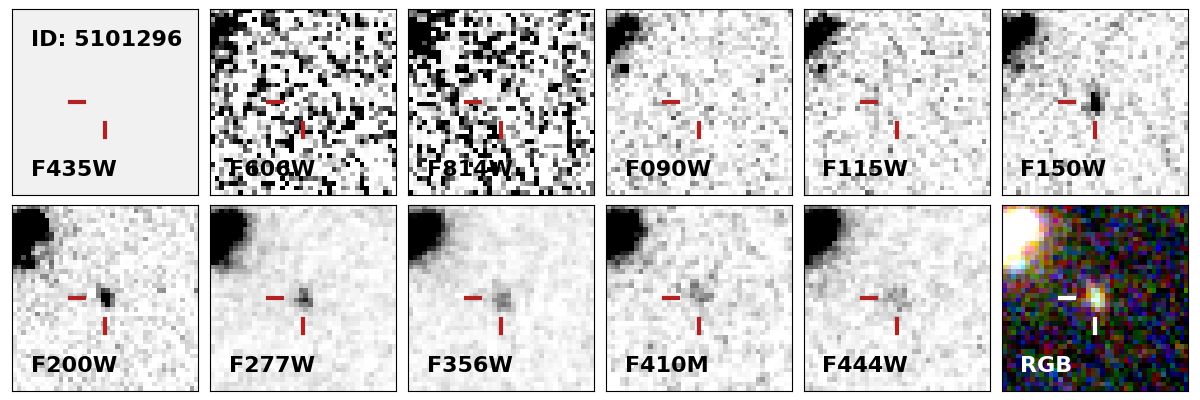}}\\
    \includegraphics[width=0.48\linewidth]{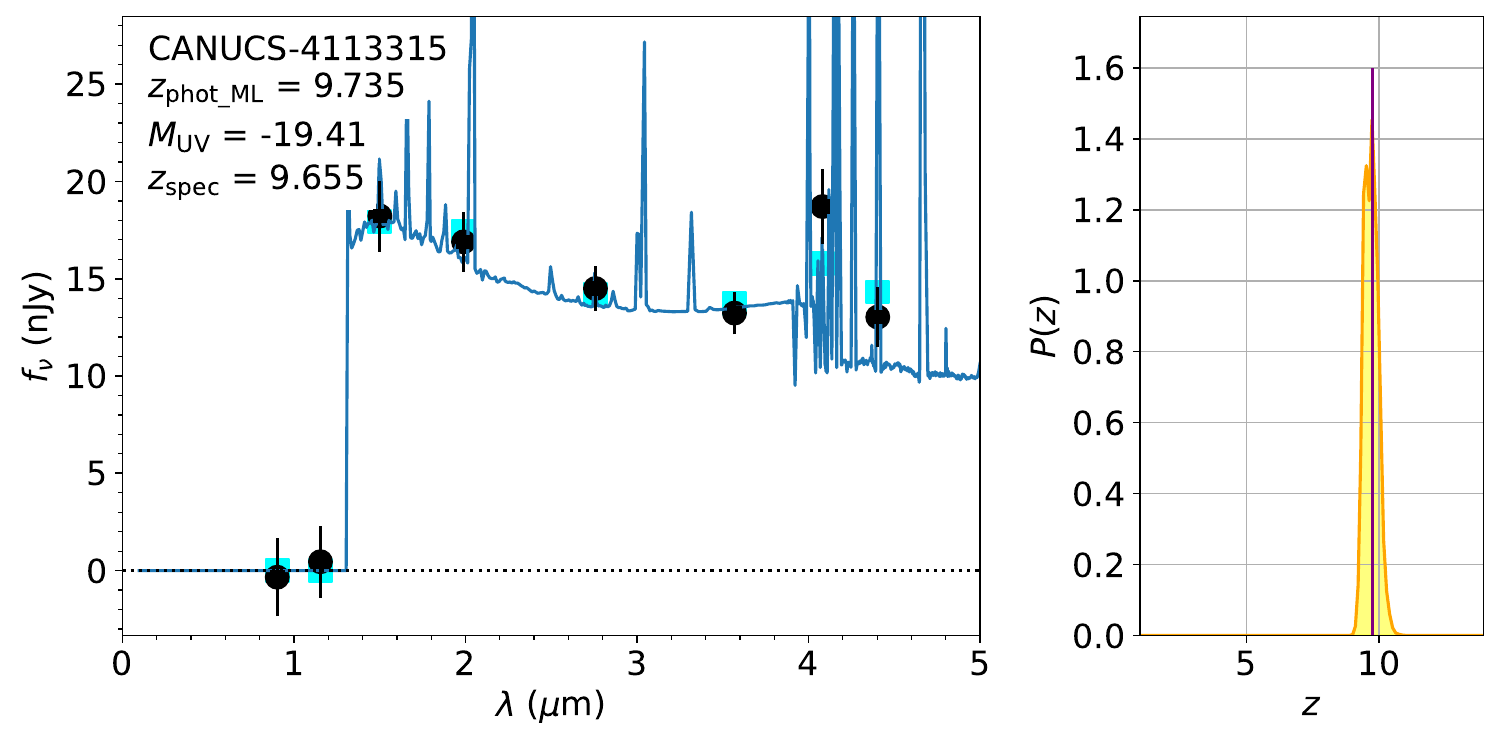}
    \raisebox{1.0cm}{\includegraphics[width=0.48\linewidth]{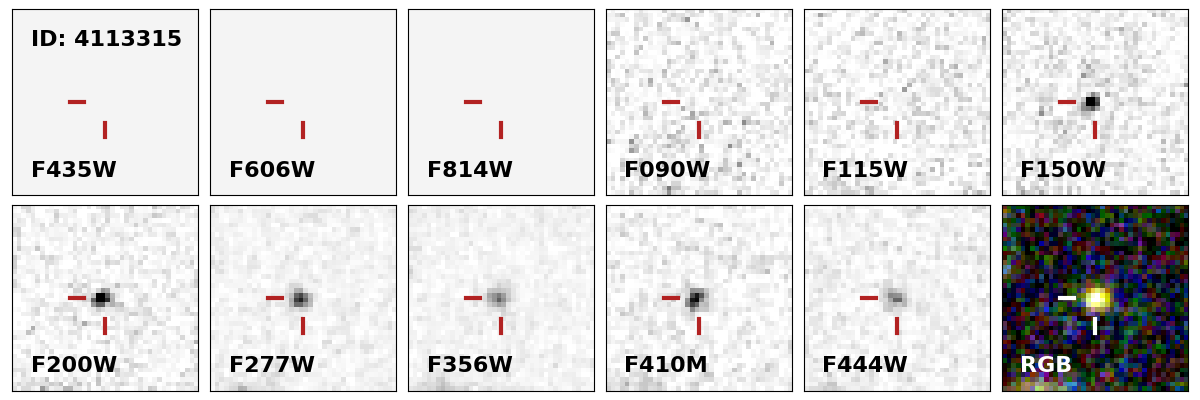}}\\
    \includegraphics[width=0.48\linewidth]{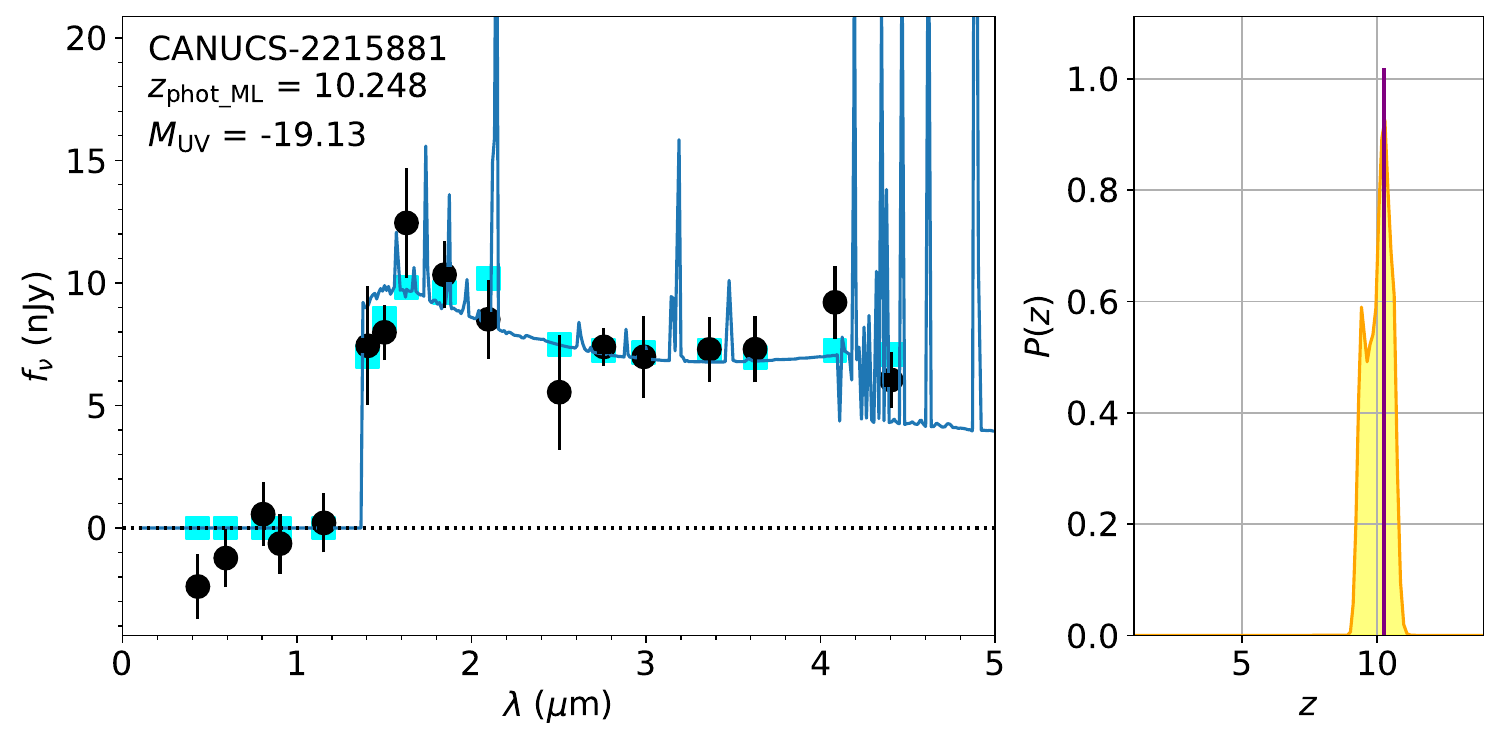}
    \includegraphics[width=0.48\linewidth]{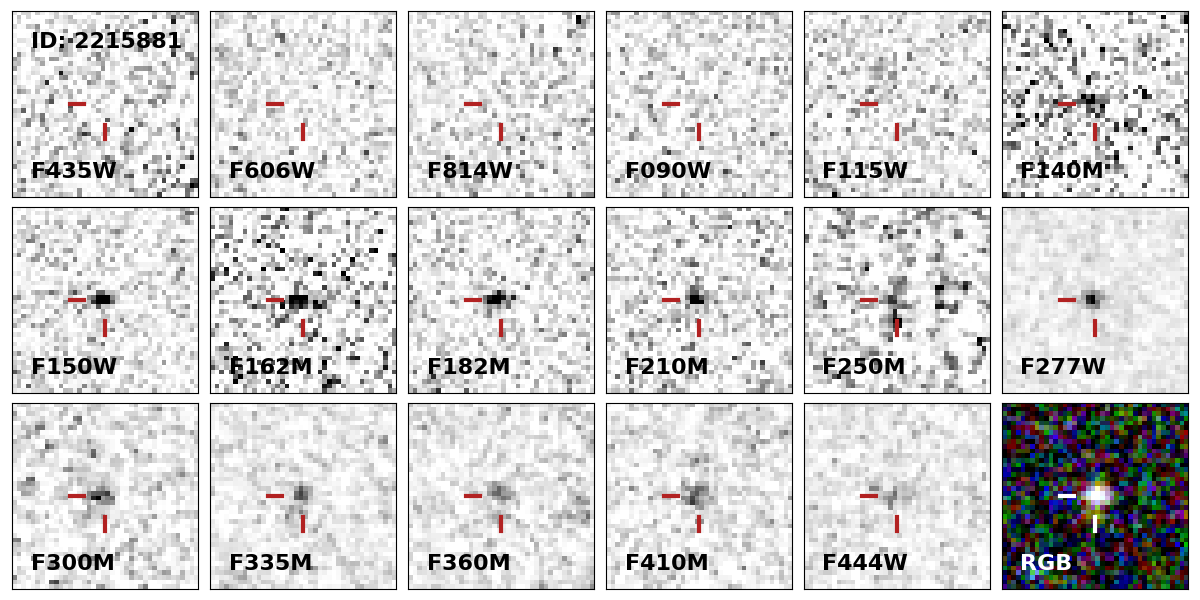}\\
    \includegraphics[width=0.48\linewidth]{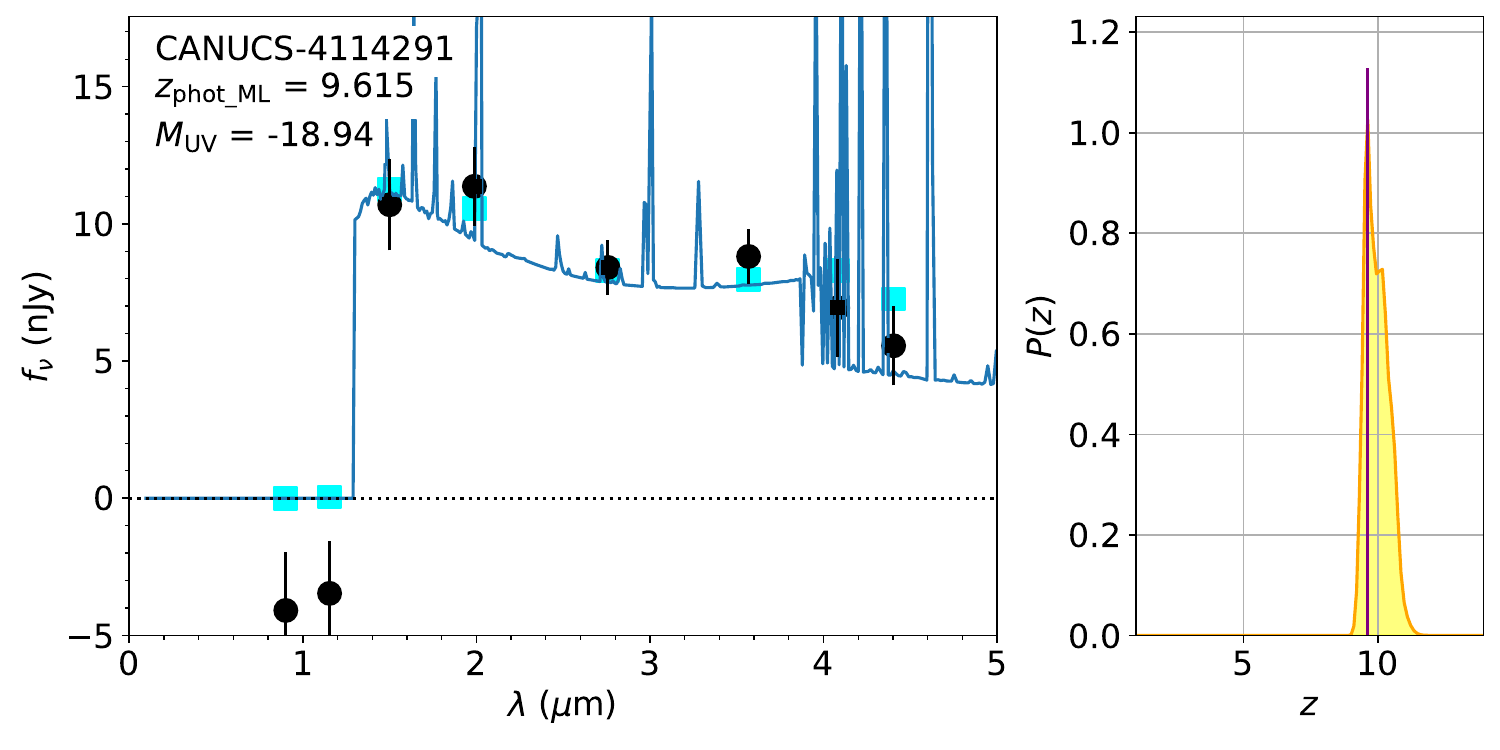}
    \raisebox{1.0cm}{\includegraphics[width=0.48\linewidth]{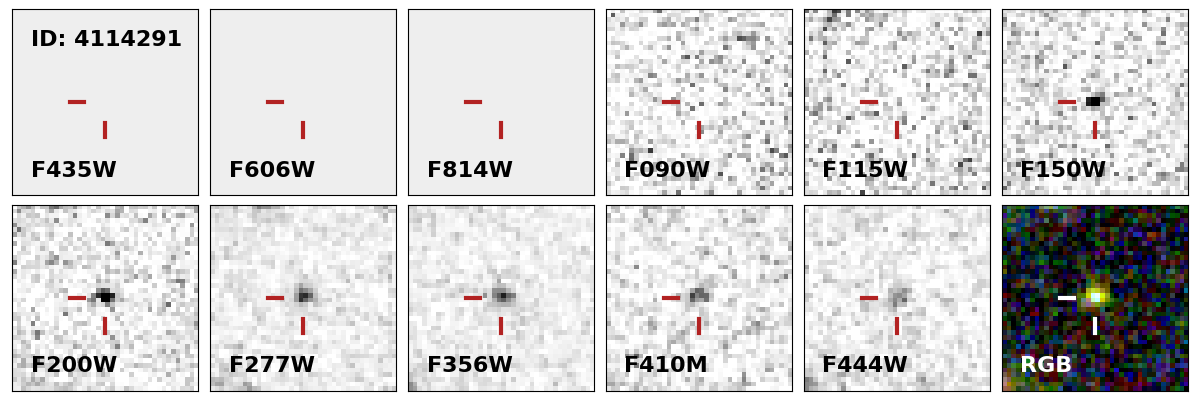}}\\
    \includegraphics[width=0.48\linewidth]{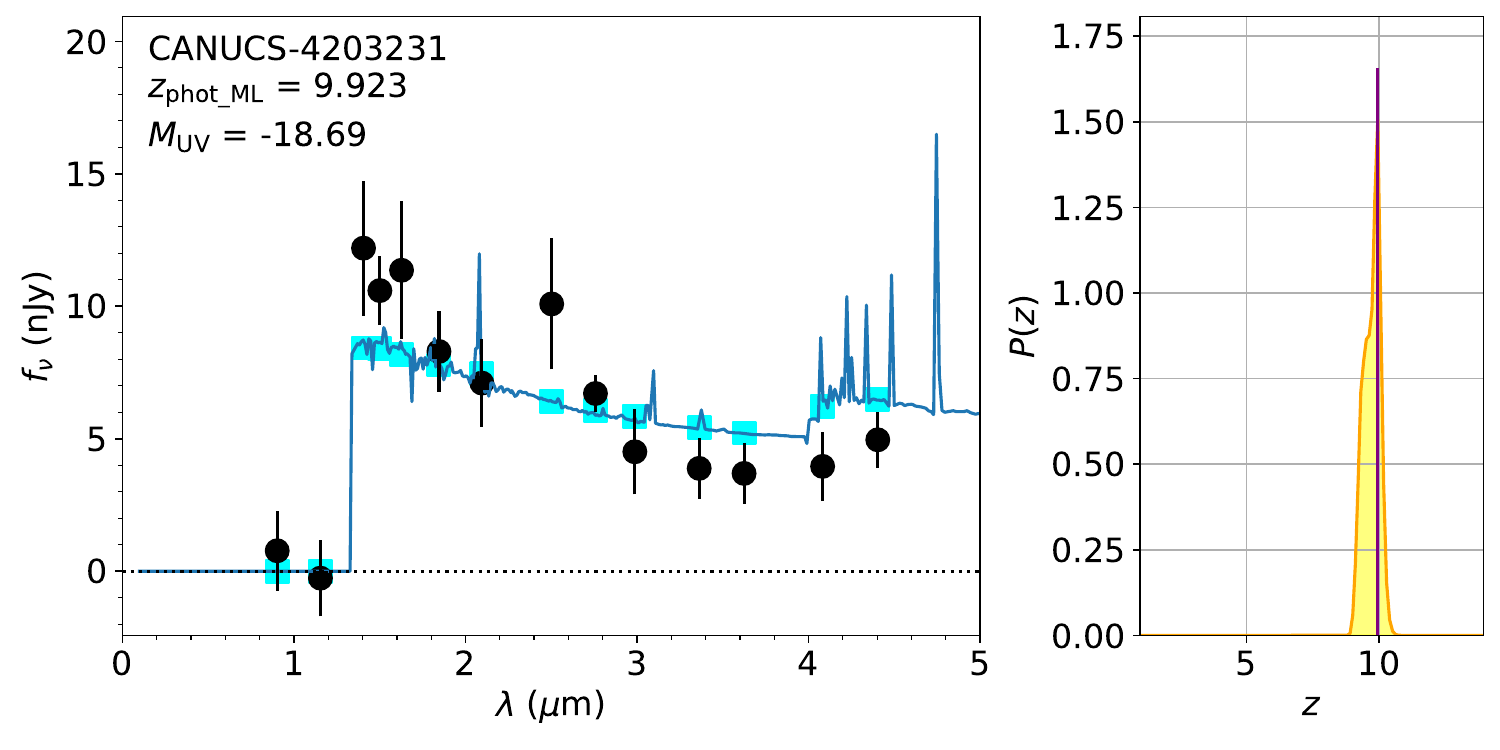}
    \includegraphics[width=0.48\linewidth]{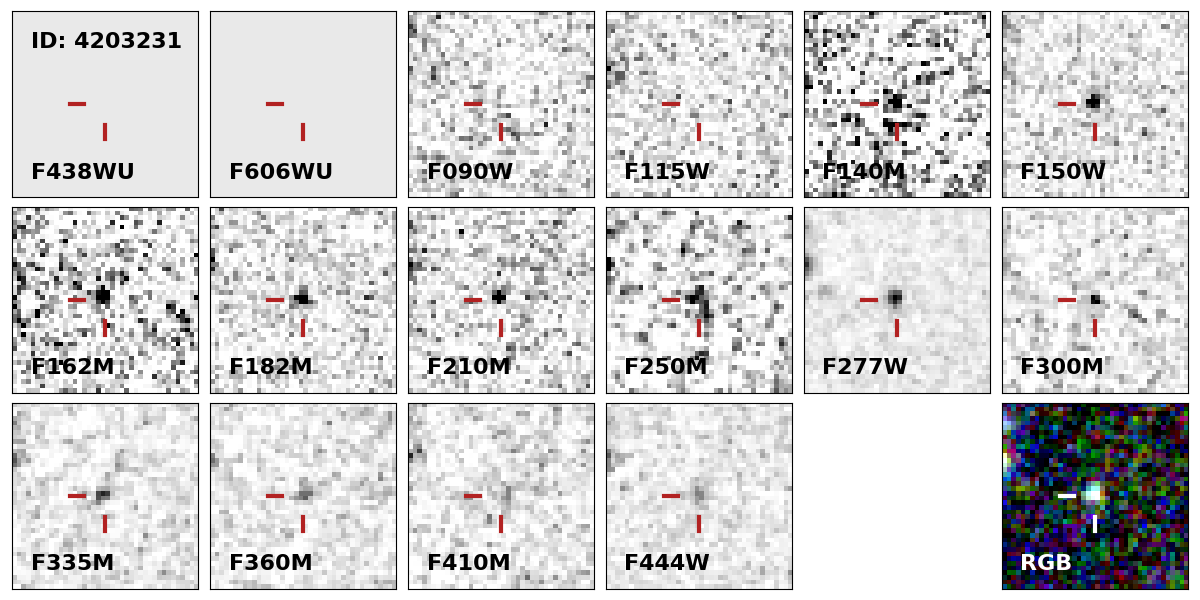}\\
    \caption{Example SEDs for CANUCS galaxies in the range $9.5<z_{\rm phot\_ML}<10.5$. Details as in Figure \ref{fig:seds7p5}.}
    \label{fig:seds9p5}
\end{figure*}

\begin{figure*}
    \centering
    \includegraphics[width=0.48\linewidth]{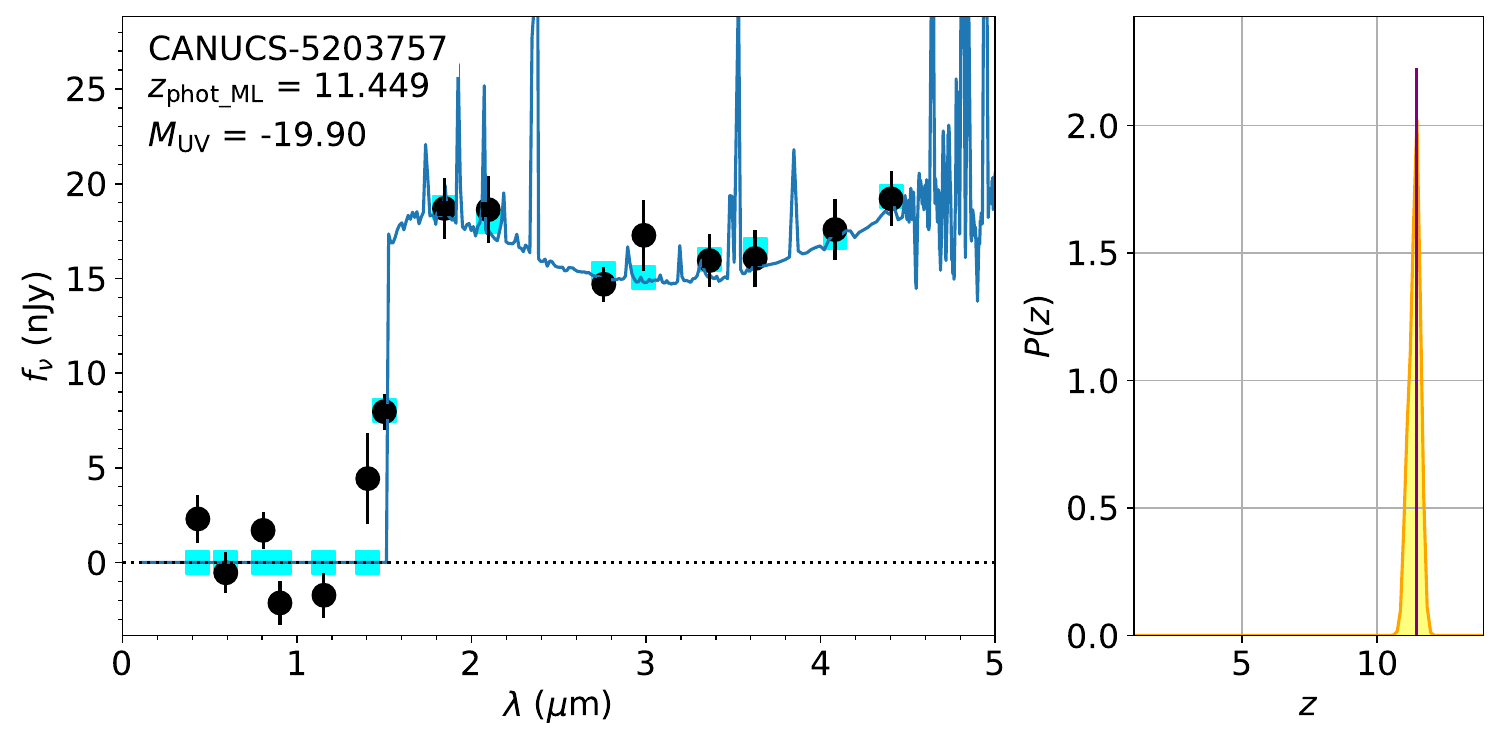}
    \includegraphics[width=0.48\linewidth]{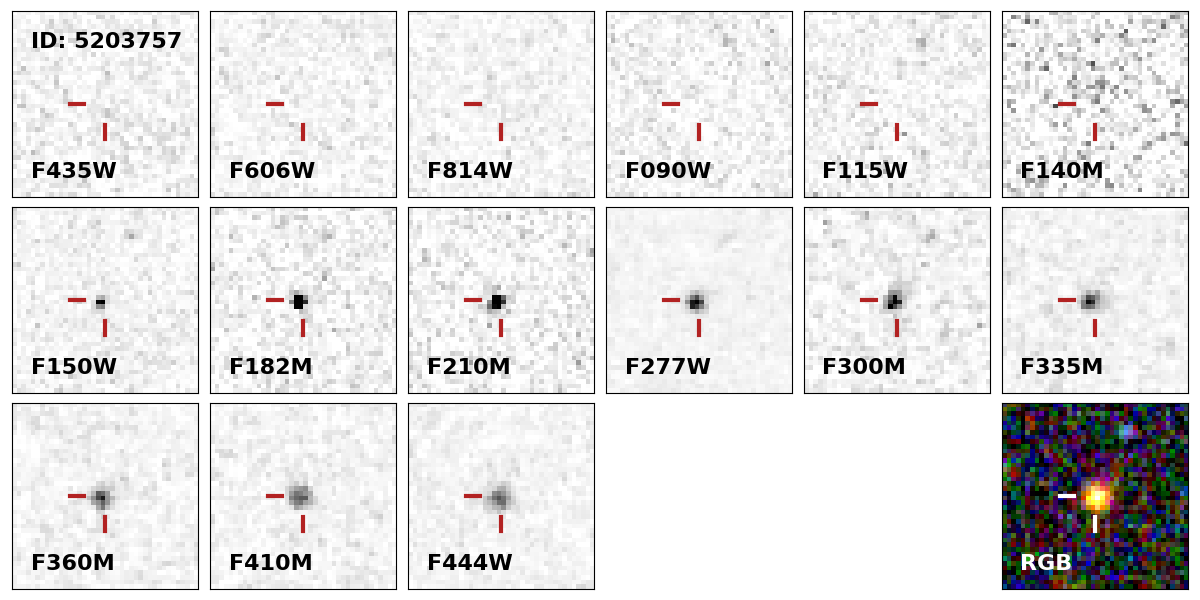}\\
    \includegraphics[width=0.48\linewidth]{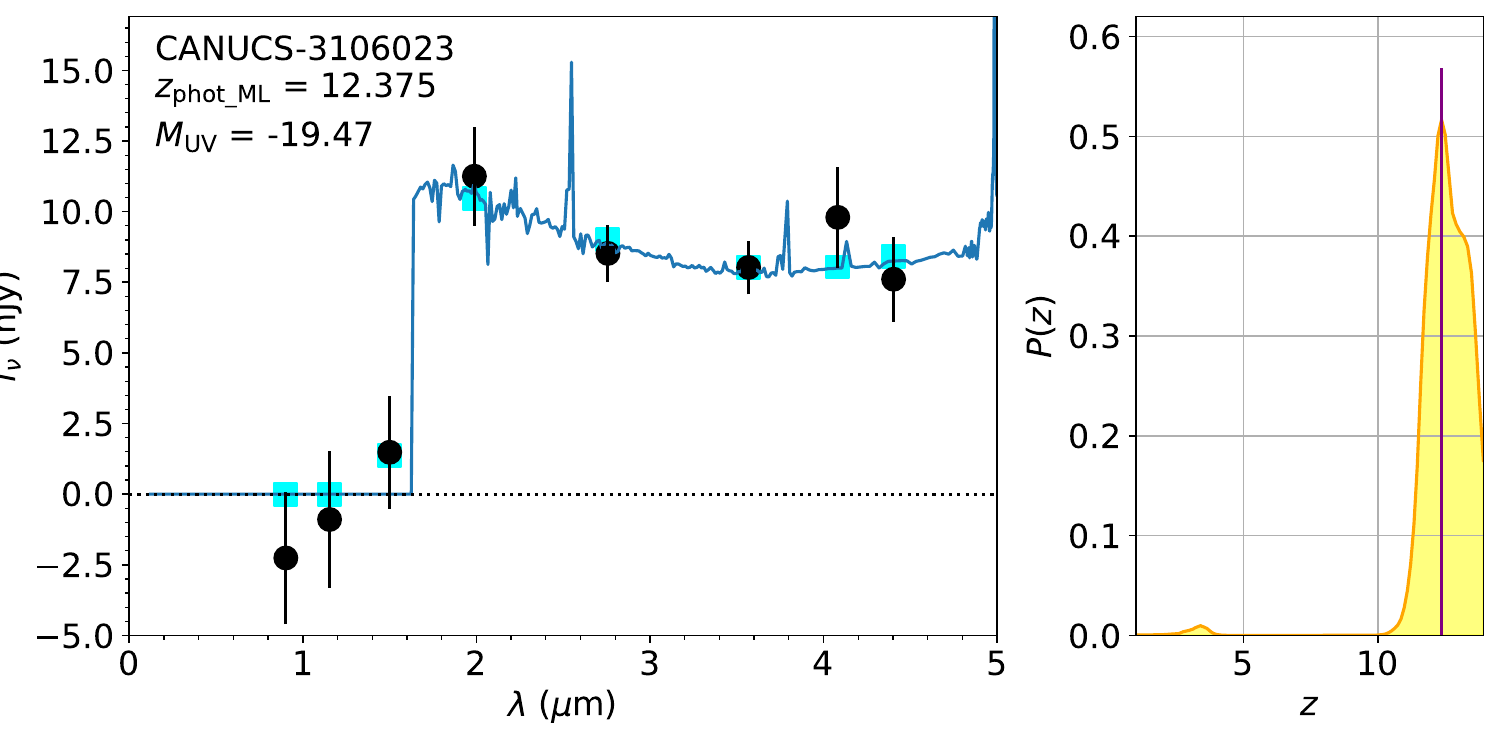}
    \raisebox{1.0cm}{\includegraphics[width=0.48\linewidth]{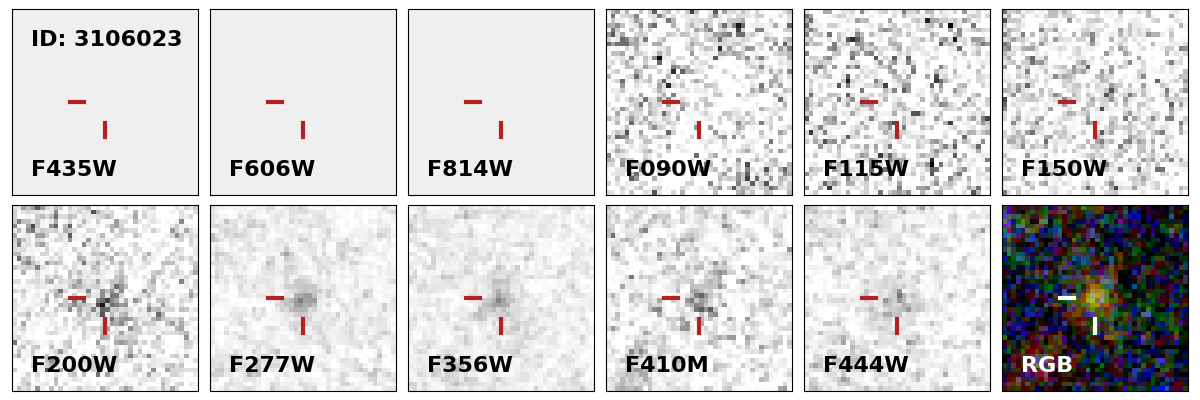}}\\
    \includegraphics[width=0.48\linewidth]{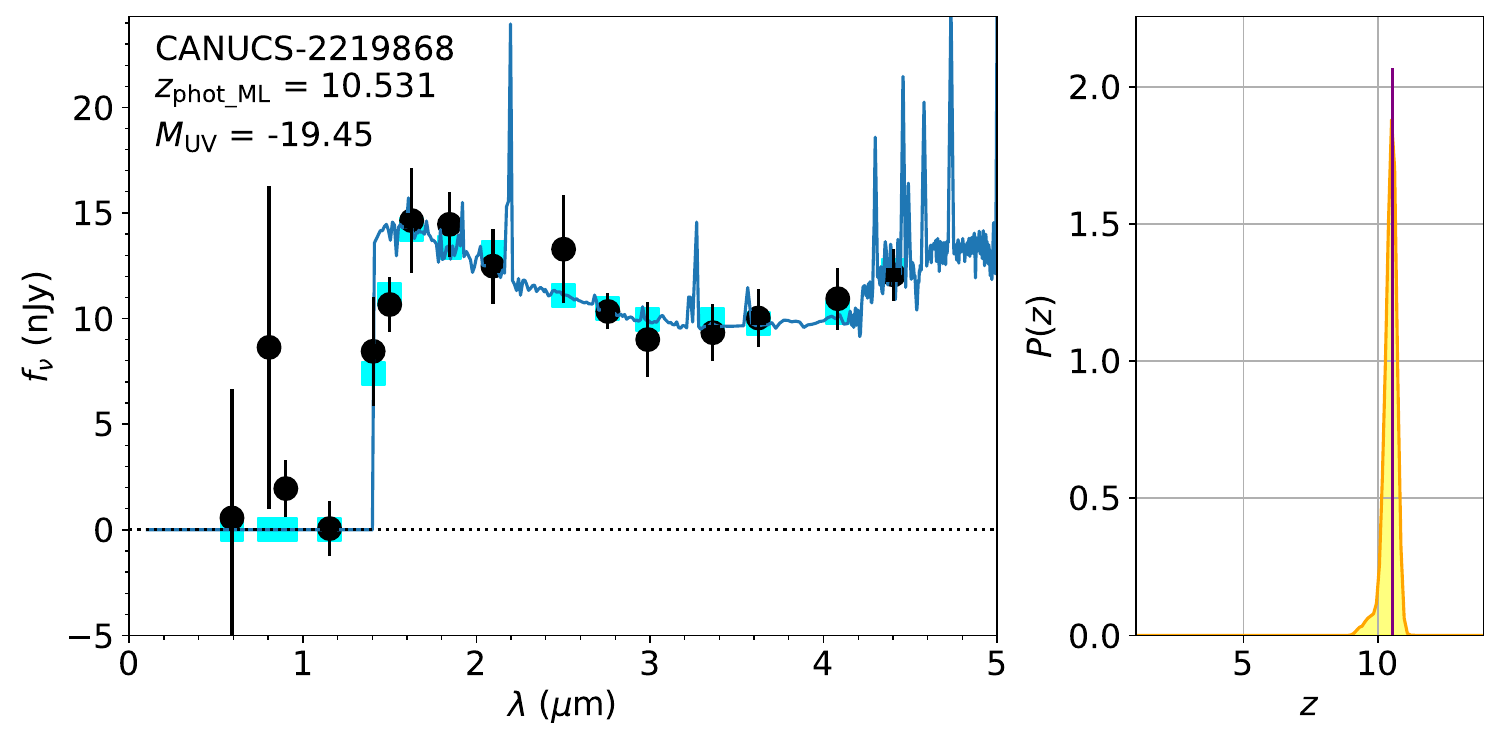}
    \includegraphics[width=0.48\linewidth]{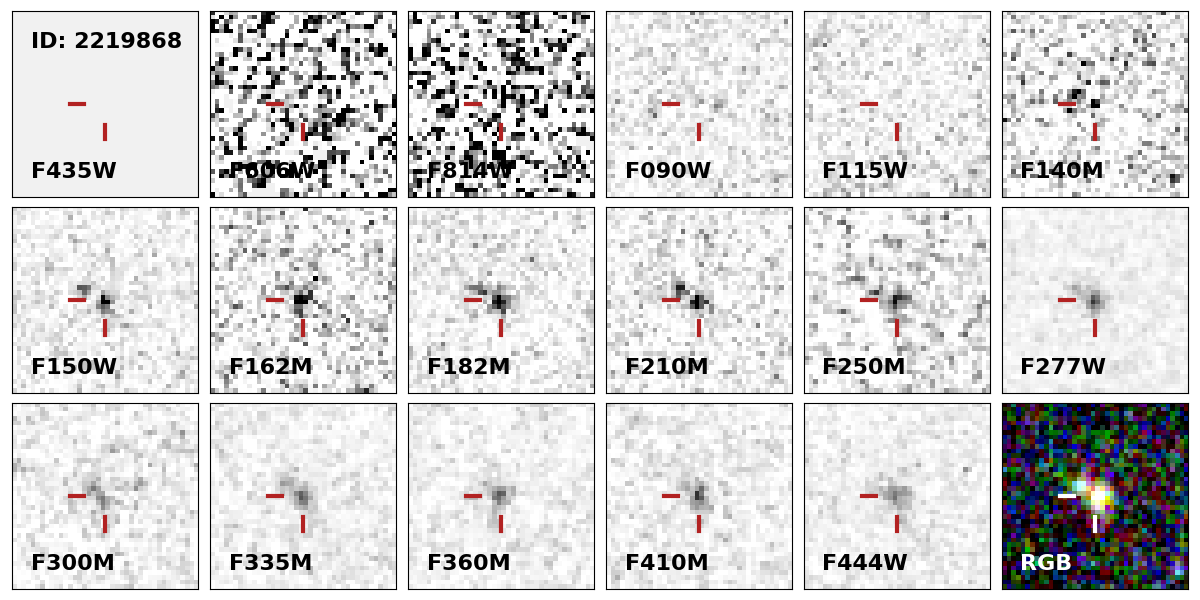}\\
    \includegraphics[width=0.48\linewidth]{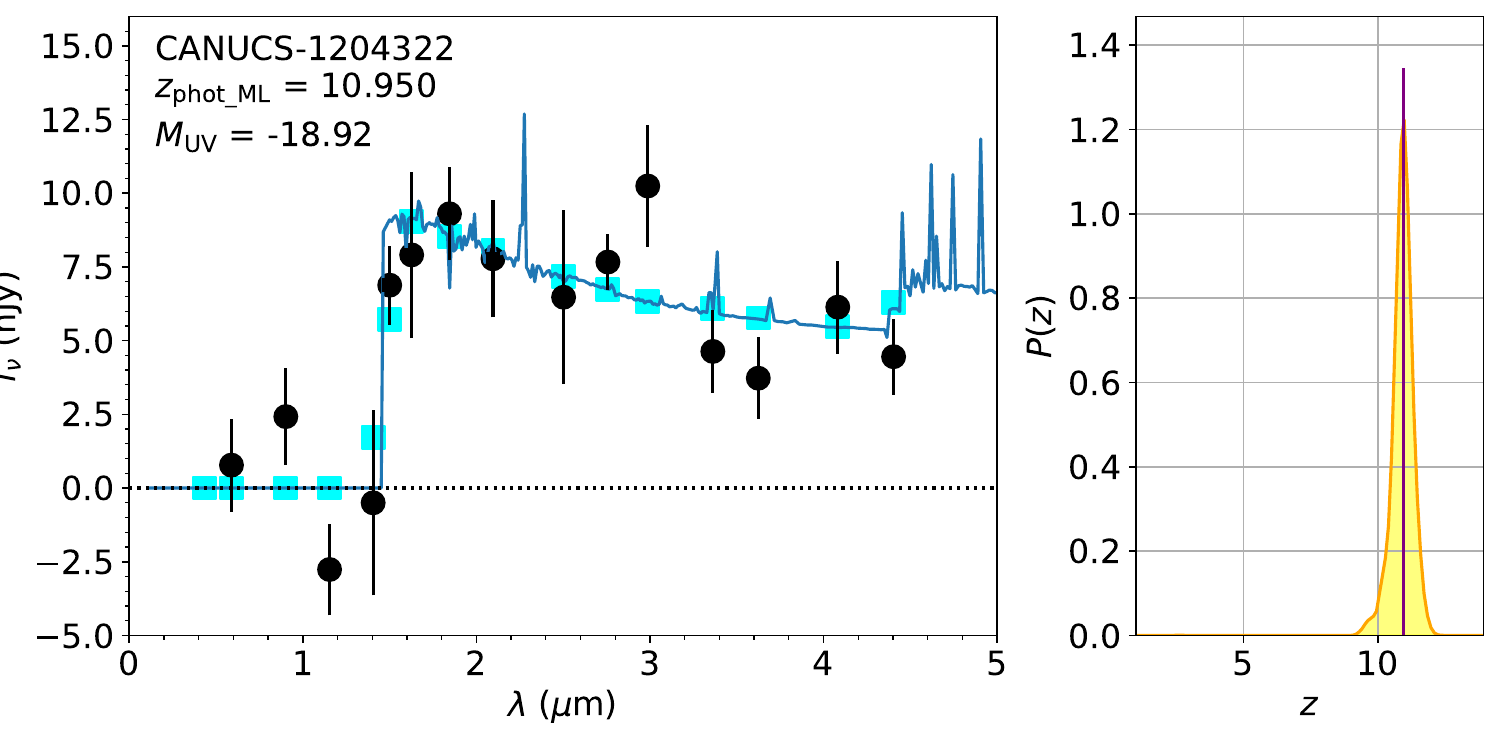}
    \includegraphics[width=0.48\linewidth]{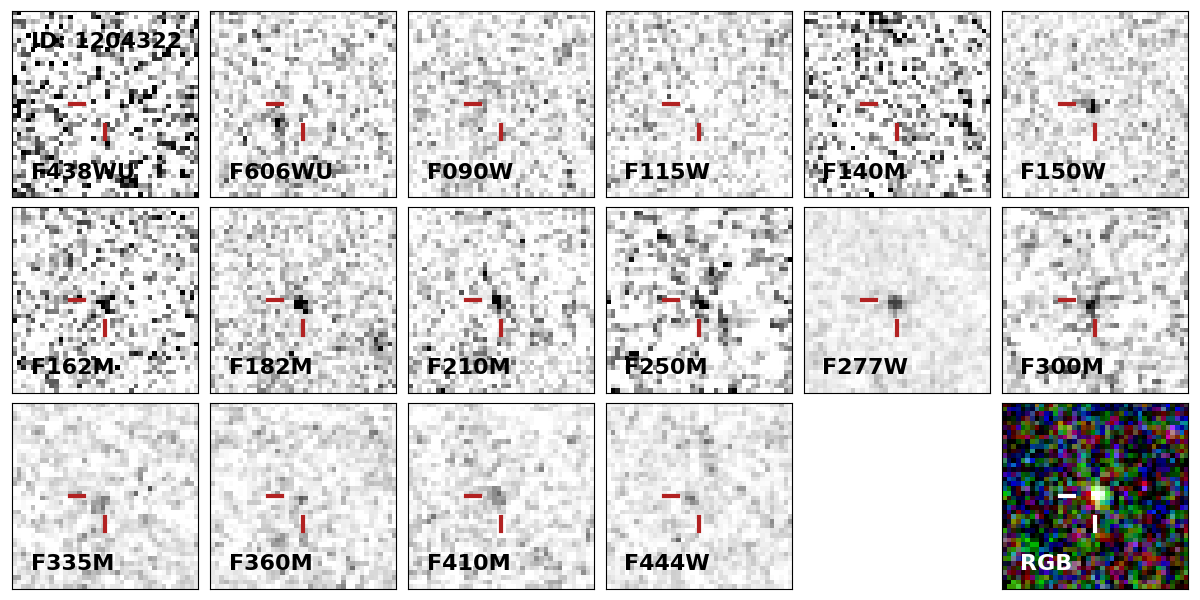}\\
    \includegraphics[width=0.48\linewidth]{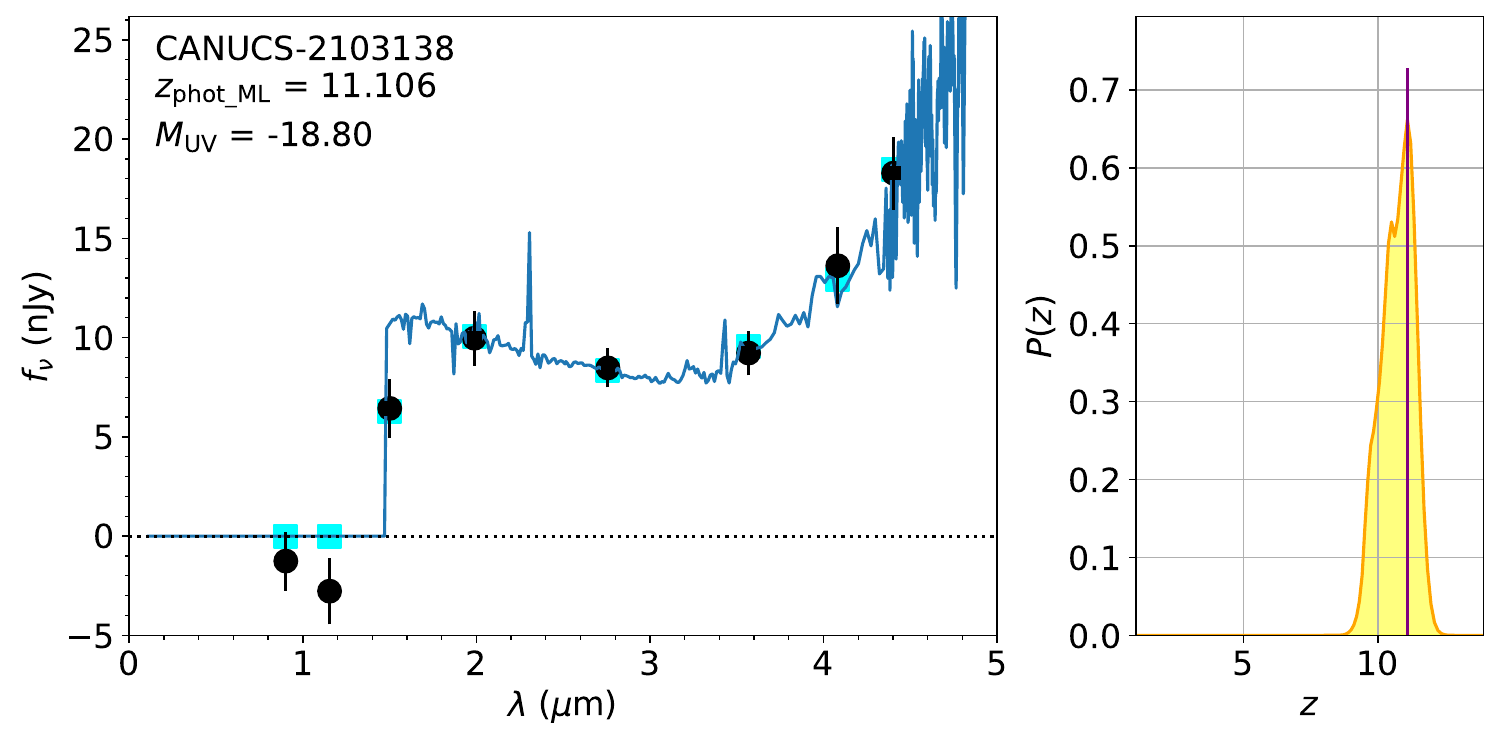}
    \raisebox{1.0cm}{\includegraphics[width=0.48\linewidth]{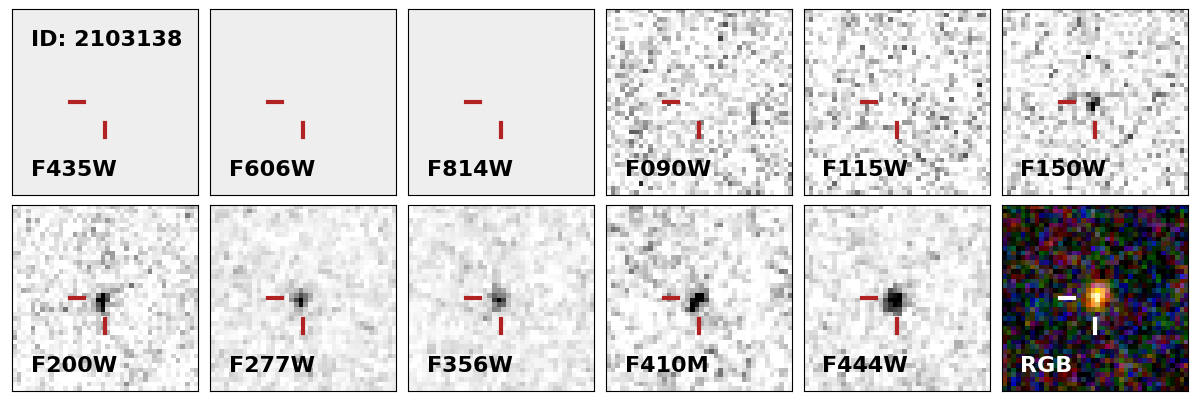}}\\
    \caption{Example SEDs for CANUCS galaxies in the range $10.5<z_{\rm phot\_ML}<12.5$. Details as in Figure \ref{fig:seds7p5}.}
    \label{fig:seds10p5}
\end{figure*}

In Figures \ref{fig:seds7p5}, \ref{fig:seds8p5}, \ref{fig:seds9p5}, \ref{fig:seds10p5} we show multi-filter images cutouts and EAZY-py photometric redshift fits for a selection of galaxies in each redshift bin. We adjust the two highest redshift bins compared to the luminosity function (Section \ref{sec:binned}) in order to show more of the $z>9.5$ galaxies. The galaxies shown span the full range of absolute magnitude of our sample.

\clearpage

\section{Stacked images}
\label{sec:stacks}

Stacked images are generated for three groups of galaxies from the luminosity function sample: A $7.5<z<9.6$ NCF (112 galaxies), B $7.5<z<9.6$ CLU (27 galaxies), C $9.6<z<12.4$ CLU+NCF (19 galaxies). For the highest redshift bin we combined the NCF and CLU samples because there are too few galaxies in either sub-sample. Since the NCF fields are not observed with the wide filters F200W and F356W, but are observed with a pair of medium filters that cover the same wavelength range, we generated pseudo images for group C in these filters by combining F182M+F210M and F335M+F360M, respectively. 

Image cutouts with sides of 1.6 arcsec are extracted at the position of each galaxy from the final background-subtracted mosaics. For each galaxy, the images of all filters are divided by the 0.3 arcsec aperture flux in the F277W filter. These normalized images in each filter are then combined using two algorithms; a mean and a median. No variance weighting or sigma-clipping is applied to ensure that all galaxies contribute equally to all filters. 

\begin{figure}
    \centering
    \includegraphics[width=0.87\linewidth]{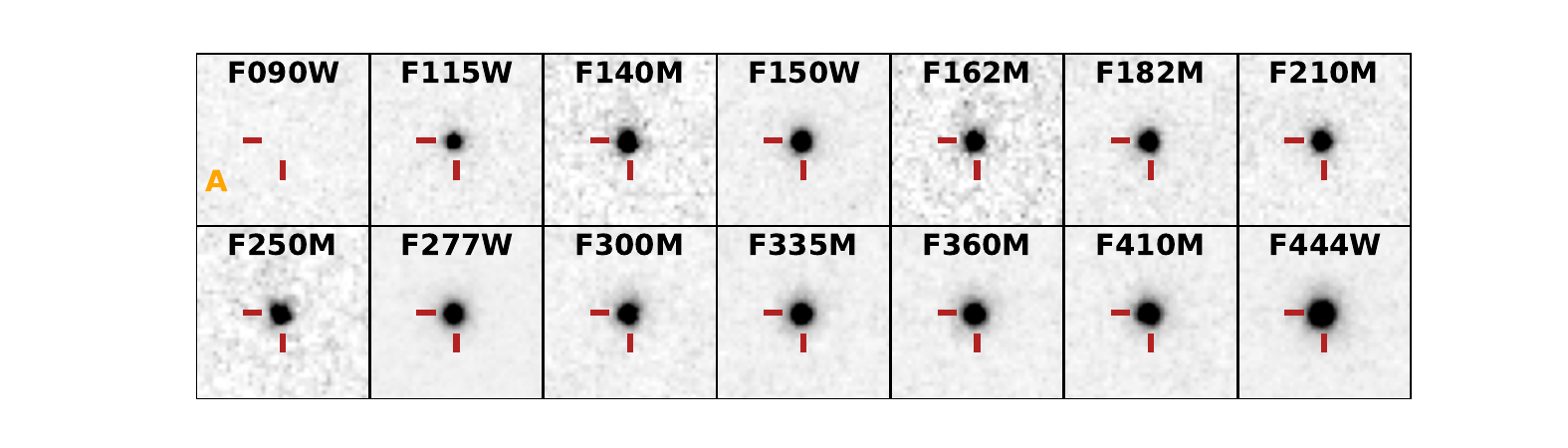}\\
    \includegraphics[width=0.99\linewidth]{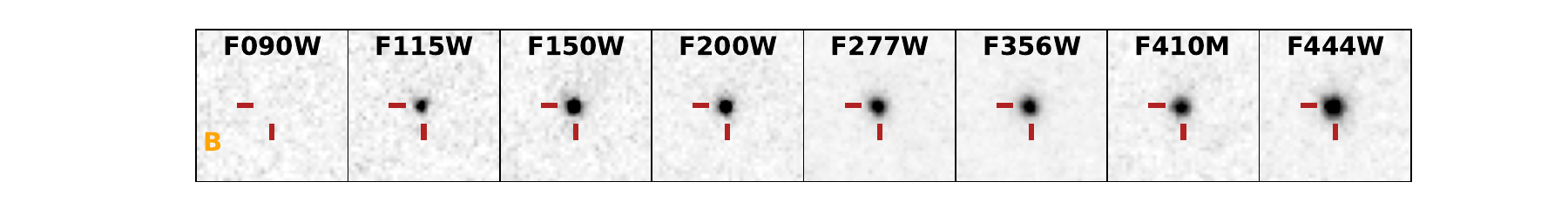}\\
    \includegraphics[width=0.99\linewidth]{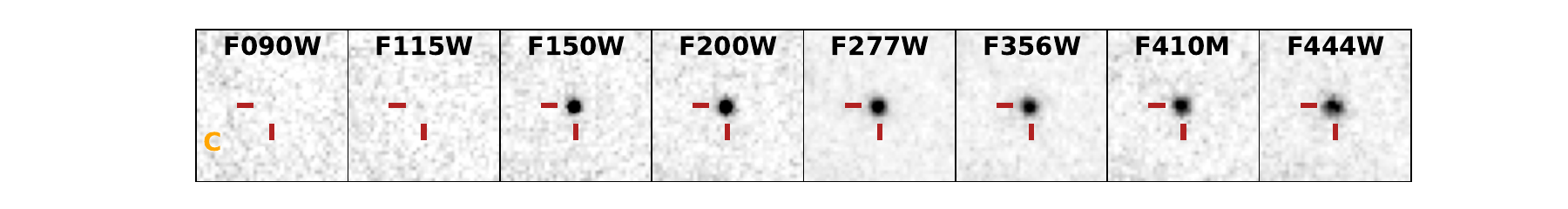}
    \caption{Median stacked NIRCam images of the CANUCS high-redshift galaxy LF sample. The galaxies have been stacked in three groups: A $7.5<z<9.6$ NCF, B $7.5<z<9.6$ CLU, C $9.6<z<12.4$ CLU+NCF. All filters for each group are shown with the same minimum and maximum greyscale limits.}
    \label{fig:stacked}
\end{figure}

In Figure \ref{fig:stacked} we show the median stacks for all filters in the three groups of galaxies. For groups A and B we observe zero flux in the F090W filter, as expected for galaxies at $z>7.5$. The F115W flux of both groups is substantially depressed compared to F140M or F150W as expected based on the Lyman break falling within F115W. For the higher redshift stack in group C there is zero flux in both F090W and F115W, as expected for this redshift range. The F150W flux is depressed compared to F200W due to the Lyman break. In all cases, these stacked images support the assigned redshifts of the CANUCS high-redshift sample.

We do not show the mean stacked images here because they contain substantial contamination in the outer regions away from the target galaxy. This is to be expected because no sigma-clipping is applied in order to avoid removal of any positive flux in the shortest wavelengths from potential low-redshift interlopers. The mean stacked images also show no flux excess at the target galaxy location in the shorter wavelength filters and support the conclusions shown with the median stacks.

\clearpage

\section{Properties of CANUCS $z>7.5$ galaxy LF sample}
\label{sec:galaxytables}

\begin{longtable}{cccccccccc}
\hline \hline
CANUCS & RA & DEC & $z_{\rm phot\_ML}$ & $z_{\rm phot\_50}$ & $z_{\rm spec}$ & $M_{\rm UV}$ & F277W Flux & Magnification & Ref. \\
ID & (J2000) &  (J2000)  &  &  &  &  & (nJy) &  &  \\
\hline
1112184 & 64.416334 & -11.868792 & 9.12 & $9.29^{+0.62}_{-0.27}$ & - & -19.47 & $25.09 \pm 2.83$ & 1.24 &  \\
1113392 & 64.427912 & -11.857791 & 8.60 & $8.60^{+0.08}_{-0.09}$ & - & -21.07 & $117.31 \pm 2.31$ & 1.12 &  \\
1114433 & 64.420766 & -11.848376 & 7.88 & $7.98^{+0.27}_{-0.15}$ & - & -19.71 & $44.09 \pm 5.19$ & 1.11 &  \\
1117924 & 64.409253 & -11.856963 & 7.81 & $7.78^{+0.17}_{-0.90}$ & - & -19.25 & $22.97 \pm 2.48$ & 1.19 &  \\
1201610 & 64.361253 & -11.850067 & 8.93 & $9.17^{+0.78}_{-0.44}$ & - & -18.37 & $12.33 \pm 1.14$ & 1.35 &  \\
1202562 & 64.364363 & -11.841447 & 8.29 & $8.24^{+0.13}_{-0.15}$ & - & -19.83 & $43.37 \pm 1.41$ & 1.24 &  \\
1202834 & 64.367898 & -11.839160 & 8.33 & $8.27^{+0.15}_{-0.18}$ & 7.8968 & -19.49 & $43.42 \pm 1.65$ & 1.22 &  \\
1204322 & 64.361722 & -11.828135 & 10.95 & $10.94^{+0.33}_{-0.36}$ & - & -18.92 & $12.46 \pm 1.42$ & 1.16 &  \\
1205347 & 64.354602 & -11.819836 & 7.87 & $7.94^{+0.28}_{-0.14}$ & - & -19.11 & $16.14 \pm 1.68$ & 1.12 &  \\
1205631 & 64.352628 & -11.816450 & 8.37 & $8.37^{+0.13}_{-0.13}$ & - & -19.50 & $28.53 \pm 2.90$ & 1.11 &  \\
1209908 & 64.388467 & -11.779490 & 7.92 & $8.00^{+0.21}_{-0.16}$ & - & -19.61 & $28.80 \pm 2.03$ & 1.03 &  \\
1209941 & 64.388020 & -11.779145 & 7.99 & $7.99^{+0.07}_{-0.07}$ & - & -20.23 & $43.98 \pm 1.30$ & 1.03 &  \\
2103138 & 39.990156 & -1.635501 & 11.11 & $10.79^{+0.55}_{-0.72}$ & - & -18.80 & $13.37 \pm 1.40$ & 1.54 &  \\
2103472 & 40.011804 & -1.632604 & 7.87 & $7.92^{+0.27}_{-0.15}$ & - & -19.43 & $30.28 \pm 2.51$ & 1.40 &  \\
2104108 & 40.018551 & -1.626878 & 8.73 & $8.69^{+0.63}_{-0.49}$ & - & -18.55 & $13.34 \pm 1.59$ & 1.36 &  \\
2104426 & 40.014535 & -1.624574 & 7.59 & $7.54^{+0.08}_{-0.29}$ & - & -19.04 & $21.07 \pm 1.43$ & 1.40 &  \\
2104427 & 40.004549 & -1.624585 & 8.23 & $8.23^{+0.13}_{-0.13}$ & - & -19.45 & $31.28 \pm 1.75$ & 1.53 &  \\
2105635 & 39.995983 & -1.616268 & 8.11 & $8.19^{+0.36}_{-0.24}$ & 7.8712 & -18.29 & $13.35 \pm 1.33$ & 1.75 &  \\
2105689 & 40.017812 & -1.616011 & 8.36 & $8.34^{+0.25}_{-0.29}$ & - & -19.55 & $31.54 \pm 2.52$ & 1.38 &  \\
2106297 & 40.013225 & -1.612119 & 8.96 & $9.04^{+0.66}_{-0.40}$ & - & -19.34 & $23.57 \pm 2.28$ & 1.44 &  \\
2106548 & 40.005395 & -1.610657 & 7.86 & $7.93^{+0.23}_{-0.13}$ & - & -18.83 & $17.49 \pm 1.28$ & 1.61 &  \\
2108376 & 40.005148 & -1.598971 & 8.11 & $8.08^{+0.16}_{-0.16}$ & 7.8623 & -19.58 & $39.46 \pm 1.97$ & 1.64 &  \\
2108442 & 40.005794 & -1.598519 & 8.35 & $8.35^{+0.19}_{-0.19}$ & - & -18.99 & $22.45 \pm 1.50$ & 1.63 &  \\
2108450 & 40.005509 & -1.598455 & 8.08 & $8.09^{+0.16}_{-0.14}$ & 7.8658 & -19.29 & $29.55 \pm 1.81$ & 1.63 &  \\
2203922 & 40.063501 & -1.665635 & 8.50 & $8.52^{+0.21}_{-0.22}$ & - & -19.57 & $28.25 \pm 1.65$ & 1.15 &  \\
2204887 & 40.065163 & -1.659988 & 8.32 & $8.33^{+0.15}_{-0.12}$ & - & -19.01 & $17.72 \pm 1.11$ & 1.15 &  \\
2207469 & 40.069591 & -1.648977 & 7.54 & $7.51^{+0.18}_{-0.23}$ & - & -18.41 & $13.24 \pm 1.41$ & 1.14 &  \\
2207986 & 40.081332 & -1.646965 & 7.73 & $7.87^{+0.34}_{-0.16}$ & - & -18.78 & $15.74 \pm 1.36$ & 1.13 &  \\
2208848 & 40.089019 & -1.643428 & 7.59 & $7.23^{+0.33}_{-0.22}$ & - & -18.27 & $11.79 \pm 0.97$ & 1.12 &  \\
2209594 & 40.081390 & -1.640279 & 9.34 & $9.41^{+0.67}_{-0.64}$ & - & -18.95 & $15.03 \pm 1.22$ & 1.13 &  \\
2209855 & 40.049752 & -1.639260 & 9.29 & $9.33^{+0.36}_{-0.17}$ & - & -19.35 & $22.12 \pm 1.33$ & 1.19 & (6) \\
2211160 & 40.074841 & -1.633946 & 8.08 & $8.11^{+0.20}_{-0.16}$ & - & -18.56 & $12.47 \pm 1.39$ & 1.14 &  \\
2215881 & 40.056131 & -1.615683 & 10.25 & $10.11^{+0.40}_{-0.59}$ & - & -19.13 & $13.63 \pm 1.28$ & 1.17 &  \\
2217040 & 40.036499 & -1.611575 & 7.99 & $8.07^{+0.23}_{-0.14}$ & - & -18.92 & $19.76 \pm 1.38$ & 1.24 &  \\
2219289 & 40.045520 & -1.603048 & 8.76 & $8.77^{+0.24}_{-0.24}$ & - & -18.97 & $13.85 \pm 1.27$ & 1.21 &  \\
2219868 & 40.032578 & -1.600549 & 10.53 & $10.49^{+0.19}_{-0.25}$ & - & -19.45 & $19.73 \pm 1.29$ & 1.26 &  \\
2221983 & 40.049026 & -1.589292 & 8.15 & $8.16^{+0.17}_{-0.17}$ & - & -19.85 & $54.32 \pm 1.51$ & 1.19 &  \\
2222200 & 40.046236 & -1.588088 & 7.68 & $7.70^{+0.28}_{-0.07}$ & - & -19.06 & $21.63 \pm 1.51$ & 1.20 &  \\
2224402 & 40.060774 & -1.660736 & 7.90 & $7.97^{+0.48}_{-0.15}$ & - & -18.24 & $10.79 \pm 1.26$ & 1.16 &  \\
3105745 & 64.065850 & -24.103364 & 8.16 & $8.42^{+1.96}_{-0.36}$ & - & -18.83 & $17.03 \pm 2.01$ & 1.19 &  \\
3106023 & 64.086469 & -24.101399 & 12.37 & $12.56^{+0.82}_{-0.72}$ & - & -19.47 & $16.00 \pm 1.74$ & 1.13 &  \\
3106386 & 64.063649 & -24.098755 & 9.28 & $9.36^{+0.42}_{-0.23}$ & 9.0337 & -19.16 & $18.25 \pm 2.04$ & 1.23 &  \\
3107548 & 64.063612 & -24.090512 & 7.64 & $7.65^{+0.04}_{-0.03}$ & - & -20.00 & $42.57 \pm 2.33$ & 1.28 &  \\
3205919 & 64.154649 & -24.125183 & 8.70 & $8.79^{+0.39}_{-0.37}$ & - & -18.50 & $9.14 \pm 0.97$ & 1.04 &  \\
3206020 & 64.162673 & -24.124664 & 9.19 & $9.20^{+0.11}_{-0.10}$ & - & -19.60 & $18.13 \pm 1.53$ & 1.04 &  \\
3207111 & 64.141566 & -24.119282 & 7.82 & $7.87^{+0.18}_{-0.12}$ & - & -18.67 & $11.08 \pm 1.37$ & 1.05 &  \\
3207153 & 64.156598 & -24.119097 & 7.92 & $7.94^{+0.42}_{-0.49}$ & - & -18.28 & $11.98 \pm 1.50$ & 1.04 &  \\
3208332 & 64.149907 & -24.113386 & 8.92 & $8.90^{+0.12}_{-0.17}$ & - & -19.53 & $24.31 \pm 1.10$ & 1.05 & (1,4,5,6) \\
3208685 & 64.147946 & -24.111762 & 8.14 & $8.13^{+0.27}_{-0.29}$ & - & -19.41 & $23.72 \pm 1.68$ & 1.05 & (3,6) \\
3209366 & 64.151798 & -24.108482 & 8.18 & $8.23^{+0.23}_{-0.21}$ & - & -19.32 & $20.70 \pm 1.40$ & 1.05 & (1,3,5,6) \\
3209675 & 64.156672 & -24.106984 & 7.50 & $7.36^{+0.17}_{-0.31}$ & - & -19.13 & $16.37 \pm 1.47$ & 1.05 & (4) \\
3210267 & 64.164034 & -24.103908 & 8.30 & $8.31^{+0.29}_{-0.28}$ & - & -18.74 & $14.45 \pm 1.35$ & 1.04 &  \\
3210426 & 64.155708 & -24.103033 & 8.15 & $8.21^{+0.22}_{-0.16}$ & - & -18.87 & $13.15 \pm 0.97$ & 1.05 & (6) \\
3210983 & 64.126832 & -24.100357 & 8.31 & $8.29^{+0.18}_{-0.20}$ & - & -19.45 & $22.67 \pm 1.81$ & 1.07 & (1,2,3,5,6) \\
3211032 & 64.126916 & -24.100100 & 7.88 & $7.89^{+0.14}_{-0.12}$ & - & -19.37 & $17.84 \pm 1.49$ & 1.07 & (1,2,3,6) \\
3211653 & 64.168271 & -24.097003 & 9.38 & $9.40^{+0.28}_{-0.10}$ & - & -19.52 & $21.99 \pm 1.65$ & 1.04 &  \\
3215595 & 64.145670 & -24.073460 & 9.09 & $9.14^{+0.42}_{-0.30}$ & - & -18.94 & $14.36 \pm 1.36$ & 1.07 &  \\
3216165 & 64.110315 & -24.069308 & 8.63 & $8.66^{+0.42}_{-0.37}$ & - & -18.33 & $8.59 \pm 0.94$ & 1.13 &  \\
3216756 & 64.114333 & -24.064572 & 7.90 & $8.02^{+0.24}_{-0.16}$ & - & -19.45 & $25.16 \pm 1.52$ & 1.13 &  \\
3218263 & 64.156883 & -24.127047 & 9.33 & $9.24^{+0.24}_{-1.24}$ & - & -19.22 & $15.72 \pm 1.47$ & 1.04 &  \\
3218264 & 64.156887 & -24.126996 & 9.19 & $9.45^{+0.42}_{-0.31}$ & - & -19.53 & $19.94 \pm 1.43$ & 1.04 &  \\
3218914 & 64.136973 & -24.111670 & 8.27 & $8.25^{+0.11}_{-0.12}$ & - & -20.59 & $58.52 \pm 2.21$ & 1.06 & (3,6) \\
3219059 & 64.151687 & -24.108256 & 8.25 & $8.28^{+0.22}_{-0.16}$ & - & -18.86 & $18.00 \pm 1.36$ & 1.05 & (1,3,5,6) \\
3219591 & 64.165151 & -24.094526 & 8.92 & $8.92^{+0.22}_{-0.22}$ & - & -19.19 & $20.89 \pm 1.09$ & 1.05 &  \\
4112911 & 215.958426 & 24.118182 & 8.31 & $8.31^{+0.29}_{-0.29}$ & - & -19.15 & $25.00 \pm 1.89$ & 1.35 &  \\
4113315 & 215.939736 & 24.121318 & 9.73 & $9.70^{+0.27}_{-0.26}$ & 9.655 & -19.41 & $20.71 \pm 1.29$ & 1.18 &  \\
4113384 & 215.938375 & 24.121828 & 8.86 & $9.03^{+0.71}_{-0.42}$ & - & -18.86 & $12.60 \pm 1.36$ & 1.17 &  \\
4113832 & 215.939441 & 24.126061 & 10.31 & $10.05^{+0.62}_{-1.20}$ & - & -19.27 & $21.41 \pm 1.89$ & 1.15 &  \\
4114040 & 215.922381 & 24.127850 & 7.86 & $7.96^{+0.30}_{-0.16}$ & - & -18.88 & $15.02 \pm 1.51$ & 1.08 &  \\
4114206 & 215.923251 & 24.128951 & 8.68 & $8.64^{+0.24}_{-0.26}$ & - & -19.52 & $24.30 \pm 1.74$ & 1.08 &  \\
4114291 & 215.940737 & 24.129747 & 9.62 & $9.93^{+0.51}_{-0.39}$ & - & -18.94 & $12.01 \pm 1.31$ & 1.13 &  \\
4114965 & 215.950579 & 24.135571 & 8.06 & $8.07^{+0.16}_{-0.14}$ & - & -19.60 & $31.68 \pm 2.07$ & 1.13 &  \\
4118874 & 215.934747 & 24.110364 & 7.62 & $7.61^{+0.05}_{-0.07}$ & - & -18.93 & $16.45 \pm 1.87$ & 1.23 &  \\
4201498 & 215.890787 & 24.108984 & 7.80 & $7.86^{+0.27}_{-0.22}$ & - & -18.63 & $11.71 \pm 1.15$ & 1.04 &  \\
4201636 & 215.873846 & 24.110101 & 7.91 & $7.91^{+0.08}_{-0.08}$ & - & -19.27 & $18.01 \pm 1.22$ & 1.03 &  \\
4201776 & 215.889971 & 24.111038 & 8.30 & $8.24^{+0.14}_{-0.16}$ & - & -19.72 & $25.87 \pm 1.13$ & 1.04 &  \\
4203231 & 215.893737 & 24.118722 & 9.92 & $9.77^{+0.26}_{-0.39}$ & - & -18.69 & $9.58 \pm 0.86$ & 1.04 &  \\
4205103 & 215.870856 & 24.128550 & 8.06 & $8.07^{+0.07}_{-0.08}$ & - & -19.18 & $17.99 \pm 0.88$ & 1.02 &  \\
4205222 & 215.891816 & 24.129343 & 8.07 & $8.07^{+0.07}_{-0.07}$ & - & -19.63 & $24.50 \pm 0.93$ & 1.03 &  \\
4207023 & 215.876638 & 24.140908 & 8.11 & $8.11^{+0.08}_{-0.07}$ & - & -21.22 & $122.45 \pm 1.33$ & 1.02 &  \\
4210412 & 215.873331 & 24.160027 & 7.90 & $8.13^{+0.36}_{-0.32}$ & - & -18.25 & $8.69 \pm 1.06$ & 1.01 &  \\
4210846 & 215.862895 & 24.162029 & 7.97 & $8.01^{+0.21}_{-0.13}$ & - & -19.19 & $17.11 \pm 0.93$ & 1.01 &  \\
4210883 & 215.874228 & 24.162117 & 7.63 & $7.61^{+0.10}_{-0.51}$ & - & -19.75 & $36.61 \pm 1.16$ & 1.01 &  \\
4211130 & 215.883412 & 24.163096 & 8.31 & $8.38^{+1.93}_{-0.38}$ & - & -18.08 & $8.68 \pm 0.88$ & 1.01 &  \\
4211820 & 215.875738 & 24.165635 & 8.93 & $8.84^{+0.17}_{-0.34}$ & - & -19.07 & $13.91 \pm 0.96$ & 1.01 &  \\
4212044 & 215.886091 & 24.166451 & 8.03 & $8.02^{+0.08}_{-0.10}$ & - & -19.64 & $24.51 \pm 0.95$ & 1.01 &  \\
4212544 & 215.882484 & 24.168284 & 7.97 & $9.27^{+1.04}_{-1.10}$ & - & -18.05 & $9.55 \pm 0.89$ & 1.01 &  \\
4212771 & 215.884224 & 24.169069 & 7.70 & $7.75^{+0.31}_{-0.24}$ & - & -18.28 & $9.04 \pm 0.89$ & 1.01 &  \\
4212932 & 215.864791 & 24.169628 & 7.81 & $7.85^{+0.18}_{-0.13}$ & - & -19.22 & $20.31 \pm 1.33$ & 1.01 &  \\
4214058 & 215.886237 & 24.173433 & 8.79 & $8.80^{+0.12}_{-0.11}$ & - & -20.60 & $73.84 \pm 1.68$ & 1.01 &  \\
4214066 & 215.853540 & 24.173379 & 7.98 & $8.85^{+1.68}_{-0.81}$ & - & -18.35 & $9.60 \pm 0.92$ & 1.01 &  \\
4214507 & 215.882922 & 24.174927 & 8.97 & $9.10^{+1.49}_{-0.50}$ & - & -18.99 & $19.04 \pm 1.44$ & 1.01 &  \\
4214522 & 215.882684 & 24.174998 & 8.70 & $8.70^{+0.12}_{-0.12}$ & - & -20.58 & $58.72 \pm 1.41$ & 1.01 &  \\
4214564 & 215.882719 & 24.175127 & 8.64 & $8.66^{+0.38}_{-0.36}$ & - & -18.76 & $11.10 \pm 1.24$ & 1.01 &  \\
4215268 & 215.859036 & 24.177485 & 7.64 & $7.64^{+0.11}_{-0.17}$ & - & -19.15 & $17.10 \pm 1.05$ & 1.01 &  \\
4215592 & 215.856683 & 24.178440 & 7.61 & $7.41^{+0.30}_{-0.39}$ & - & -18.59 & $10.09 \pm 0.90$ & 1.01 &  \\
4215682 & 215.884639 & 24.178729 & 9.31 & $9.34^{+0.36}_{-0.17}$ & - & -19.44 & $19.60 \pm 1.05$ & 1.01 &  \\
4217489 & 215.869228 & 24.184467 & 7.58 & $7.52^{+0.21}_{-0.27}$ & - & -18.77 & $14.18 \pm 1.17$ & 1.01 &  \\
4217773 & 215.860830 & 24.185518 & 7.75 & $7.87^{+0.25}_{-0.19}$ & - & -18.80 & $17.20 \pm 1.09$ & 1.01 &  \\
4218166 & 215.877709 & 24.187381 & 7.50 & $7.39^{+0.16}_{-0.21}$ & - & -19.40 & $26.20 \pm 1.35$ & 1.01 &  \\
4218306 & 215.884780 & 24.188228 & 8.88 & $8.80^{+0.12}_{-0.17}$ & - & -20.21 & $52.23 \pm 1.21$ & 1.01 &  \\
4218762 & 215.895004 & 24.110801 & 8.30 & $8.33^{+0.19}_{-0.15}$ & - & -19.48 & $17.46 \pm 1.42$ & 1.05 &  \\
4218763 & 215.895056 & 24.110877 & 8.30 & $8.31^{+0.12}_{-0.09}$ & - & -20.20 & $39.04 \pm 1.90$ & 1.05 &  \\
4218764 & 215.895106 & 24.110848 & 8.28 & $8.38^{+0.28}_{-0.15}$ & - & -19.25 & $17.18 \pm 1.17$ & 1.05 &  \\
4218854 & 215.863002 & 24.114390 & 9.00 & $9.00^{+0.07}_{-0.07}$ & - & -20.12 & $36.75 \pm 0.95$ & 1.02 &  \\
4218855 & 215.863059 & 24.114449 & 9.08 & $9.08^{+0.08}_{-0.07}$ & - & -20.49 & $48.85 \pm 1.10$ & 1.02 &  \\
4220075 & 215.860252 & 24.156177 & 8.47 & $8.47^{+0.14}_{-0.14}$ & - & -19.95 & $32.95 \pm 1.59$ & 1.01 &  \\
4220076 & 215.860227 & 24.156280 & 8.32 & $8.37^{+0.19}_{-0.12}$ & - & -19.57 & $27.48 \pm 1.50$ & 1.01 &  \\
4220147 & 215.892082 & 24.157992 & 8.44 & $8.62^{+0.44}_{-0.38}$ & - & -18.99 & $12.87 \pm 1.34$ & 1.02 &  \\
4220148 & 215.892211 & 24.158065 & 8.27 & $8.30^{+0.17}_{-0.16}$ & - & -19.50 & $30.92 \pm 1.77$ & 1.02 &  \\
4220638 & 215.859528 & 24.174061 & 7.60 & $7.61^{+0.17}_{-0.39}$ & - & -19.12 & $19.85 \pm 1.15$ & 1.01 &  \\
4220639 & 215.859597 & 24.174106 & 7.60 & $7.61^{+0.02}_{-0.02}$ & - & -20.25 & $47.91 \pm 1.15$ & 1.01 &  \\
4220749 & 215.859113 & 24.177734 & 7.62 & $7.62^{+0.04}_{-0.03}$ & - & -20.80 & $82.86 \pm 1.20$ & 1.01 &  \\
5101296 & 177.379427 & 22.345013 & 9.62 & $9.77^{+0.38}_{-0.30}$ & - & -19.47 & $19.14 \pm 1.68$ & 1.04 &  \\
5101971 & 177.347572 & 22.351317 & 9.57 & $9.76^{+0.51}_{-0.44}$ & - & -19.30 & $15.02 \pm 1.65$ & 1.03 &  \\
5104911 & 177.359468 & 22.375532 & 8.61 & $8.61^{+0.40}_{-0.40}$ & - & -19.52 & $30.38 \pm 2.60$ & 1.09 &  \\
5112687 & 177.390929 & 22.349767 & 8.73 & $8.74^{+0.16}_{-0.15}$ & - & -20.50 & $71.38 \pm 1.79$ & 1.06 &  \\
5200958 & 177.401686 & 22.263621 & 8.63 & $8.62^{+0.28}_{-0.28}$ & - & -19.05 & $13.17 \pm 1.02$ & 1.00 &  \\
5201413 & 177.389141 & 22.268134 & 10.52 & $10.54^{+0.46}_{-0.60}$ & - & -19.00 & $9.92 \pm 1.17$ & 1.00 &  \\
5201525 & 177.381010 & 22.269275 & 9.20 & $9.20^{+0.09}_{-0.07}$ & - & -19.78 & $22.88 \pm 1.03$ & 1.00 &  \\
5201697 & 177.391878 & 22.270752 & 8.34 & $8.64^{+1.30}_{-0.46}$ & - & -18.50 & $9.43 \pm 1.14$ & 1.00 &  \\
5201859 & 177.373166 & 22.272067 & 7.71 & $7.74^{+0.15}_{-0.10}$ & - & -19.44 & $22.21 \pm 1.26$ & 1.00 &  \\
5202022 & 177.396431 & 22.273188 & 8.12 & $8.14^{+0.30}_{-0.29}$ & - & -18.95 & $12.41 \pm 1.00$ & 1.00 &  \\
5202132 & 177.367460 & 22.274077 & 7.88 & $7.91^{+0.32}_{-0.65}$ & - & -18.14 & $8.01 \pm 0.92$ & 1.00 &  \\
5202535 & 177.374516 & 22.277168 & 8.28 & $8.28^{+0.09}_{-0.09}$ & - & -20.76 & $69.42 \pm 1.12$ & 1.00 &  \\
5202881 & 177.375474 & 22.279778 & 8.53 & $8.54^{+0.18}_{-0.16}$ & - & -19.55 & $22.95 \pm 1.15$ & 1.00 &  \\
5203078 & 177.374286 & 22.281325 & 8.17 & $8.18^{+0.19}_{-0.18}$ & - & -18.93 & $12.05 \pm 0.89$ & 1.00 &  \\
5203250 & 177.382348 & 22.282596 & 7.63 & $7.62^{+0.03}_{-0.02}$ & - & -20.82 & $76.69 \pm 0.98$ & 1.01 &  \\
5203428 & 177.388358 & 22.283808 & 7.72 & $7.79^{+0.23}_{-0.15}$ & - & -18.52 & $8.73 \pm 0.79$ & 1.01 &  \\
5203757 & 177.408239 & 22.286122 & 11.45 & $11.41^{+0.19}_{-0.25}$ & - & -19.90 & $21.75 \pm 0.83$ & 1.01 &  \\
5204117 & 177.384174 & 22.288789 & 7.79 & $7.78^{+0.08}_{-0.04}$ & - & -19.41 & $21.44 \pm 0.95$ & 1.01 &  \\
5204130 & 177.398746 & 22.288870 & 9.45 & $9.59^{+0.30}_{-0.21}$ & - & -19.40 & $16.70 \pm 1.15$ & 1.01 &  \\
5204940 & 177.396918 & 22.294806 & 8.07 & $8.11^{+0.21}_{-0.15}$ & - & -18.84 & $14.38 \pm 0.96$ & 1.01 &  \\
5205612 & 177.390921 & 22.299783 & 9.44 & $9.45^{+0.38}_{-0.36}$ & - & -18.94 & $11.17 \pm 1.13$ & 1.01 &  \\
5206609 & 177.407145 & 22.307713 & 9.22 & $9.26^{+0.58}_{-0.37}$ & - & -18.79 & $12.37 \pm 1.11$ & 1.01 &  \\
5207945 & 177.402297 & 22.318279 & 8.35 & $8.31^{+0.35}_{-0.29}$ & - & -18.86 & $15.60 \pm 1.01$ & 1.02 &  \\
5208093 & 177.433843 & 22.319529 & 9.02 & $8.95^{+0.19}_{-0.31}$ & - & -18.89 & $11.30 \pm 0.80$ & 1.02 &  \\
5208921 & 177.421132 & 22.327848 & 8.03 & $8.04^{+0.11}_{-0.10}$ & - & -19.62 & $39.18 \pm 1.10$ & 1.03 &  \\
5208946 & 177.413678 & 22.328065 & 8.91 & $9.53^{+0.71}_{-0.74}$ & - & -18.97 & $14.87 \pm 1.30$ & 1.03 &  \\
5209073 & 177.435346 & 22.329424 & 7.86 & $7.87^{+0.09}_{-0.07}$ & - & -19.74 & $28.12 \pm 1.08$ & 1.03 &  \\
5209178 & 177.419894 & 22.330538 & 7.73 & $8.01^{+0.37}_{-0.31}$ & - & -18.35 & $8.30 \pm 0.80$ & 1.03 &  \\
5209189 & 177.431494 & 22.330610 & 8.06 & $8.21^{+0.44}_{-0.30}$ & - & -18.47 & $9.87 \pm 1.11$ & 1.03 &  \\
5209212 & 177.420093 & 22.330898 & 8.17 & $8.18^{+0.11}_{-0.10}$ & - & -20.24 & $41.77 \pm 1.26$ & 1.03 &  \\
5209218 & 177.420252 & 22.330900 & 8.13 & $8.32^{+0.55}_{-0.30}$ & - & -18.53 & $10.36 \pm 1.27$ & 1.03 &  \\
5209312 & 177.421043 & 22.331985 & 7.85 & $7.89^{+0.32}_{-0.26}$ & - & -18.23 & $8.89 \pm 1.03$ & 1.03 &  \\
5209389 & 177.427298 & 22.332885 & 7.94 & $8.00^{+0.25}_{-0.20}$ & - & -18.49 & $9.25 \pm 0.98$ & 1.04 &  \\
5209485 & 177.427114 & 22.334085 & 7.63 & $7.50^{+0.30}_{-0.38}$ & - & -18.22 & $8.40 \pm 0.82$ & 1.04 &  \\
5209557 & 177.425380 & 22.335121 & 8.38 & $8.33^{+0.12}_{-0.17}$ & - & -20.03 & $47.71 \pm 1.35$ & 1.04 &  \\
5209629 & 177.424176 & 22.335956 & 8.12 & $8.13^{+0.16}_{-0.13}$ & - & -20.14 & $42.67 \pm 0.95$ & 1.04 &  \\
5209975 & 177.419641 & 22.340956 & 7.70 & $7.69^{+0.08}_{-0.05}$ & - & -19.85 & $34.79 \pm 1.25$ & 1.05 &  \\
5212161 & 177.406111 & 22.280154 & 9.33 & $9.33^{+0.07}_{-0.09}$ & - & -20.40 & $58.40 \pm 1.86$ & 1.01 & (6) \\
5212846 & 177.386361 & 22.292660 & 8.38 & $8.30^{+0.46}_{-0.31}$ & - & -19.34 & $20.71 \pm 1.36$ & 1.01 &  \\
5213494 & 177.422920 & 22.304999 & 8.02 & $8.03^{+0.17}_{-0.14}$ & - & -19.51 & $25.66 \pm 1.27$ & 1.01 & (6) \\
5213496 & 177.413900 & 22.305118 & 7.61 & $8.20^{+0.64}_{-1.10}$ & - & -18.65 & $13.87 \pm 1.07$ & 1.01 &  \\
5213805 & 177.406367 & 22.311639 & 8.32 & $8.34^{+0.37}_{-0.28}$ & - & -19.27 & $30.16 \pm 1.84$ & 1.01 &  \\
5214565 & 177.409300 & 22.329284 & 7.69 & $7.64^{+0.25}_{-0.73}$ & - & -18.83 & $13.79 \pm 1.11$ & 1.03 &  \\
5220098 & 177.391054 & 22.290679 & 9.80 & $9.81^{+0.37}_{-0.35}$ & - & -19.04 & $11.39 \pm 1.36$ & 1.01 &  \\
5220925 & 177.384506 & 22.300022 & 8.44 & $8.44^{+0.42}_{-0.43}$ & - & -18.54 & $9.50 \pm 0.99$ & 1.01 & \\ 
\hline
\caption{Basic information for all 158 galaxies at $z>7.5$ in the CANUCS LF sample (CLU module A and NCF). The first two digits of the ID number encode the cluster and field information as follows; first digit: 1 MACS0417, 2 Abell370, 3 MACS0416, 4 MACS1423, 5 MACS1149, second digit: 1 CLU field, 2 NCF field.   $z_{\rm phot\_ML}$ is the EAZY-py photometric redshift maximum likelihood value. The $z_{\rm phot\_50}$ column gives the EAZY-py probability distribution median with 16\% and 84\% uncertainties. $z_{\rm spec}$ gives the spectroscopic redshift for those sources observed with NIRSpec. $M_{\rm UV}$ is the UV absolute magnitude that has been corrected for gravitational lensing magnification given in the magnification column. The F277W fluxes are total observed fluxes corrected from a 0.3 arcsec aperture as described in Section \ref{sec:data}. A small fraction of the galaxies have been previously identified to be high-redshift based on Hubble Frontier Field Parallel imaging. References for these galaxies are (1) \cite{Coe2015}, (2) \cite{Finkelstein2015}, (3) \cite{Infante2015}, (4) \cite{Harikane2016}, (5) \cite{McLeod2016}, (6) \cite{Ishigaki2018}.}
\label{tab:sample}
\end{longtable}

\end{document}